\documentclass[useAMS,usenatbib,usegraphicx]{mn2e}

\pdfpagewidth=\paperwidth
\pdfpageheight=\paperheight
 
\newcommand{\ion}[2]{\mbox{#1\,\textsc{#2}}}
\newcommand{\forbid}[3]{\mbox{[#1\,\textsc{#2}]\ensuremath{_{#3}}}}
\newcommand{\Halpha}{\mbox{H\ensuremath{\alpha}}}
\newcommand{\Hbeta}{\mbox{H\ensuremath{\beta}}}

\usepackage[T1]{fontenc}
\usepackage{aecompl}

\usepackage{xfrac}
\usepackage{color}
\usepackage{amsmath}
\definecolor{grey}{gray}{0.4}
\definecolor{plum}{RGB}{146,38,143}
\usepackage{multicol}

\pdfpageattr{/Group <</S /Transparency /I true /CS /DeviceRGB>>}

\title[Bluedisk: Gas-phase metallicity profiles]{Gas-phase metallicity profiles of the Bluedisk galaxies: Is metallicity in a local star-formation regulated equilibrium?}
\author[David Carton et al.]{
David~Carton$^{1}$\thanks{E-mail:carton@strw.leidenuniv.nl},
Jarle~Brinchmann$^{1}$,
Jing~Wang$^{2}$,
Frank~Bigiel$^{3}$,
Diane~Cormier$^{3}$,\newauthor{}
Thijs~van~der~Hulst$^{4}$,
Gyula~I.~G.~J\'{o}zsa$^{5}$,
Paolo~Serra$^{2}$,
Marc~A.~W.~Verheijen$^{4}$\\
$^{1}$Leiden Observatory, Leiden University, PO Box 9513, 2300 RA, Leiden, The Netherlands\\
$^{2}$CSIRO Astronomy and Space Science, Australia Telescope National Facility, PO Box 76, Epping, NSW 1710, Australia\\
$^{3}$Institut f\"{u}r theoretische Astrophysik, Zentrum f\"{u}r Astronomie der Universit\"{a}t Heidelberg, Albert-\"{U}berle-Str. 2, 69120, Heidelberg, Germany\\
$^{4}$Kapteyn Astronomical Institute, University of Groningen, PO Box 800, 9700 AV, Groningen, The Netherlands\\
$^{5}$SKA South Africa, 3rd Floor, The Park, Park Road, Pinelands, 7405, South Africa\\
}
\begin{document}

\date{Accepted 2015 April 29. Received 2015 April 29; in original form 2015 January 28}

\pagerange{\pageref{firstpage}--\pageref{lastpage}} \pubyear{YEAR}

\maketitle

\label{firstpage}

\begin{abstract}
As part of the Bluedisk survey we analyse the radial gas-phase metallicity profiles of 50 late-type galaxies.
We compare the metallicity profiles of a sample of \ion{H}{i}-rich galaxies against a control sample of \ion{H}{i}-`normal' galaxies.
We find the metallicity gradient of a galaxy to be strongly correlated with its \ion{H}{i} mass fraction ($\textrm{M(H\,\textsc{i})} / \textrm{M}_{\ast}$).
We note that some galaxies exhibit a steeper metallicity profile in the outer disc than in the inner disc.
These galaxies are found in both the \ion{H}{i}-rich and control samples. This contradicts a previous indication that these outer drops are exclusive to \ion{H}{i}-rich galaxies.
These effects are not driven by bars, although we do find some indication that barred galaxies have flatter metallicity profiles.
By applying a simple analytical model we are able to account for the variety of metallicity profiles that the two samples present.
The success of this model implies that the metallicity in these isolated galaxies may be in a local equilibrium, regulated by star formation.
This insight could provide an explanation of the observed local mass-metallicity relation.
\end{abstract}


\section{Introduction}

Galaxy formation has been much studied over the past decades, but despite significant successes in this endeavour, it remains unclear exactly how disc galaxies evolve at late times.
A particular stumbling block has been determining exactly through which processes a galaxy acquires its cold gas.
One of the current leading scenarios is the accretion of gas into the galaxy halo.
This halo gas subsequently cools to form a gas disc, from which stars will form \citep{1978MNRAS.183..341W, 1980MNRAS.193..189F, 1998MNRAS.295..319M}.
It is predicted that the angular momentum of the accreting gas will grow over time.
The angular momentum of the gas disc will therefore also increase, and thus gas cooling from this will settle at increasing radii.
This paradigm is commonly referred to as ``inside-out'' growth \citep{2011MNRAS.418.2493P, 2013ApJ...769...74S}.

The study of the formation and evolution of disc galaxies is complicated by the complex nature of star formation and the cycle of gas within the interstellar medium of galaxies.
Nevertheless, with three fundamental observables, namely stellar mass, gas-phase metallicity\footnote{By gas-phase metallicity we refer to the oxygen abundance ($12 + \log_{10}\left(\mathrm{O}/\mathrm{H}\right)$) of the interstellar medium (ISM).} and the star-formation rate (SFR), we can begin to unravel the life of galaxies.
The gas-phase metallicity, herein simply referred to as metallicity, is of particular interest since it is not simply a result of star formation integrated through time, but it is also strongly affected by gas flows into and out from the galaxy.
With the advent of large spectroscopic fibre surveys, such as the Sloan Digital Sky Survey (SDSS) \citep{2000AJ....120.1579Y} and Galaxy And Mass Assembly project (GAMA) \citep{2011MNRAS.413..971D}, a host of studies have explored the relationships between these three aforementioned parameters.
Of particular interest is the mass-metallicity relation \citep{2004ApJ...613..898T, 2012A&A...547A..79F}, which shows the most massive galaxies to be also the most metal rich.
This correlation is commonly attributed to either a downsizing scenario, whereby the most massive galaxies are more efficiently forming stars, or alternatively that galactic-scale winds are more effective at expelling metals from lower mass galaxies.
For a more in depth discussion of these and other mechanisms we refer the reader to \citet{2008ApJ...672L.107E}.

More recently studies have reported a secondary correlation of the mass-metallicity relation with the SFR, forming the so-called fundamental metallicity relation (FMR) \citep{2010MNRAS.408.2115M}.
The FMR presents an anti-correlation of metallicity with the SFR, which has been attributed to either inflows that dilute metallicity and/or outflows that remove metals.
We hasten to point out, however, that the FMR is not without contest, with some debate over its origin or existence \citep{2014ApJ...789L..40W, 2013A&A...554A..58S}.

While we may have copious measurements for the central metallicities of galaxies, comparatively less well studied are radial trends of metallicity in galaxies.
Early on the picture emerged that at late times ($z \la 0.1$) disc galaxies all show negative (declining radially outwards) metallicity gradients, and when expressed in terms of optical scale radii they showed remarkably similar gradients \citep{1992MNRAS.259..121V, 1994ApJ...420...87Z}.
Recently \citet{2015MNRAS.448.2030H} have shown that this common metallicity gradient can be explained by the coevolution of gas, metals and stars.

The common metallicity gradient, however, only applies to isolated galaxies.
In a study of interacting systems, \citet{2012ApJ...753....5R} showed a clear tendency towards flatter metallicity gradients, which for the early stages of interaction is consistent with simulations \citep{2010ApJ...710L.156R, 2012ApJ...746..108T}.

Despite this work it is only recently that metallicity gradients have been systematically determined for large samples of galaxies \citep{2012ApJ...745...66M, 2013A&A...554A..58S}.
With the good statistics these studies provided, these authors have shown the existence of a correlation between stellar-mass density and metallicity.
This correlation is commonly referred to the local mass-metallicity relation and, as with its global counterpart, its origin is unclear.

In the work of \citet{2012ApJ...745...66M}, whose galaxies formed part of the GALEX Arecibo SDSS Survey (GASS) \citep{2010MNRAS.403..683C}, they attempt to connect metallicity to the atomic gas (\ion{H}{i}) content of galaxies.
They show tentative hints that the most \ion{H}{i}-rich galaxies exhibit sudden drops in metallicity in their outer discs.
It is these hints that provided impetus for the work we present here.

In this paper we present resolved metallicity profiles for all 50 low redshift ($z \sim 0.025$) galaxies that form the Bluedisk survey \citep[][herein Paper I]{2013MNRAS.433..270W}.
The resolved \ion{H}{i} maps are the cornerstone of Bluedisk project, providing both the structure and kinematics of the atomic gas disc.
The goal of the Bluedisk project is to study in detail two classes of galaxies: an ``\ion{H}{i}-rich'' sample, consisting of those galaxies with stellar masses above $10^{10}\,\textrm{M}_\odot$ and with excess atomic gas, and for comparison a ``control'' sample consisting of galaxies of similar stellar mass, whose \ion{H}{i} content is normal or mildly poor.

We structure this paper as follows: in Section~\ref{sec:data} we outline the existing data of the Bluedisk galaxies.
In Section~\ref{sec:obervations} we descrive our observations and discuss our data-reduction process.
In Section~\ref{sec:analysis} we detail our spectral fitting procedures, discuss the global properties of our galaxy population.
We also explain our method for determination of metallicity, among other quantities.
In Section~\ref{sec:results} we present our results, focusing on the resolved metallicity of the Bluedisk galaxies.
We use Section~\ref{sec:discussion} to develop and apply a simple a model to explain the radial metallicity  profiles in terms of their gas and stellar mass contents.
Finally, we provide our concluding remarks in \ref{sec:conclusions}.
Throughout this paper we assume a cosmology with $H_0 = 70\ \textrm{km}\,\textrm{s}^{-1}\,\textrm{Mpc}^{-1}$, $\Omega_{\mathrm{m}} = 0.3$ and $\Omega_{\Lambda} = 0.7$.

\section{Data}\label{sec:data}

Measurements of the \ion{H}{i} content of the Bluedisk galaxies have been obtained using the Westerbork Synthesis Radio Telescope (WSRT) with observations for 49 out of the sample of 50 galaxies, including one non-detection.
SDSS images have been analysed to provide optical properties of the galaxies.
A full description of the analysis is available in Paper I.
We will make use of this data in the context of our new optical spectroscopic data.

\subsection{Bluedisk galaxy classification}

A key aspect of the Bluedisk strategy is the classification of galaxies into two well-matched \ion{H}{i}-rich and control samples.
As uncovered by \citet{2010MNRAS.403..683C} there exists a scaling relation between the \ion{H}{i} mass fraction ($f_{H\textsc{i}} = M_{H\textsc{i}} / M_\ast$), the stellar mass surface density and the observed $\textrm{NUV}-r$ color.
Using the difference between the observed and expected \ion{H}{i} mass fractions, we can bisect the Bluedisk population into \ion{H}{i}-rich and control samples.
We require the samples to contain only isolated galaxies, thus an additional category of non-isolated galaxies has been formed, namely the ``excluded'' sample.
All three samples are described at length in Paper I.
We make, however, one minor modification to the classifications listed therein, such that in this paper we consider that BD 39, formerly part of the control sample, to now be a member of the excluded sample.
Our motivation for this being that we have identified potentially a small neighbouring galaxy at the same redshift.
As a result, this leaves 23 \ion{H}{i}-rich galaxies and 18 control galaxies, which we will focus our attention on.

\subsection{Bulge-Disc-Bar decomposition}\label{subsec:bulge_disc_bar_decomposition}

We determine the inclination of the stellar disc using SDSS \textit{r} band images.
With these images we also perform a bulge-disc-bar decomposition, enabling use to measure a bulge-to-total ratio for each galaxy.

Our procedure for bulge-disc-bar decomposition mostly follows that of \citet{2009ApJ...696..411W}. The steps are as follows:
\begin{enumerate}
\item We use the \texttt{SExtractor} software \citep{1996A&AS..117..393B} to measure the position, ellipticity ($e$) and position angle (PA) for each galaxy. We also make mask images that flag all neighbouring sources using the segmentation map produced by SExtractor. These masks are used in all the following isophote and model fitting steps. 

\item Using the photometric measurements as an initial guess, we perform isophote ellipse fitting on the images, and obtain surface density, PA and $e$ profiles for the galaxies.
With the surface density profile, we accurately measure the background surface density of the images.
The shape (size, PA and $e$) of the outermost isophote is viewed as the shape of the galactic disc. 
See \citet{2012MNRAS.423.3486W} for futher details of this step.

\item We use the \texttt{GALFIT} package \citep{2002AJ....124..266P} to fit models of the bulge, disc and bar to the galaxies. We use exponential models to represent the discs, and use S\'ersic models to represent bulges and bars. The S\'ersic index is allowed to vary within 1.2 to 6 for bulge models and vary within 0.3 to 0.9 for bar models. These are typical values reported in \citet{2009MNRAS.399..621G}.
We first fit a single bulge model, followed by a bulge$+$disc model and finally bulge$+$disc$+$bar model. The model obtained from each fitting step is used as initial guess for the following fitting step.
When fitting discs and bars, the size, PA and $e$ measurements  from \mbox{step 1} are used as initial guess. During the fitting, the PA of bars and discs and are allowed to vary in a very small range ($\pm20\degr$). The $e$ of the disc is fixed, and the $e$ of bar is required to be smaller than the $e$ of disc. So in the end, we have 3 types of models (bulge, bulge$+$disc and bulge$+$disc$+$bar) for each galaxy. We choose the model with the minimum value of reduced $\chi^2$ calculated from the residual map as the best model. 
\end{enumerate}

\subsection{Stellar mass densities}\label{subsec:stellar_mass_densities}

To derive spatial resolved maps of the stellar mass density, $\Sigma_{\ast}$, we use SDSS \textit{ugriz} photometry.

We fit the five-band SDSS photometry using the composite stellar population synthesis (SPS) models of BC03, applying the procedure described in \citet{2003MNRAS.341...33K} and \citet{2005MNRAS.362...41G}.
The SPS models combine an underlying exponentially declining star formation history with random bursts of star formation superposed on this.
The modelling includes a dust component.
The flux from young stars ($<10$\,Myr) is attenuated following a dust attenuation curve of the form $\tau(\lambda) \propto \lambda^{-1.3}$.
Whereas, the flux from long-lived stars ($>10$\,Myr) is attenuated by a $\tau(\lambda) \propto \lambda^{-0.7}$ power law.
The library is described in more detail in \citet{2005MNRAS.362...41G}.
From the results of this SPS fitting we obtain a posterior distribution on $\Sigma_{\ast}$.

To perform this SPS modelling we require a good S/N across the images.
We achieve this by using the weighted Voronoi tessellation method of \citet{2006MNRAS.368..497D}, a generalization of the algorithm by \citet{2003MNRAS.342..345C}.
We define our measure of S/N from that of the $u-z$ colour maps (the $u-z$ combination typically offering the poorest S/N).
Adopting a threshold $\textrm{S/N} = 5$ we therefore ensure a good S/N in all colour maps.
As the SDSS images contain foreground objects we mask these objects manually, along with any other spurious features.

\section{Observations}\label{sec:obervations}
Optical long-slit spectroscopic observations of all 50 Bluedisk galaxies were performed in January and May 2013 using the Intermediate dispersion Spectrograph and Imaging System (ISIS) on the 4.2\,m William Herschel Telescope (WHT), in a variety of seeing conditions (0.7--1.7\,arcsec FWHM).
The ISIS spectrograph was operated in a dual arm mode using the standard 5300\,\AA\ dichroic, with the GG495 blocking filter in the red arm.
Employing the R600B and R600R gratings in the blue and red arms respectively, a discontinuous spectral coverage of 3700--5300, 5750--7200\,\AA\ was provided, with a spectral resolution of $\sim$1.7\,\AA\ FWHM constant across all wavelengths.
Each target was observed with a minimum of $3\times1200$\,s exposures.
A slit width of 3\,arcsec was used for all observations presented here, optimizing emission line signal-to-noise (S/N) at the expense of spectral resolution.

Each spectroscopic slit was positioned to coincide with the centre of the galaxy, as defined by the SDSS photometric catalogue.
The orientation of the spectroscopic slit was aligned to the kinematic major axis determined from the WSRT velocity moment maps.
Where this angle was close to that of a clear natural axis of the galaxy, the slit was more precisely aligned with this optical feature.
A final design requirement was to ensure a region observed by a 3\,arcsec fibre in the SDSS spectroscopic catalogue was included in the slit, which for all but one target (BD~31), was at the galaxy centre.
Overall, the general result was such that the slits were aligned with the optical major axis.

Standard bias frames were obtained for each night, in addition to lamp flat and twilight sky flat exposures.
After each target pointing, additional spectroscopic calibration images were obtained, which comprised \mbox{Cu-Ar+Cu-Ne} arc-lamp exposures, as well as an observation of a standard star.
The standard stars were selected from the ING spectrophotometric catalogue\footnote{http://catserver.ing.iac.es/landscape/tn065-100/workflux.php} and were observed at a similar airmass to the targeted galaxy.
Finally, for characterization of the charge-coupled devices (CCDs), a set of dark frames were also gathered.

\subsection{Reduction}

Standard \texttt{PyRAF} tasks were used to calculate the CCD bias offsets, pixel gain variations and telescope vignetting effects.
With the \texttt{imcombine} task cosmic-ray rejection was performed using a mean image combine and a $+$3\,$\sigma$ rejection.
Bad-pixel masks were constructed manually, based upon the dark and lamp-flat frames.
Wavelength calibration was performed using a custom routine, which fitted the arc-lamp spectra at multiple points along the spatial axis.
With the use a 2D-spline interpolation, the corresponding wavelength of every pixel was identified.
Subsequent sky subtraction was carried out using blank regions in slit. 

The dispersion axis of the spectrograph was not perfectly aligned with the CCD axes, this rotation, while small, induced a small ($\la$\,4\,arcsec) shift in spatial position in the spatial CCD coordinates between the wavelength extrema.
To remove this effect the centre of the target was traced by binning the spectrum in numerous wavelength bins, and fitting a symmetric profile to the spatial intensity distribution in each bin. S\'{e}rsic and Gaussian profiles were assumed for galaxy and standard star targets, respectively.
A linear fit to this produced a mapping of wavelength to object centre, from which a rectified 2D spectrum was created.

Flux calibration was performed in a two-step procedure.
Firstly an absolute flux calibration by comparing the response of the standard star against the reference spectrum.
This was refined by extracting a spectrum from a 3\,arcsec square effective aperture matched to the equivalent position of the SDSS fibre spectrum (3\,arcsec diameter).
We additionally applied a telluric correction by measuring the transmission of the standard star, adopting a linear pseudo-continuum across the affected spectral regions. 

\begin{figure}
\includegraphics[width=\linewidth]{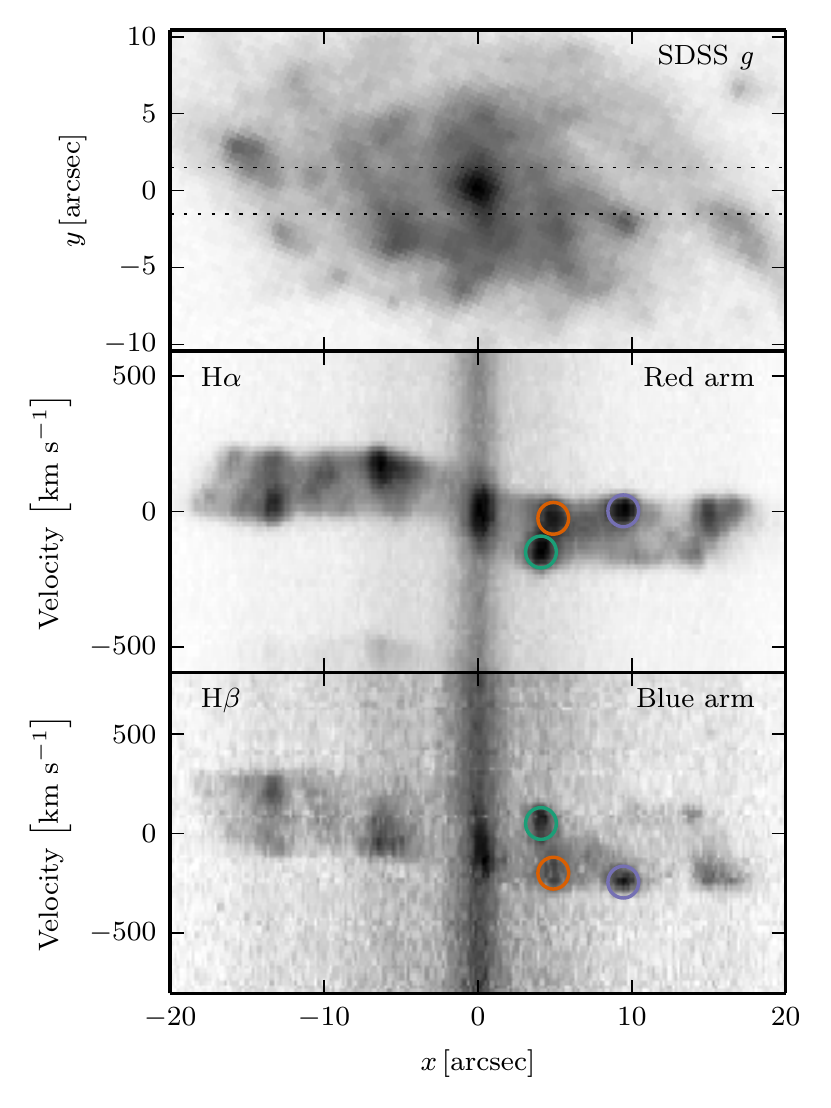}
\caption{Illustration the ``contrary offset'' effect of the ISIS instrument, due to the opposite alignment of the dispersion axes of the red and blue spectral arms. (Top) SDSS \textit{g} band image of \mbox{BD 20} with 3\,arcsec wide slit indicated by the dotted horizontal lines. (Middle and Bottom) Corresponding 2D spectra centred about the \Halpha{} and \Hbeta{} emission line features of the red and blue arms. While the overall velocity curve is preserved, the vertical position in the slit, i.e. in the y-direction, of emitting region is convolved with the velocity information. The coloured circles highlight emission line clumps where this contrary spatial convolution effect is clearly seen.}
\label{fig:contrarymotion}
\end{figure}

On inspection of the data, spectral information was significantly convolved with the spatial profile of the slit, this is an expected consequence of using a spectroscopic slit wider than the seeing disc.
For clumpy emission-line regions, asymmetric line profiles will be produced and as a result, erroneous velocities will be inferred.
Worthy of note is the optical design of the ISIS instrument, which has the dispersion axes in opposite directions for the red and blue arms.
This results in ``contrary offsets'' in the two arms, as shown in Fig.~\ref{fig:contrarymotion}.
We must properly account for these in our analysis.

\section{Analysis}\label{sec:analysis}

\begin{figure*}
\includegraphics[width=\linewidth, trim=0.25cm 0.25cm 0cm 0cm]{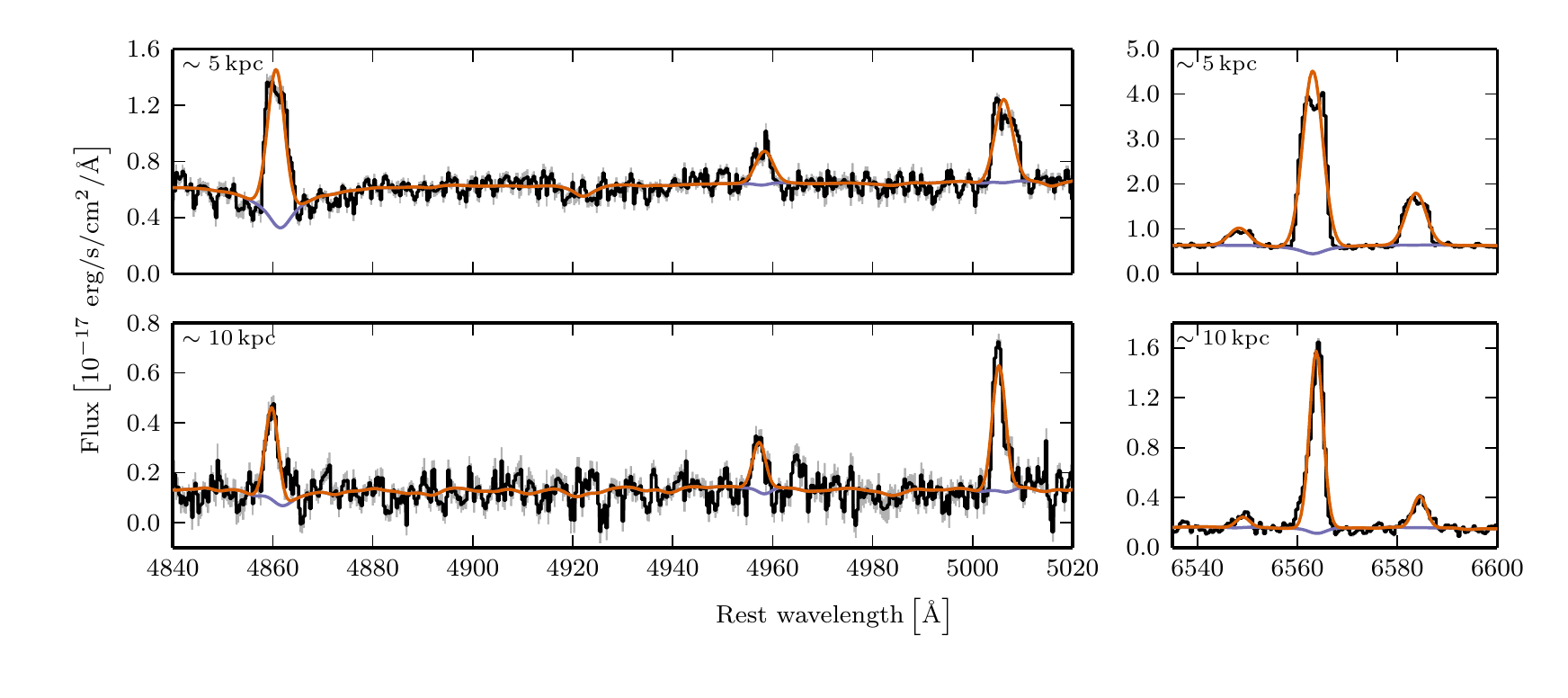}
\caption{Example of \texttt{platefit} spectral fitting to emission lines for galaxy BD 5, shown at two radii, $\sim$5\,kpc (top) and $\sim$10\,kpc (bottom). Observed spectrum and its error indicated by black line and shaded area, respectively. The best-fitting model solution is shown in orange, the continuum component is shown in blue. (Left) Spectra from blue arm covering region containing \Hbeta{}, \forbid{O}{iii}{\lambda4959} and \forbid{O}{iii}{\lambda5007} lines. (Right) Spectra from red arm covering region containing \forbid{N}{ii}{\lambda6548}, \Halpha{} and \forbid{N}{ii}{\lambda6584} lines.}
\label{fig:spectral-fit}
\end{figure*}

A standard approach to emission line modelling is to assume the lines can each be approximated by a single Gaussian function all with the same velocity offset and dispersion.
We preform this spectral fitting using the SDSS \texttt{platefit} spectral fitting routine \citep{2004ApJ...613..898T, 2004MNRAS.351.1151B}, which first fits a continuum to the spectrum with the emission-line features masked, before fitting a sum of Gaussian functions to the residual spectrum. The velocity offsets of the continuum and emission-line components are not tied together. The velocity of the emission-line component may vary up to $\pm500\,\textrm{km}\,\textrm{s}^{-1}$ from that of the continuum component.
The initial continuum fitting was performed using stellar population synthesis templates from \citet[][hererin BC03]{2003MNRAS.344.1000B}, with a fixed velocity offset given by the SDSS redshift.
We therefore update with the velocity determined from the emission line fitting, and again recompute both continuum and line fitting steps.
Due to the discontinuous wavelength coverage and the relatively low spectral resolution, the velocity dispersion of the stellar continuum is difficult to determine, we therefore adopt the velocity dispersion calculated from the SDSS fibre spectrum.
Typical values being twice that of the effect the spectral resolution at 5500\,\AA.
We assume that the stellar velocity dispersion is constant across the whole galaxy, whilst this is not ideal we note that it produces a visibly acceptable result, see Fig.~\ref{fig:spectral-fit}.
Errors on measured line fluxes are determined by the Levenberg-Marquardt least-squares fitting, however, these formal errors are often an underestimate of the real errors of the line fluxes.
Following a procedure derived from SDSS duplicate observations, as discussed by \citet[][herein B13]{2013MNRAS.432.2112B}, we can translate our formal uncertainties to more representative values.
As a result of the aforementioned contrary offsets in the red and blue arms we modify the standard Gaussian fitting by tying the velocity offsets of the blue and red instrument arms separately. 

Spatially binning the 2D spectrum is necessary to optimally extract emission line fluxes.
In order to avoid the line broadening effects caused by co-adding spectra with different velocity offsets, we adopt a similar approach to \citet{2010ApJ...720.1126M}, whereby a two-stage binning strategy is applied.
Firstly we adopt a simple binning process, working from the centre of the galaxy outwards, accreting spectra until a minimum continuum S/N of 6\,\AA$^{-1}$\ is reached.
If a bin spans more than 10\,arcsec before this threshold is reached, then the binning is terminated.
Using the spectral fitting routine, we extract the velocity of the emission lines in the red-arm, where \Halpha{} is dominant.
To this velocity we fit the rotation curve using the following parametrization of \citet{2004A&A...420...97B}
\begin{equation}
V(r) = V_\mathrm{max} \frac{r}{\left(r^a + r_0^a\right)^{1/a}} + V_0,
\label{eq:vel_model}
\end{equation} 
where $r$ is the radius, $V_\mathrm{max}$ is the maximum velocity at $r \gg r_0$, $V_0$ is a constant offset velocity, $a$ and $r_0$ control the shape of the profile.
By using a model we can interpolate the rotation velocity at any position along the spectroscopic slit in a numerically stable fashion.
Weighting the velocity measurements by \Halpha{} S/N, this model provides a reasonable approximation to the true rotation curve, within the limitations imposed by clumpy emission smaller than the slit.
Using this velocity fit, the 2D spectrum was shifted to a common rest frame. 

Since we are interested in the emission line properties it is ideal to bin spectra on emission line criteria, as opposed to the stellar continuum criteria used previously.
Therefore the best-fit model continuum is first subtracted from the rest-frame shifted spectra, before we apply a custom binning algorithm.
Due to the clumpy nature of the emission, any binning algorithm must account for this.
With this in mind, we apply a moving boxcar bin to the unbinned spectra, working from the galaxy centre outwards.
On each subsequent pass the boxcar is incrementally increased in size.
The S/N of each each boxcar bin is determined from a least-squares fit of a Gaussian function to the \Halpha{} line, where a successful bin is when S/N in \Halpha{} exceeds 6.
After the boxcar binning is completed, any remaining unbinned spectra are accreted into the nearest bins, provided their contribution boosts the S/N.
To reduce the statistical dependency between neighbouring bins we impose a minimum bin size of $\sim$1.6\,arcsec, roughly equal to that of the worst FWHM seeing of our observations.
The bin centre is defined by the \Halpha{}-weighted contribution of each 1D spectrum to its respective bin.
As the spectroscopic slits were not necessarily aligned with the measured semi-major axes of the galaxies, we additionally deproject these radii, assuming a thin disc and adopting the inclinations and position angles of the galaxies, as derived in Section~\ref{subsec:bulge_disc_bar_decomposition}.

With the new binning, we apply again our full spectral fitting procedure to the rest-frame shifted spectra.
All results in the following are derived from the resulting outputs.

\subsection{Bluedisk SDSS properties}

\begin{figure*}
\includegraphics[width=\linewidth]{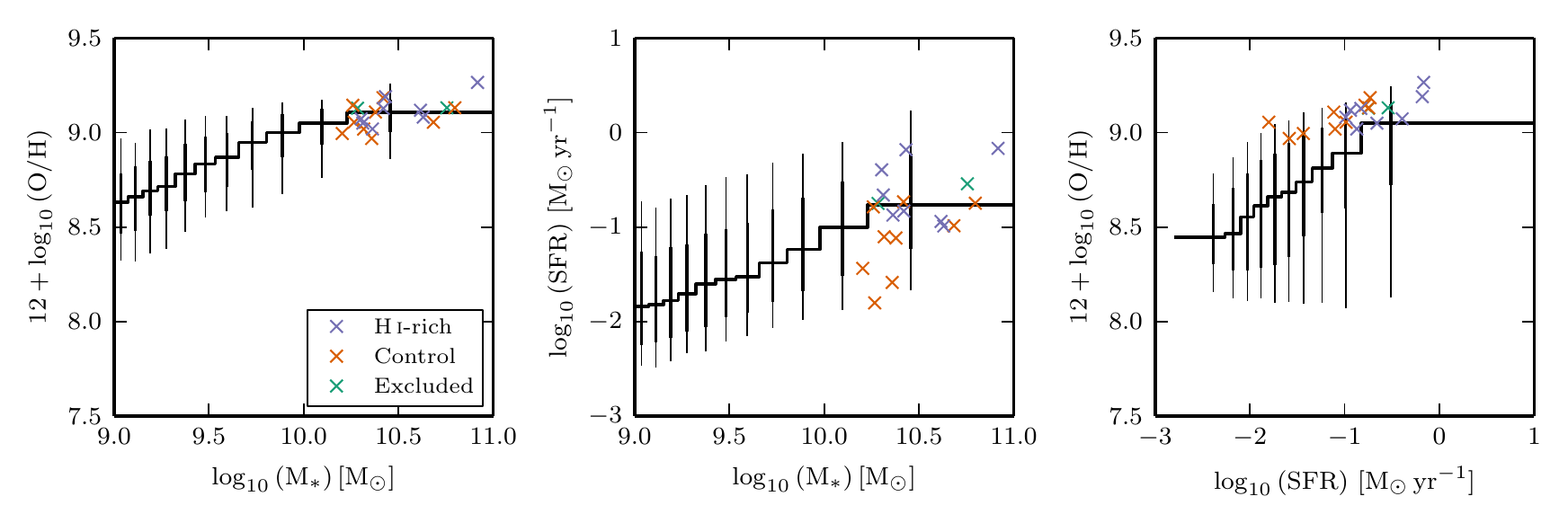}
\caption{Comparison of Bluedisk galaxy properties with the whole SDSS DR7 population that has all three quantities determined (total galaxy stellar mass, fibre SFR, and gas-phase metallicity). The SDSS DR7 sample is selected to span a similar redshift range as the Bluedisk sample (0.02\,$\le$\,z\,$\le$\,0.03). In each plot, crosses indicate SDSS measurements of the Bluedisk \ion{H}{i}-rich, control and excluded  samples. In each plot a stepped-curve represents the median ordinate value of given abscissal bin. (Left and Middle) Sample divided by mass into vigintile bins and (Right) Sample divided by SFR into decile bins. Thick and thin vertical lines represent the 68\% (1\,$\sigma$) and 95\% (2\,$\sigma$) ranges respectively.}
\label{fig:sdss_m_sfr_z}
\end{figure*}

In Fig.~\ref{fig:sdss_m_sfr_z} we compare the central metallicities, central SFRs and total stellar masses of the Bluedisk sample to other galaxies drawn from the 7\textsuperscript{th} data release (DR7) of the SDSS.
We discuss the estimation of metallicity in Section~\ref{subsec:inferring_metallicities} below, here it suffices to say that the same estimator has been used for SDSS and the Bluedisk spectra.
The Bluedisk sample lies on the same mass-metallicity relation and mass-SFR relation of the DR7 population at a similar redshift.
The galaxies have noticeably elevated metallicities with respect to their counterparts of similar SFRs, however, this is to be expected since they are amongst the most massive $\sim$10\% of galaxies at their epoch.
When observed as a function of SFR, a clear split is observed between the \ion{H}{i}-rich and control populations, with \ion{H}{i}-rich galaxies more actively forming stars at their centres.
While not an explicit selection criteria of the survey, the two populations do not differ significantly in central metallicity.

\subsection{Contamination from non-star-forming sources}

To avoid deriving erroneous metallicities, we must take care to exclude spectra contaminated by significant line emission from active galactic nuclei (AGN) or low-ionization nuclear emission-line regions (LINERs).

\begin{figure}
\includegraphics[width=\linewidth]{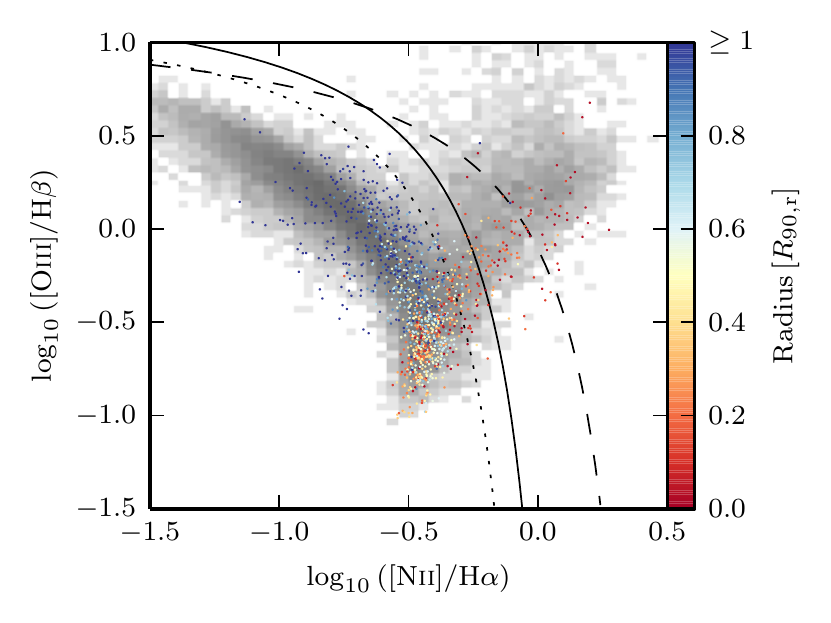}
\caption{The BPT diagnostic diagram for all binned galaxy data points with \Halpha{} S/N\,$\ge$\,10 and S/N\,$\ge$\,3 in the other three emission lines. Colours indicate radius from galaxy centre in units of $\textrm{R}_{90,\mathrm{r}}$, the radius containing 90\% of the SDSS \textit{r} band luminosity. Shaded underlay shows BPT histogram of SDSS DR7 fibres in the same redshift range (0.02\,$\le$\,z\,$\le$\,0.03). The dashed line indicates the division between pure AGN and ``composite'' galaxies \citep{2001ApJ...556..121K}. The dotted line represents the nominal \citet{2003MNRAS.346.1055K} SF diagnostic line, but for our classification we use a modified form of this offset $+$0.1\,dex in both axes (solid line).}
\label{fig:bpt}
\end{figure}

Following the prescription of \citet{2004MNRAS.351.1151B} we classify the spectra into five catagories.
Namely star-forming (SF), AGN/LINER, ``composite'' SF + AGN/LINER, low S/N AGN/LINER and low S/N SF.
For this we use the diagnostic criteria of \citet{2003MNRAS.346.1055K} and \citet{2001ApJ...556..121K}, applied to the (\forbid{O}{iii}{\lambda5007}/\Hbeta{}), and (\forbid{N}{ii}{\lambda6584}/\Halpha{}) emission-line ratios.
However, we find the \citet{2003MNRAS.346.1055K} division between SF and composite emission is too aggressive, resulting in the rejection of data points at large radii.
Following \citet{2012ApJ...745...66M}, we therefore loosen our criteria by offsetting this diagnostic line diagnostic $+$0.1\,dex in both \forbid{O}{iii}{}/\Hbeta{} and \forbid{N}{ii}{}/\Halpha{} line ratios.
Fig.~\ref{fig:bpt} shows the \citet*{1981PASP...93....5B} (BPT) diagnostic diagram for the co-added spectra.
From this we can see that almost all excluded data points are found in the inner regions of the galaxies where we could expect contamination by AGN and shock heated gas.

We ultimately apply an \Halpha{} S/N\,$\ge$\,10 cut to our data. Although we do not apply a equivalent width (EW) cut, we note that 96\% of our spectral bins have $\textrm{EW}\left(\Halpha{}\right)>3$. This EW limit was recommended by \citet{2011MNRAS.413.1687C} to identify pure star-forming galaxies, where \Halpha{} emission is associated with \ion{H}{ii} regions rather than post-asymptotic giant branch stars.

\subsection{Inferring Metallicities}\label{subsec:inferring_metallicities}

We derive gas-phase metallicities using the method developed by B13.
This method applies a Bayesian framework to a grid of  photoionization models.
We shall pair the B13 methodology with the photoionization models of \citet[][herein CL01]{2001MNRAS.323..887C}.
To correct for dust, the B13 procedure uses a two-component dust-absorption model of \citet{2000ApJ...539..718C}, with a wavelength dependant attenuation curve of the form $\tau(\lambda) \propto \lambda^{-1.3}$.
For a set of emission-line fluxes\footnote{The B13 analysis was performed using emission lines: \forbid{O}{ii}{\lambda3727}, \mbox{H$\delta$}, \mbox{H$\gamma$}, \Hbeta{}, \forbid{O}{iii}{\lambda4959}, \forbid{O}{iii}{\lambda5007}, \Halpha{}, \forbid{N}{ii}{\lambda6584}, \forbid{S}{ii}{\lambda6716} and \forbid{S}{ii}{\lambda6731}.}, we obtain a posterior probability distribution on the metallicity.
However, to test our CL01 derived metallicities we will now outline a series of other metallicity determination methods.

A common set of methods for inferring metallicities is to use line-ratio diagnostics that have directly calibrated to oxygen abundances of \ion{H}{ii} regions (either theoretical or observed).
We use two such methods, one using the theoretically derived relations of \citet[][herein KK04]{2004ApJ...617..240K}, and another method using the empirically derived relations of \citet{2011MNRAS.412.1145P} known as the NS calibrator.
In addition we check the sensitivity of our CL01 metallicities to the choice of photoionization models, by applying the models of \citet[][herein D13]{2013ApJS..208...10D} within the B13 framework.
It should be noted that the B13 method is similar in principle to others such as \texttt{IZI} \citep{2015ApJ...798...99B} and \texttt{HII-CHI-MISTRY} \citep{2014MNRAS.441.2663P}.

Dust attenuation affects line ratios, therefore we must correct for dust before we apply the KK04 and NS calibrators.
We adopt the same $\tau(\lambda) \propto \lambda^{-1.3}$ attenuation curve and calculate its normalization by assuming an intrinsic \Halpha{}/\Hbeta{} Case-B ratio of 2.85 (temperature, $T=10^4$\,K, and electron number density, $n_e=10^4$\,cm$^{-3}$) \citep[][p. 78]{2006agna.book.....O}.
We note, however, that by assuming such physical properties of the \ion{H}{ii} regions we must a priori assume a metallicity.
With the B13 approach we avoid this assumption by simultaneously correcting for dust attenuation when inferring metallicity.
Nevertheless, for spectra with high S/N, both the B13 and empirical Case-B approaches yield similar results for the strength of the attenuation.

Uncertainties on the metallicity determinations for the KK04 and NS methods are determined using Monte Carlo simulations.
We assume the true line fluxes to be normally distributed about the measured line flux, with a standard deviation equal to the error in the measured value.
Of these many realisations, we take the median as the metallicity value, and the symmetrized $\pm$1\,$\sigma$ quantiles to be its associated error.
For the CL01 and D13 models, we can extract the median and its error directly from the cumulative posterior probability of the metallicity parameter.
Due to the finite sampling of the metallicity parameter in these models, we impose an additional minimum uncertainly of $\pm$0.05\,dex \citep{2004MNRAS.351.1151B}.
We additionally apply the same minimum error to the KK04 and NS methods.

In Appendix~\ref{app:comparing_indicators} we compare the different metallicity methods.
We show that the CL01 method produces results consistent with the other three methods.
We adopt CL01 as our default method for metallicity determination.
Herein for simplicity when we refer to metallicity, we are referring to that which is derived from the CL01 models.

Note that in contrast to some studies of abundance gradients our spatial bins include line emission from both \ion{H}{ii} regions and diffuse emission.
Since previous studies have found that the diffuse emission is powered by radiation escaping from \ion{H}{ii} regions \citep[e.g.][]{2003ApJ...586..902H, 2013ApJ...775..109K}, we can approximately treat the combined line emission as coming from an HII region with larger volume and hence a lower ionization parameter. 
The CL01 models are well suited for this as they cover a range in ionization parameters.
In addition, note that while the CL01 models are not specifically optimised for spatially resolved regions in galaxies, tests in B13 showed that the CL01 models perform well in this case.

\subsection{Estimating gas mass densities}\label{gas_mass_densitites}

To determine gas surface mass density at the same resolution as our metallicities, we estimate the gas surface mass densities directly from our spectra. 
In B13 it was shown that when most of the strong lines in the optical spectrum are available it is possible to use photoionization models with a flexible treatment of metal depletion to place constraints on the gas surface mass density of galaxies. The application shown in B13 used the CL01 models which we also use here.
By jointly fitting the strong optical lines B13 showed that the total gas surface mass densities can be estimated through
\begin{equation}
  \label{eq:sigma_gas_1}
  \Sigma_{\mathrm{gas}} = 0.2\frac{\tau_V}{\xi Z}\ \textrm{M}_\odot\,\textrm{pc}^{-2},
\end{equation}
where $\tau_V$ is the optical depth in the V-band, $\xi$ the dust-to-metal ratio of the ionised gas, and $Z$ the metallicity.

They compare the result of applying this relation to spectra from the SDSS to total mass densities measured from \ion{H}{i} and $\textrm{H}_2$ mass maps from the THINGS \citep{2008AJ....136.2563W} and HERACLES \citep{2009AJ....137.4670L} surveys.
This point-by-point comparison showed that the spectroscopic method is in excellent agreement with the \ion{H}{i}+$\textrm{H}_2$ mass maps, except at the very highest gas surface densities, $\Sigma_{\mathrm{gas}} > 75\ \textrm{M}_\odot\,\textrm{pc}^{-2}$.

For the present paper we note that we use the same set of emission lines used by B13 in their study with comparable signal-to-noise, so we expect this result to carry over to our study.
This means that our spectroscopic gas densities are likely to be underestimated in the central regions of the galaxies. 
We expand on this and discuss this method more in Appendix~\ref{app:gas_surface_density_estimates}.

\section{Results}\label{sec:results}

\subsection{The local mass-metallicity relation and radial mass profiles}\label{subsec:local_mass_metallicity}

From basic analytical arguments one expects the metallicity of a system to depend on the stellar and gas mass budgets \citep{1997nceg.book.....P}.
Indeed, recent works have uncovered a correlation between stellar-mass surface density and metallicity, known as the local mass-metallicity relation \citep{2012ApJ...745...66M, 2012ApJ...756L..31R, 2013A&A...554A..58S}. 
However, it is worth noting \citet{2014AJ....148..134P} find that although local surface brightness and metallicity are correlated, there is no unique relation between the two that holds at all radii in galaxies.
Nevertheless, we will now test whether the local mass-metallicity relation holds for the Bluedisk galaxies.

\begin{figure*}
\includegraphics[width=\linewidth]{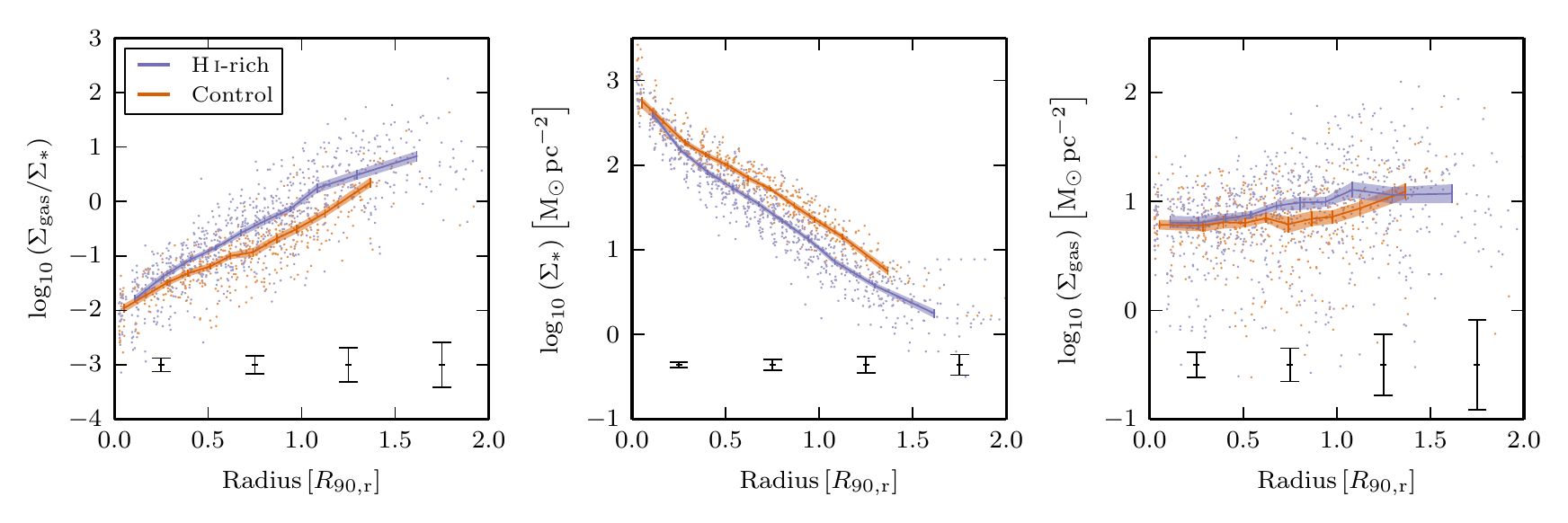}
\caption{Radial profiles of gas-to-stellar mass ratio, $r_\mathrm{gas} = \Sigma_{\mathrm{gas}} / \Sigma_{\ast}$, stellar mass density, $\Sigma_{\ast}$, and gas mass density, $\Sigma_{\mathrm{gas}}$, are shown from left to right. Data is shown for both Bluedisk \ion{H}{i}-rich and control samples, coloured blue and orange respectively. We plot the individual data points, as well as the binned median trends in each plot. Shaded regions indicate $\pm$1\,$\sigma$ errors on the trends, as determined by bootstrapped Monte-Carlo realisations. Radius is in scale units of $\textrm{R}_{90,\mathrm{r}}$, the radius containing 90\% of the SDSS \textit{r} band luminosity. Black vertical bars indicate median error in each 0.5\,$\textrm{R}_{90,\mathrm{r}}$ division.}
\label{fig:stochmod_profile}
\end{figure*}

In Fig.~\ref{fig:stochmod_profile} we present radial profiles of our spectroscopic dust-to-gas estimates of gas-mass surface densities, $\Sigma_{\mathrm{gas}}$, as well as our stellar-mass surface densities, $\Sigma_{\ast}$, which are matched in aperture.
The median trends for $\Sigma_{\ast}$ exhibit clear differences between the \ion{H}{i}-rich and control samples, with \ion{H}{i}-rich being consistently less massive at a given scale radius.
By contrast, $\Sigma_{\mathrm{gas}}$ shows no significant distinction between the samples.
We caution that although there appears to be a slight upward trend in $\Sigma_{\mathrm{gas}}$ with radius, which may be artificial (see Section~\ref{gas_mass_densitites}).

In Fig.~\ref{fig:stochmod_profile} we also show the radial profiles of the gas-to-stellar mass ratio, $r_\mathrm{gas} = \Sigma_{\mathrm{gas}} / \Sigma_{\ast}$.
Here we find that the \ion{H}{i}-rich galaxies exhibit enhanced $r_\mathrm{gas}$ ratios at all but the very centre the stellar disk.
We note that this is primarily driven by radial differences in $\Sigma_{\ast}$ rather than $\Sigma_{\mathrm{gas}}$.
In other words, at a fixed $\Sigma_{\ast}$ the \ion{H}{i}-rich and control samples are indistinguishable in terms of $r_\mathrm{gas}$.

\begin{figure*}
\includegraphics[width=\linewidth]{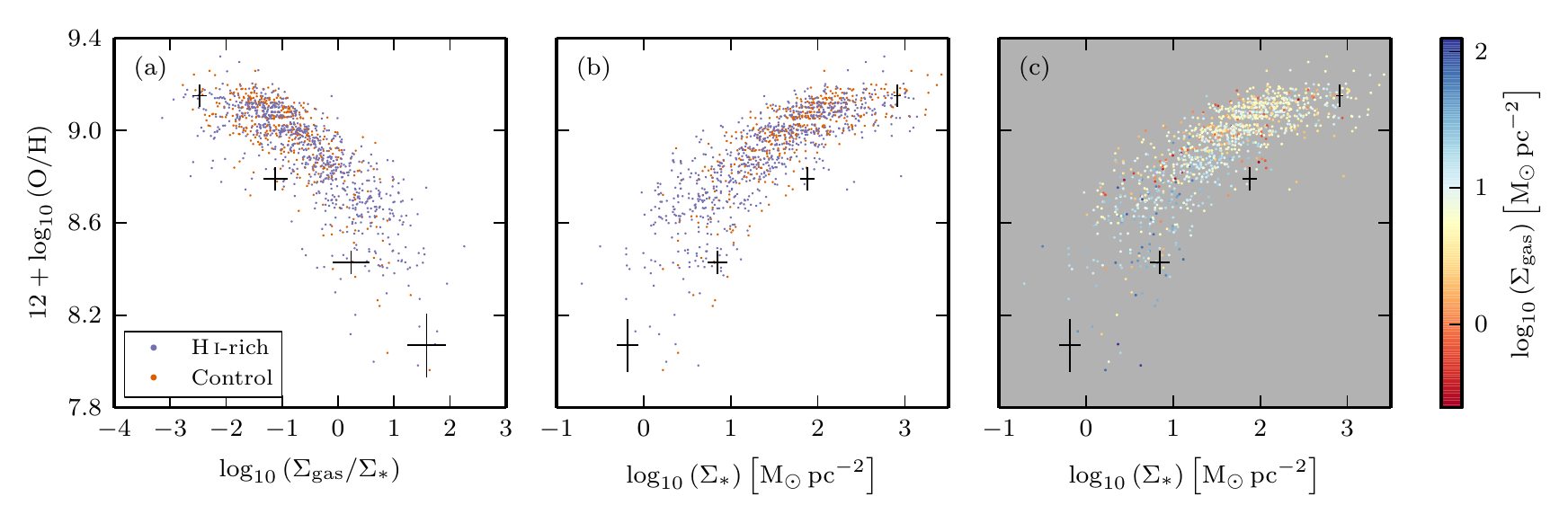}
\caption{Correlations between metallicity and surface mass densities. (a) Metallicity against gas-to-stellar mass ratio . (b),(c) both show metallicity against stellar surface mass density $\Sigma_{\ast}$ i.e. the local mass-metallicity relation, but are colour-coded in differently. In (a),(b) we colour according to Bluedisk sample, \ion{H}{i}-rich and control samples, distinguished by blue and orange data points respectively. In (c) we colour the local mass-metallicity relation by gas surface mass density, $\Sigma_{\mathrm{gas}}$.  Black crosses indicate median error within equally spaced bins.}
\label{fig:lMZ}
\end{figure*}

In Fig.~\ref{fig:lMZ} we show the local mass-metallicity relation for the Bluedisk galaxies.
We also plot the correlation between $r_\mathrm{gas}$ and metallicity.
Crucially, neither of these correlations show any strong offsets between the \ion{H}{i}-rich and control samples, implying that the processes that govern these parameters are similar in both classes of galaxies.
We observe that the correlation of metallicity with $r_{\mathrm{gas}}$ is not visibly tighter than that with $\Sigma_{\ast}$.
But, as exemplified in Fig.~\mbox{\ref{fig:lMZ}(c)} we note that at the lowest stellar mass densities ($\log_{10}\left(\Sigma_{\ast}\right) \la 1.5\ \textrm{M}_\odot\,\textrm{pc}^{-2}$) a significant portion of the metallicity variation can be explained by changes in $\Sigma_{\mathrm{gas}}$.
We must caution, however, that metallicity and $\Sigma_{\mathrm{gas}}$ are not independently derived.
The nature of the modelling will introduce a small intrinsic correlation between these two parameters. The magnitude of this effect is not easily quantified.

Having shown that both \ion{H}{i}-rich and control samples form a consistent local mass-metallicity relation, we shall explore the radial metallicity profiles of the Bluedisk galaxies.

\subsection{Metallicity profiles of the Bluedisk galaxies}

\begin{figure*}
\includegraphics[width=\linewidth]{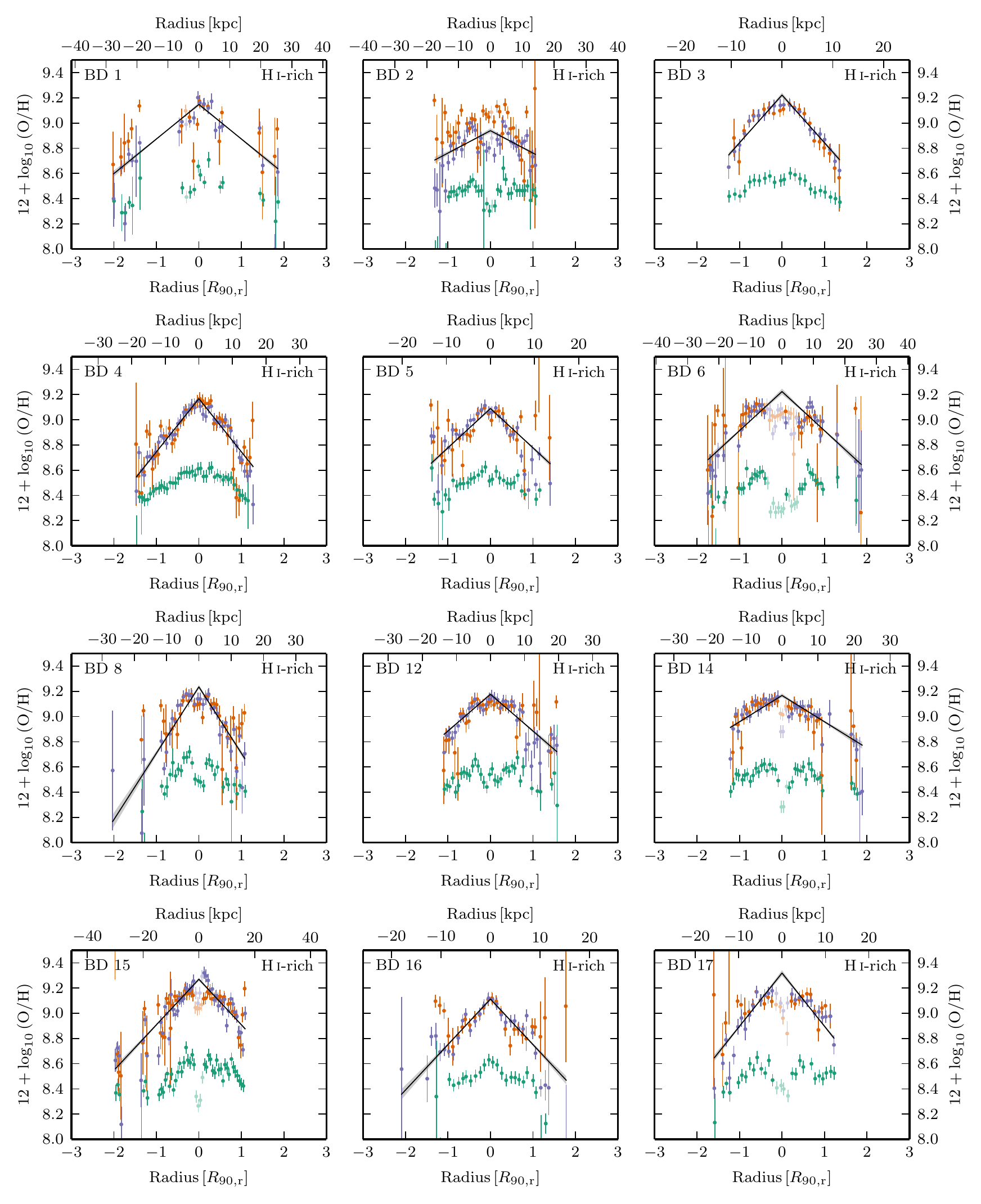}
\caption{Metallicity profiles of the 23 galaxies in the Bluedisk \ion{H}{i}-rich sample. The metallicity is inferred using the CL01, KK04 and NS methods, plotted as blue, orange and green respectively. For the CL01 metallicities we show the best-fitting straight-line model (black), where the shaded area indicates its associated $\pm$1\,$\sigma$ error in gradient. We also show in pale colours data points masked from the fitting due to their non-SF emission characteristics. All plotted data has an minimum \Halpha{} S/N$\ge$10. Radius is plotted in both units of a scale radius, $\textrm{R}_{90,\mathrm{r}}$, and physical size in kpc.}
\label{fig:profiles_blue}
\end{figure*}

We present the metallicity profiles of the Bluedisk galaxies in Figs.~\ref{fig:profiles_blue}, \ref{fig:profiles_cont} and \ref{fig:profiles_excl}, divided into their \ion{H}{i}-rich, control and excluded samples respectively.\footnote{Larger versions of the metallicity profiles in conjunction their SDSS \textit{gri} composite images are available in Appendix~\ref{app:atlas} (online-only).}
By visual inspection alone there is no clear distinction between \ion{H}{i}-rich and control samples in terms of the profile shapes.
The control sample does appear to be more radially truncated, however, this is to be expected.
It has been shown by \citet{2011MNRAS.412.1081W} that \ion{H}{i} mass fraction is correlated with the $g-i$ colour gradient in galaxies.
This implies that the \ion{H}{i}-rich galaxies have higher rates of star formation in their outer discs.
Therefore we expect the control sample to have less extended star formation, ultimately limiting the radius out to which we can robustly detect emission lines.
For a more quantitative analysis we must study the measured gradients of the linear model.

\addtocounter{figure}{-1}
\begin{figure*}
\includegraphics[width=\linewidth]{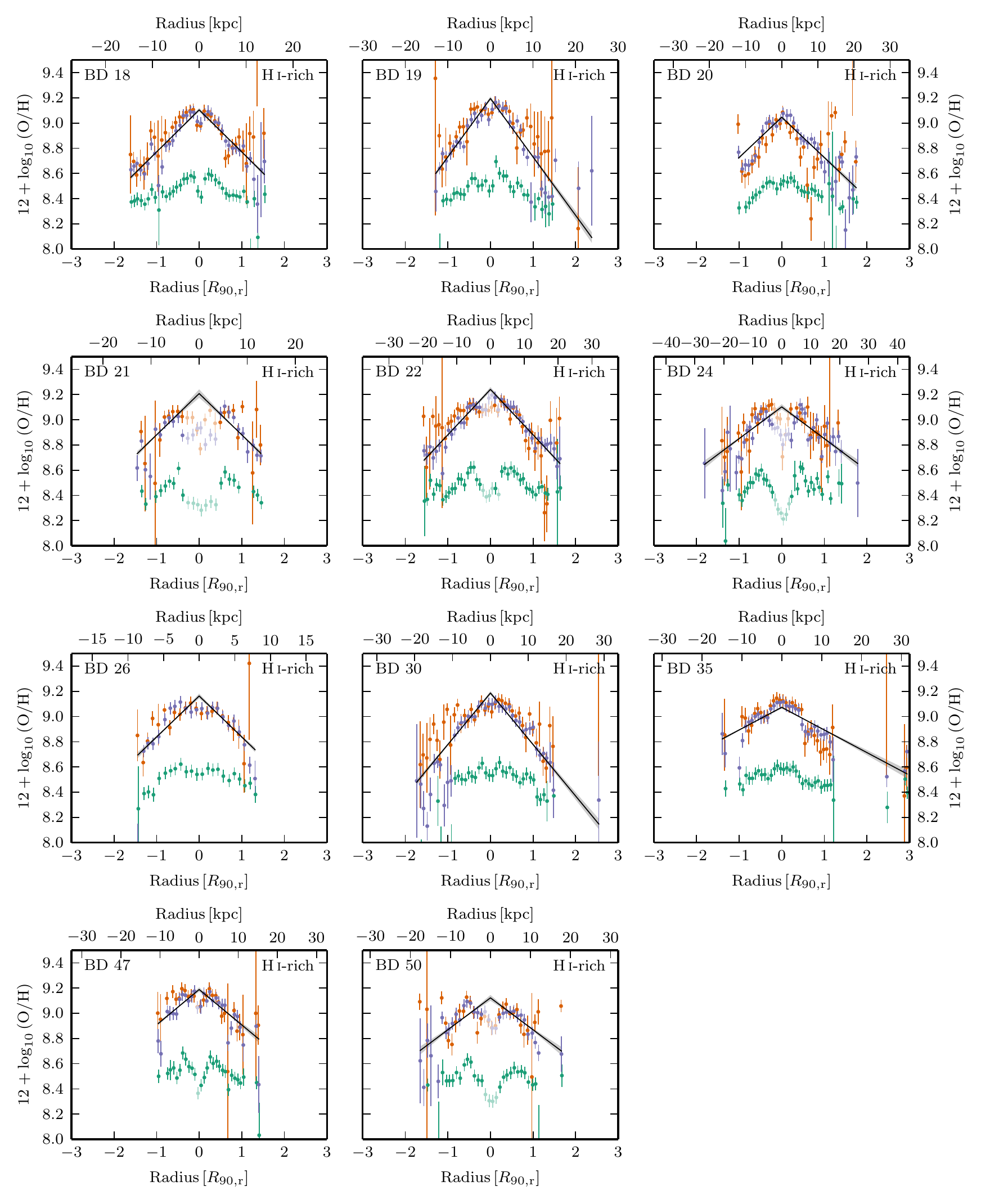}
\caption{Metallicity profiles of the 23 galaxies in the Bluedisk \ion{H}{i}-rich sample -- \textit{continued}.}
\end{figure*}

\begin{figure*}
\includegraphics[width=\linewidth]{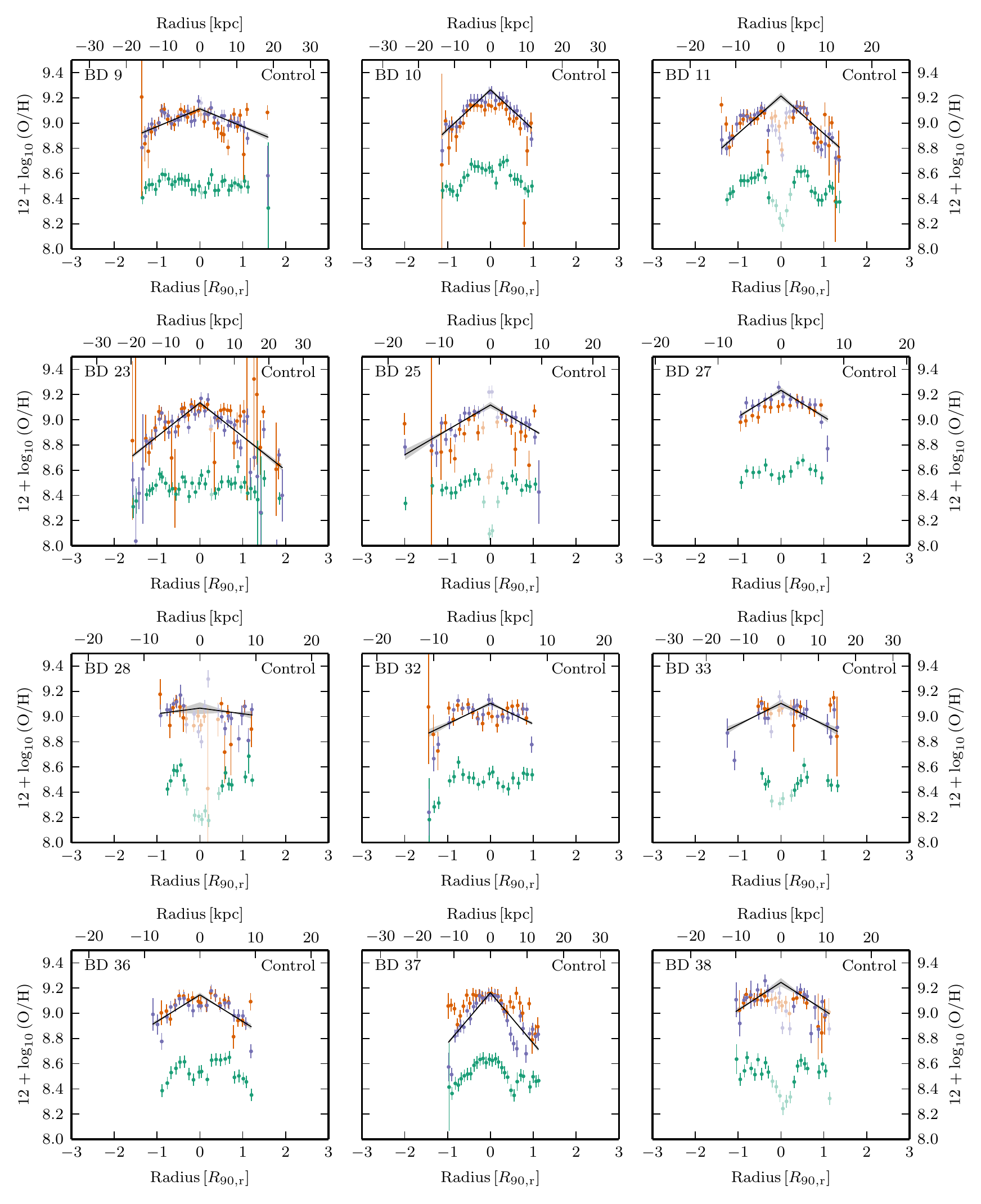}
\caption{Metallicity profiles of the 18 galaxies in the Bluedisk control sample, see Fig.~\ref{fig:profiles_blue} for details.}
\label{fig:profiles_cont}
\end{figure*}

\addtocounter{figure}{-1}
\begin{figure*}
\includegraphics[width=\linewidth]{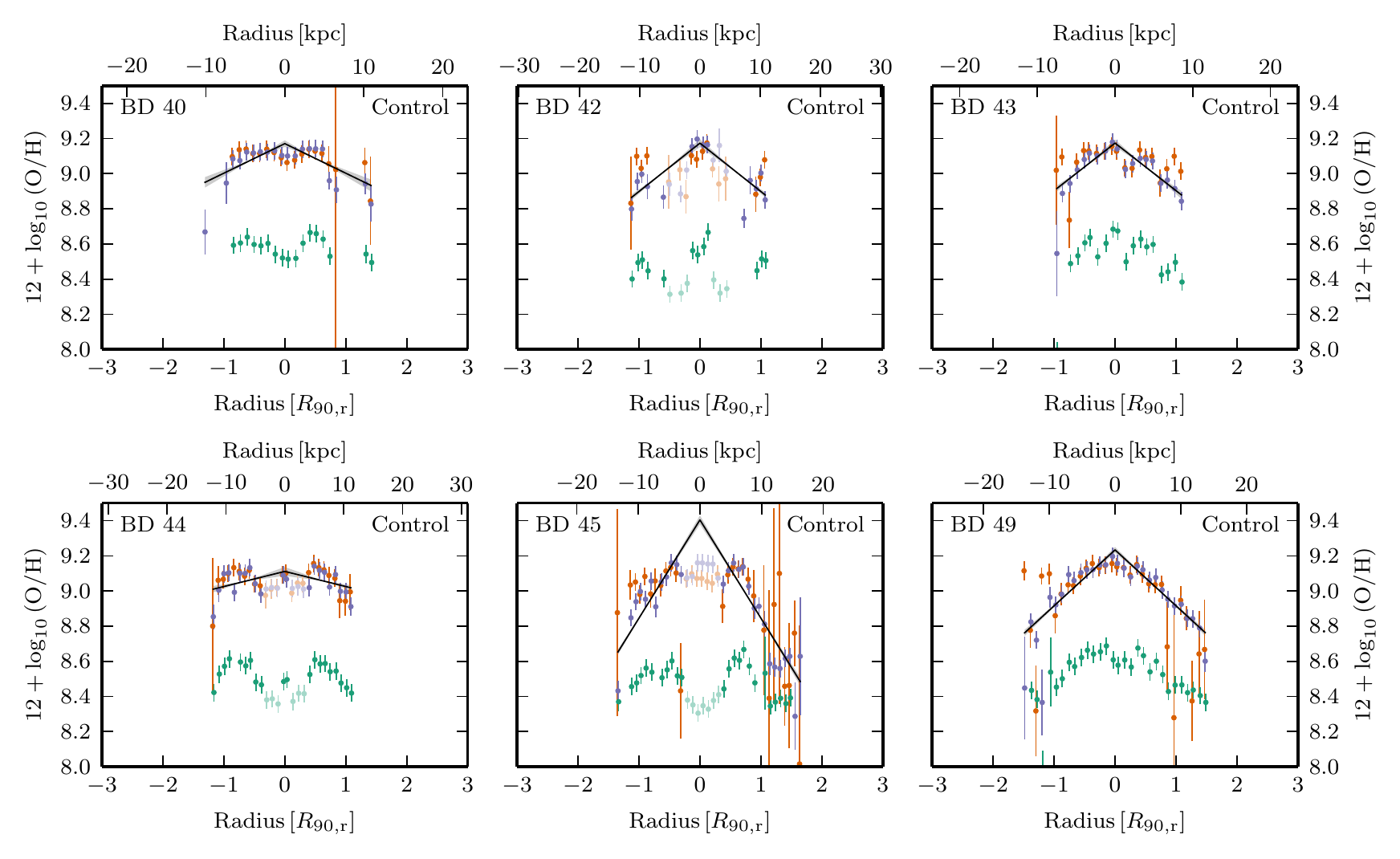}
\caption{Metallicity profiles of the 18 galaxies in the Bluedisk control sample -- \textit{continued}.}
\end{figure*}

\begin{figure*}
\includegraphics[width=\linewidth]{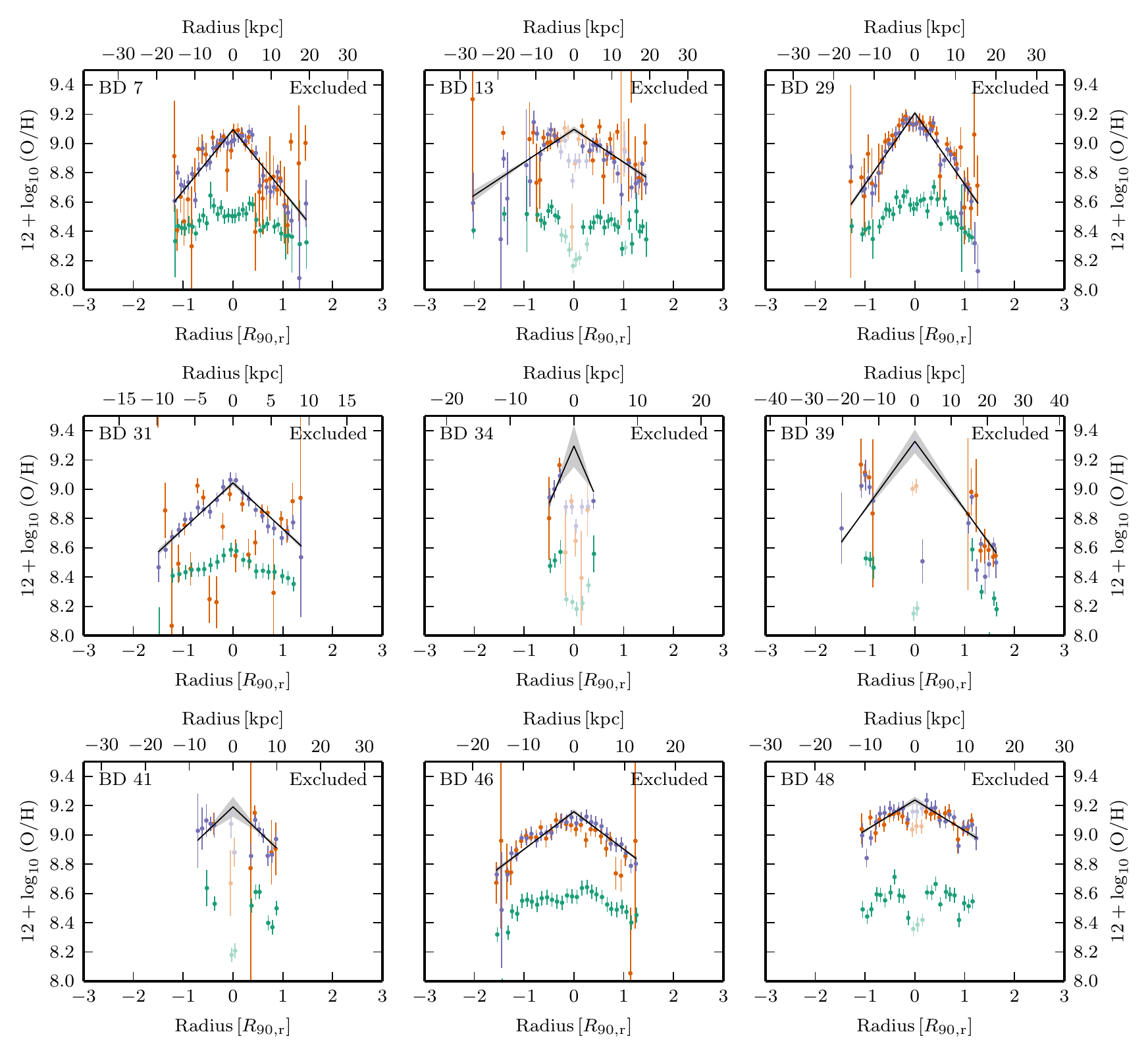}
\caption{Metallicity profiles of the 9 galaxies in the excluded Bluedisk sample, see Fig.~\ref{fig:profiles_blue} for details.}
\label{fig:profiles_excl}
\end{figure*}

\begin{figure*}
\includegraphics[width=\linewidth]{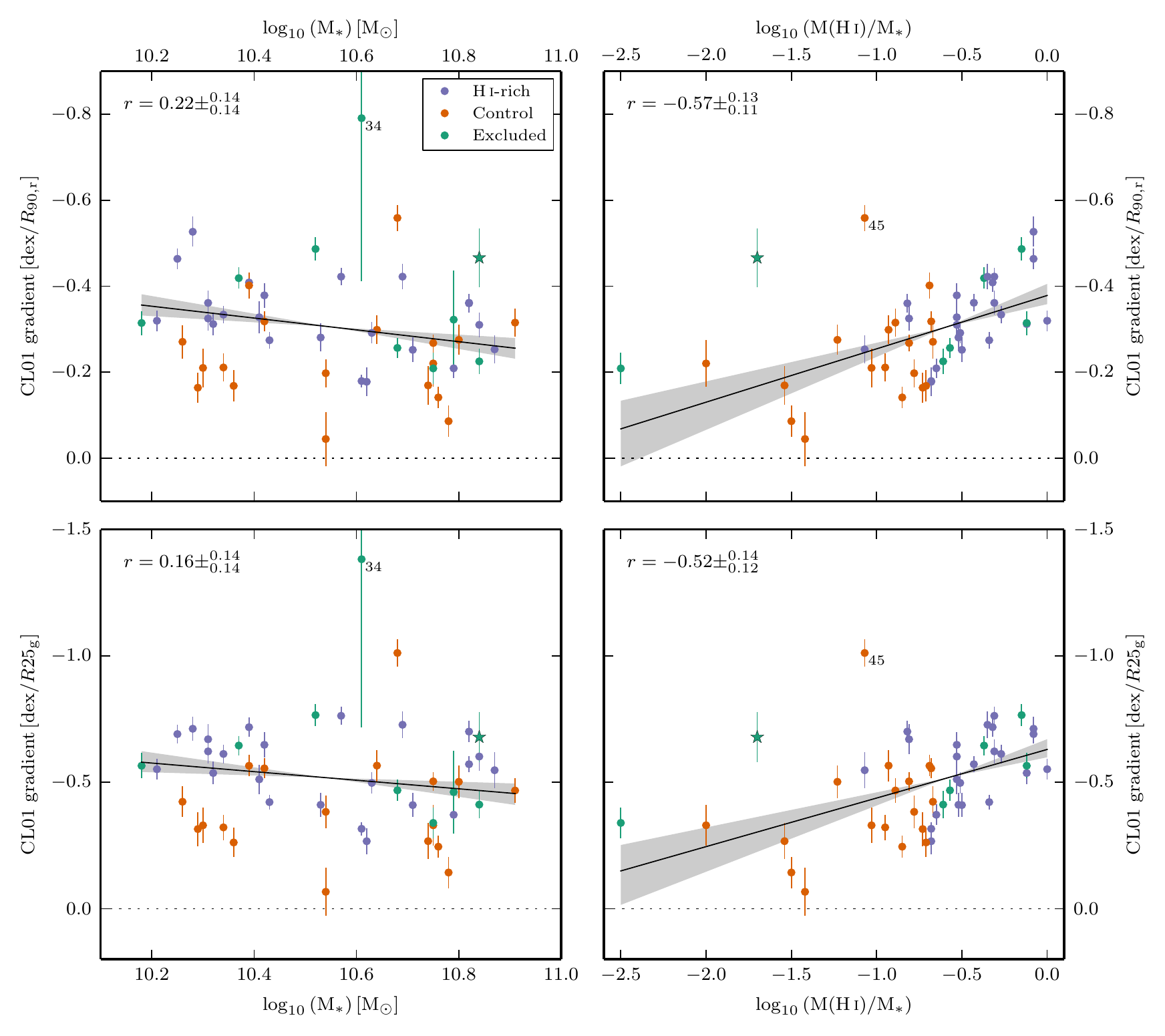}
\caption{CL01 metallicity gradients as a function of total stellar mass (left) and \ion{H}{i} mass faction (right).
We report gradients using two different scale radii, $\textrm{R}_{90,\mathrm{r}}$ (top) and $\textrm{R}25_\mathrm{g}$ (bottom).
Colours indicate Bluedisk sample classification \ion{H}{i}-rich (blue), control (orange) and excluded (green).
Best linear fit and its error in slope are indicated by the solid lines and the shaded regions respectively. The Spearman's rank correlation coefficient is given in the top-left corner of each figure. A star indicates BD 39, which is excluded from the regression and the r-statistic computation. Numbers label individual galaxies with especially steep metallicity gradients that we reference in the text. For reference, $r = 0.29$ is the two-tailed Spearman's $r$-value at a $\alpha = 0.05$ significance level.}
\label{fig:grad_vs_mass}
\end{figure*}

As is common in the literature we approximate the metallicity profiles with a symmetric linear best fit.
In most cases this simple functional form encapsulates the overall change in metallicity from the centre to the outskirts of the galaxy.
In Fig.~\ref{fig:grad_vs_mass} we show the dependence of the metallicity gradient on both stellar mass and the \ion{H}{i} mass fraction.
There exists a significant correlation between \ion{H}{i} mass faction and metallicity gradient.
Galaxies with larger \ion{H}{i} mass fractions typically have steeper metallicity gradients.
Whereas we find no significant correlation between stellar mass and metallicity gradient.
However, this is unsurprising given the narrow range that we span ($10.2 \la \log_{10}\left(M_\ast/\textrm{M}_\odot\right) \la 11.0$).
We note these results remain unchanged when we adopt a different scale radius, $\textrm{R}25_\mathrm{g}$ (the radius at which the SDSS \textit{g} band surface brightness reaches 25\,mag/arcsec$^2$).
We highlight two galaxies with especially steep metallicity gradients. Firstly, BD 34 which shows very large errors in its measured metallicity gradient.
This gradient is measured from only four valid data points and is poorly constrained.
Secondly, BD 45 whose metallicity profile shows some hints of asymmetry, but otherwise offers no explanation for the excessively steep inferred metallicity gradient.
Regardless, we retain both these outlying galaxies in our analysis.

From visual inspection, describing some of these galaxies with a straight-line model appears to poorly reflect the true metallicity profile.
In a number of galaxies the metallicity gradient appears to be increasing with radius.
As such, a gradient measured from the outer disc would be much steeper than one measured from the inner disc.
With long-slit spectra we only measure metallicity along one dimension of the galaxy.
So if significant azimuthal metallicity variations are present, our metallicity measurements may not be indicative the whole galaxy at a given radius.
Simulations of \citet{2014arXiv1411.7585P} indicate, however, that azimuthal variations decay are expected to decay on timescales shorter than the orbital period of the galaxy.
Indeed, observationally there is little support for strong azimuthal variations, with \citet{2015A&A...573A.105S} reporting only modest ($< 0.05$\,dex) azimuthal variations.

In the following, by stacking the metallicity profiles we shall attempt to produce average metallicity profiles.


\subsubsection{Stacked average metallicity profiles}\label{subsec:profile_stack}

\begin{figure*}
\includegraphics[width=\linewidth]{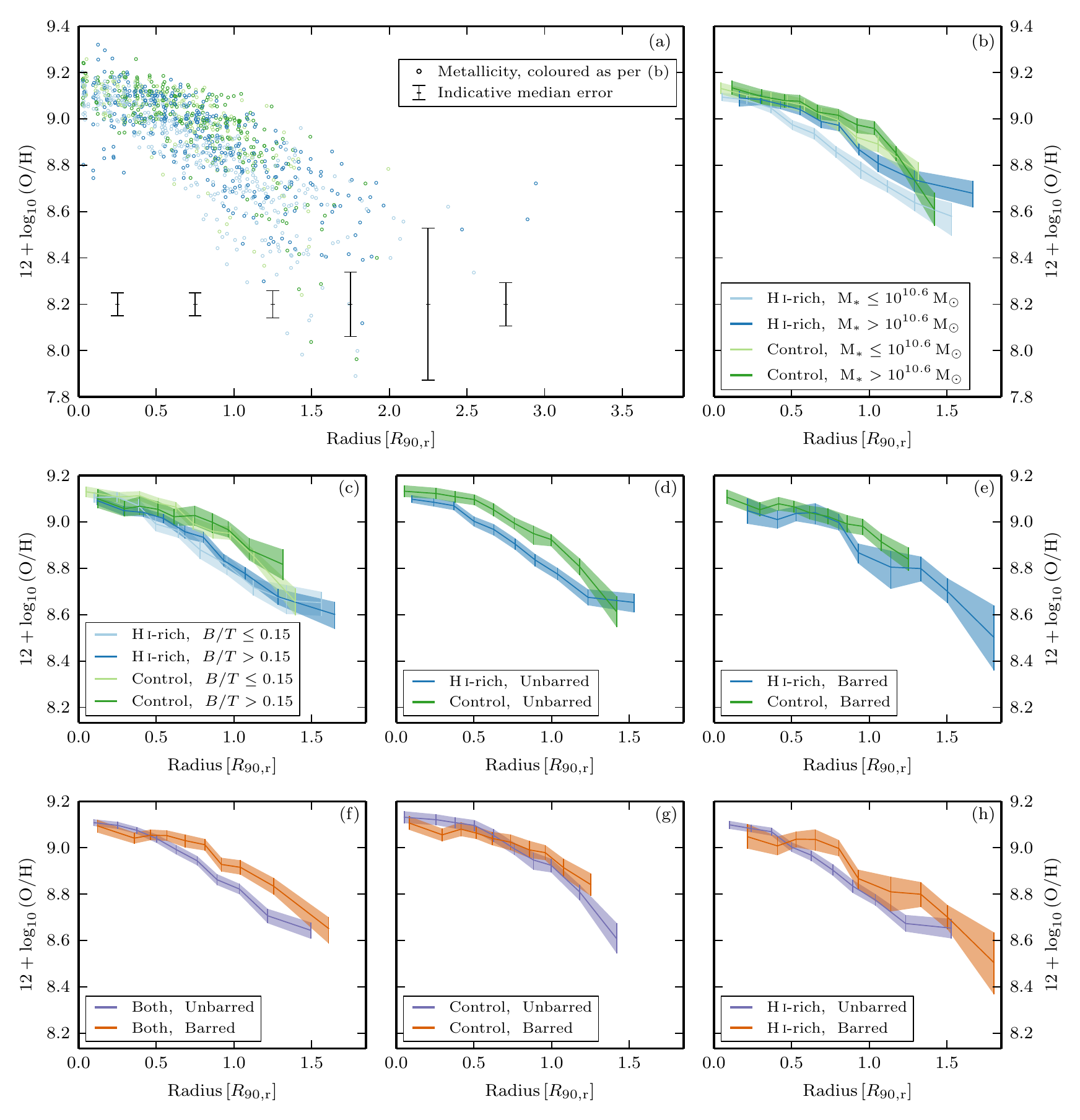}
\caption{Results from stacking metallicity profiles. (a),(b) Metallicity of the Bluedisk galaxies, colour distinguishing \ion{H}{i}-rich and control sample, as well as low and high stellar mass. (a) The individual metallicity data points. Vertical bars indicate median error in each 0.5\,$\textrm{R}_{90,\mathrm{r}}$ division. (b) Median value of galaxies stacked in decile radial bins. Shaded regions indicate $\pm$1\,$\sigma$ errors determined by bootstrapping Monte-Carlo realisations. (c) Similarly stacked profiles, however, split by B/T ratio instead of stellar mass. Dividing the \ion{H}{i}-rich and control galaxies into barred and unbarred samples we generate the panels (d)-(h).} 
\label{fig:profile_stack}
\end{figure*}

To study the metallicity profiles for the Bluedisk galaxies further, we stack the individual metallicity data points into equal mass decile radial bins.
Tracing the median metallicity of the bins we construct the average metallicity profiles, which are shown in Fig.~\ref{fig:profile_stack}.
We caution that since the metallicity data points are equally weighted, the outermost bin of each stack might be considered unreliable (see Appendix~\ref{app:weighted_profile_stack}).

In Fig.~\mbox{\ref{fig:profile_stack}(b)} we show the stacked profiles of the \ion{H}{i}-rich and control samples.
We also bisect each sample by total stellar mass.
We observe that all galaxies have similar central metallicities, but different profile shapes.
We note that the outermost bin of the high-mass control profile has a spuriously low metallicity and should be ignored (see Appendix~\ref{app:weighted_profile_stack}).
Putting this aside, the stacked profiles appear to indicate a shallower inner gradient and a steeper outer gradient.
It is difficult to define the characteristic radius at which this transition happens. However, by eye it seems that the transition occurs at a smaller radius in the \ion{H}{i}-rich galaxies than in the control galaxies.
Overall we note that the transitions in the stacked profiles do not appear as abrupt as they do in the unstacked profiles.
This would imply that using $\textrm{R}_{90,\mathrm{r}}$ as a radial coordinate is not ideal for expressing this turnover.
Indeed, since we observe a local mass-metallicity relation, a scale radius based on stellar mass density would perhaps be more appropriate.

Beyond \ion{H}{i} characteristics there are other aspects which may affect metallicity profiles.
Using semi-analytical models \citet{2013MNRAS.434.1531F} predict metallicity gradients should be correlated with bulge-to-total (B/T) ratio.
In these models galaxies with more prominent bulges are expected to have shallower gas phase metallicity gradients as the gas distribution in these galaxies is set by later infall of gas.
We show the stacked profiles of the galaxy samples bisected by B/T light ratio in Fig.~\mbox{\ref{fig:profile_stack}(c)}.
We observe no apparent connection between metallicity profiles and bulge prominence, however, given the low bisecting threshold (B/T=0.15) we find our results to only be in mild tension with these predictions.
These observations are consistent with \citet{2014A&A...563A..49S} who observe no correlation between metallicity gradient and morphological galaxy type.

It may be possible that the differences we observe between our \ion{H}{i}-rich and control populations are drivmen by the effects of bars.
It has been established that there exists an anticorrelation between \ion{H}{i} mass fraction and the bar occurrence fraction \citep{2004A&A...416..515D,2012MNRAS.424.2180M}.
Numerical simulations have also shown that the presence of galaxy bars can drive enhance gas mixing, flattening the metallicity profile \citep{1994ApJ...430L.105F,2011A&A...527A.147M}.
These effects of bars on metallicity been borne out by observations \citep[e.g][]{1994ApJ...424..599M, 1999ApJ...516...62D}.
However, it has also been shown that when metallicity gradients are expressed units of effective disk radius, rather than physical distance, bars show no significant impact on the metallicity gradient \citep{2014A&A...563A..49S}.
Even so, it is prudent when comparing the metallicity profiles of the \ion{H}{i}-rich and control populations that we take care to exclude the potential impact of bars.

By visual inspection we classify 30\% of the Bluedisk galaxies to be strongly barred (with four galaxies indeterminate due to high inclination).
This rate is consistent with galaxies of the same stellar mass \citep{2012MNRAS.423.1485S}.
In Fig.~\mbox{\ref{fig:profile_stack}(d)} we show stacked metallicity profiles of the unbarred galaxies.
The distinction between \ion{H}{i}-rich and control samples clearly remains after excluding barred galaxies.
In Fig.~\mbox{\ref{fig:profile_stack}(f)} we show the effects of bars on the whole sample, and in Fig.~\mbox{\ref{fig:profile_stack}(g)} the effects of bars exclusively on the control sample.
From both of these figures we notice some flattening of the metallicity profile in galaxies with bars.
We caution the reader that our sample of \ion{H}{i}-rich galaxies with bars is very small.

We have repeated the stacking analyses for the other metallicity calibrators. We see similar effects when using the KK04 method, but we do not observe the outer metallicity drops when using the NS method. It should be noted, however, that the NS method does not allow for variations in the ionization parameter. Methods that do not include this extra dimensionality may not be best suited for the study we present here.

Finally, we note in Fig.~\mbox{\ref{fig:profile_stack}(a)} there appears to be a significant amount of scatter in the inner regions of the galaxies.
We identify galaxies that harbour AGN using the (\ion{O}{iii}/\Hbeta{}) and (\ion{N}{ii}/\Halpha{}) emission-line ratios from the centre of each galaxy.
We adopt the criterion of \citet{2003MNRAS.346.1055K} and identify galaxies with central non-SF emission.
We find that galaxies with central non-SF emission exhibit a two-fold increase in the scatter of the metallicities in the inner region ($r < 0.3$\,$\textrm{R}_{90,\mathrm{r}}$).
This might be an indication of AGN interacting with the central environment.
However, it is equally plausible that we are not sufficiently excluding non-SF contaminated data points, producing erroneous metallicity estimates.
Our long-slit spectroscopic observations are not ideal for such study of metallicity scatter, integral field spectroscopy with good spatial resolution may provide sufficient data to study both the radial and azimuthal metallicity scatter in the inner regions of galaxies.

\subsection{Summary of results}

In the next section we shall construct a simple analytical model to explain the metallicity profiles we have observed. But first we shall briefly summarize our results:

\begin{itemize}
\item We reproduce the recently reported local mass-metallicity relation. However, we highlight that at low stellar-mass densities there appears to be a residual correlation of metallicity with gas-mass density.
\item The metallicity gradients of a galaxy is strongly correlated with its \ion{H}{i} mass fraction.
\item We stack the metallicity of the galaxies and derive average profile shapes. We find different average profiles for the \ion{H}{i}-rich and control galaxy samples.
\item We find galaxies in both samples that exhibit transitions from shallower inner metallicity gradients to steeper outer metallicity gradients.
\item Barred galaxies appear to have flatter metallicity profiles, but this effect does not drive the difference observed between the Bluedisk samples.
\end{itemize}



\section{Discussion}\label{sec:discussion}

Up to this point we have mainly concerned ourselves with the similarities and differences between our Bluedisk samples.
We have, however, not yet suitably tackled the complex issue regarding the origin of the metallicity profile itself.
Exploring quantitatively the interplay of the many potential mechanisms is challenging. However recent
years have seen the emergence of a class of simple ``reservoir'' models \citep{2010ApJ...718.1001B} in which stars 
form from a gas reservoir regulated by the star formation and gas flows in and out of the system. While simple, these
models are able to provide simple descriptions of the (central) metal content of galaxies at low  redshift \citep{2008MNRAS.385.2181F,2012MNRAS.421...98D,2013ApJ...772..119L}. The models are also naturally interpreted as a result of galaxies being generally close to equilibrium between star formation, outflows and inflows \citep{2012MNRAS.421...98D}.

\subsection{Modelling resolved galaxies as local gas regulators}\label{sec:deriveLGR}

The reservoir models generally consider the galaxies to be spatially unresolved. Here we therefore will 
develop a simple  extension of these models to a resolved galaxy. In particular we will extend  the  ``gas regulator''
model \citep[][herein L13]{2013ApJ...772..119L} which has been shown to successfully fit the 
central metallicites of star forming galaxies in the SDSS.  Our approach will be to minimally extend this model to 2D to see whether such a simple extension is sufficient to describe the metallicity profiles of our galaxies.

To do this we envisage our disc galaxy divided into a set of radial zones. We then assume that the mean properties of each radial zone can be described by individual gas reservoir models. Alternatively one might take this to mean that we assume that each radial zone individually is in an equilibrium between inflow, outflow and star formation -- a detailed balance principle which is not required by the reservoir models in general. Note that we do not assume that such an equilibrium holds at each point but rather in an average sense across a radial bin.

Generally there might be radial mass transfer between these zones, but in the following we will make the simplifying assumption that radial mass transfer can be ignored.
Since, semi-analytic models of \citet{2013MNRAS.434.1531F} have argued that gas flows are of minor importance. It should be noted, however, that the simulations of \citet{2011A&A...527A.147M} have shown that in the presence of bars, gas can be efficiently transported resulting in flattened metallicity gradients.

In addition to gas flows, long-lived stars are expected to migrate from their original radius, particularly in the presence of bars \citep{2008ApJ...684L..79R,2013A&A...553A.102D}. However, in the model we will assume that the mass of stars observed at a given radius represents the total mass of stars formed that given radius.
Or rather, we assume the stars remain associated to the gas from which they form.

We now outline how we adapt and apply the gas regulator model to our data. We refer the reader to L13 for a full treatment and derivation of the model.

\subsubsection{Transport of gas}

The underlying equation describing the rate of change of the reservoir gas mass (in each radial bin) is
\begin{equation}
\dot{m}_\mathrm{gas} = \dot{m}_\mathrm{in} - \dot{m}_\mathrm{out} - \dot{m}_\ast + \dot{m}_\mathrm{return} + \dot{m}_\mathrm{radial} .
\label{eq:reg_gas_basic}
\end{equation}
The components are as follows:
\begin{itemize}

\item $\dot{m}_\mathrm{in}$ is the rate of metal-poor gas inflowing from the halo to the reservoir.
We do not explicitly parametrize $\dot{m}_\mathrm{in}$ and it shall be eliminated in due course.

\item $\dot{m}_\mathrm{out}$ is the rate at which gas flows out from the reservoir and into the halo or beyond.
Since we are mostly concerned with the star forming disk, this is assumed to be driven by winds from massive stars, and we therefore consider $\dot{m}_\mathrm{out}$ to be linearly proportional linearly proportional SFR, i.e. $\dot{m}_\mathrm{out} = \lambda \cdot  \textrm{SFR}$, where $\lambda$ is the mass-loading factor.
In Section \ref{subsubsec:estimate_mass_loading} we attempt to estimate this mass-loading factor. 

\item $\dot{m}_\ast$ is the rate at which gas is converted into stars. In other words $\dot{m}_\ast = \textrm{SFR}$.
The SFR is itself assumed to be linearly proportional to the current mass of the reservoir, $\textrm{SFR} = \epsilon \cdot m_\mathrm{gas}$, where $\epsilon$ is the star-formation efficiency. This link between $m_\mathrm{gas}$ a SFR provides the regulatory aspect of the model.

\item $\dot{m}_\mathrm{return}$ is the rate at which enriched gas is returned from short-lived high-mass stars.
A fraction $R$ of the mass converted into stars is assumed to be instantaneously recycled back into the reservoir ($\dot{m}_\mathrm{return} = R \cdot \textrm{SFR}$).

Following L13 we adopt a fixed value of the return fraction $R=0.4$.
As shown by BC03 this is the mid-range value over a variety of initial mass functions (IMF), for a 10\,Gyr stellar population.
Provided there are no strong age gradients across the galaxies it is reasonable to adopt a radially constant return fraction.
The exact value of the return fraction will depend on the choice of IMF. We note, however, that our conclusions are not sensitive the exact value we adopt for $R$. This insensitivity results from the degeneracy of $R$ with parameters that we shall fit (see Section~\ref{subsubsec:fitting_reg_model}).

\item $\dot{m}_\mathrm{radial}$ is the rate at which radial flows within the disc change the gas content of the reservoir.
For simplicity we assume $\dot{m}_\mathrm{radial} = 0$, neglecting the effects of radial flows.

\end{itemize}

With these principle assumptions, equation \ref{eq:reg_gas_basic} can be written as
\begin{equation}
\dot{m}_\mathrm{gas} = \dot{m}_\mathrm{in} - (1 - R + \lambda) \textrm{SFR} .
\label{eq:reg_gas}
\end{equation}

Furthermore L13 show that by introducing the variable $r_\mathrm{gas} = m_\mathrm{gas} / m_\ast$, the ratio of gas-to-stellar mass, equation \ref{eq:reg_gas} can be conveniently rewritten as
\begin{equation}
\dot{m}_\mathrm{in} = \left( (1 - R) (1 + r_\mathrm{gas}) + \lambda + \epsilon^{-1} \frac{d\ln(r_\mathrm{gas})}{dt}\right) \cdot \textrm{SFR},
\label{eq:reg_gas_rearranged}
\end{equation}
which makes the regulatory link between the star formation rate and gas inflow explicit. We will assume that this holds in the mean in each radial bin.

\subsubsection{Transport of metals}

We now consider the flow of metals into and out from the reservoir.
In the absence of radial flows, analogously to equation \ref{eq:reg_gas_basic} we can write the rate of change of metals in the reservoir as
\begin{equation}
\dot{m}_{Z,\mathrm{gas}} = \dot{m}_{Z,\mathrm{in}} - \dot{m}_{Z,\mathrm{out}} - \dot{m}_{Z,\ast} + \dot{m}_{Z,\mathrm{return}}.
\label{eq:reg_metal_basic}
\end{equation}
This contains two source terms and two sink terms.
The components are as follows:
\begin{itemize}

\item $\dot{m}_{Z,\mathrm{in}}$ represents the metals introduced from the metal-poor halo.
We define this gas to have a typical metallicity $Z_0$.

\item $\dot{m}_{Z,\mathrm{out}}$ is the metal mass entrained in wind driven outflows.
The metallicity of this gas is that of the reservoir, $Z$.

\item $\dot{m}_{Z,\ast}$ represents the mass locked into long-lived stars, removing gas with metallicity $Z$. 

\item $\dot{m}_{Z,\mathrm{return}}$ is the metal enrichment resulting from star formation.
The characteristic yield, $y$, is defined as the metal mass returned per unit mass in long-lived stars.

\end{itemize}

With these principle assumptions, equation \ref{eq:reg_metal_basic} can be expressed as
\begin{equation}
\dot{m}_{Z,\mathrm{gas}} = Z_0 \dot{m}_\mathrm{in} - Z (1 - R + \lambda) \textrm{SFR} + y (1 - R) \textrm{SFR}.
\label{eq:reg_metal}
\end{equation}

The rate of change of reservoir metallicity can be written
\begin{equation}
\dot{Z} = \sfrac{1}{m_\textrm{gas}} \left(\dot{m}_{Z,\mathrm{gas}} - Z \dot{m}_{\mathrm{gas}} \right) .
\end{equation}

L13 find that the metallicity of such a system will approach equilibrium on a timescale shorter than the depletion timescale (i.e. $\le t_\mathrm{dep} =  \epsilon^{-1}$).
In which case they show the equilibrium metallicity to be
\begin{equation}
Z_\mathrm{eq} = Z_{0} + \frac{y}{1 + r_\mathrm{gas} + (1-R)^{-1} \left(\lambda + \epsilon^{-1} \frac{d\ln(r_\mathrm{gas})}{dt}\right)}.
\label{eq:reg_equib_metal}
\end{equation}
We now have an expression for the equilibrium metallicity of the system as a function of $r_\mathrm{gas}$, a quantity we have already obtained (see Section~\ref{subsec:local_mass_metallicity}). 
We highlight that there are other conceptually interesting ways of interpreting $r_\mathrm{gas}$, which can be alternatively be written as
\begin{equation}
r_\mathrm{gas} = \frac{m_\mathrm{gas}}{m_\ast} = \frac{\epsilon^{-1}\textrm{SFR}}{m_\ast} = \epsilon^{-1}\textrm{sSFR} , 
\end{equation}
where sSFR is the specific star formation rate.

By fitting their model to star forming galaxies from the SDSS, L13 estimate $\epsilon^{-1} \frac{d\ln(r_\mathrm{gas})}{dt} \approx -0.25$, and we shall adopt this value.
We note, however, that the model is not strongly sensitive to this factor, owing to the degeneracies arising from fitting the $y$ and $Z_0$ parameters (Section~\ref{subsubsec:fitting_reg_model}).

\subsubsection{Estimating the mass-loading factor, $\lambda$}\label{subsubsec:estimate_mass_loading}

We have parametrized wind-driven outflows via $\dot{m}_\mathrm{out} = \lambda \cdot \textrm{SFR}$, where $\lambda$ is a mass-loading factor.
We shall consider two scenarios, one with winds and the other without.
We note here that this choice is not important, for we will show in Section~\ref{section:showLGR} that these two scenarios are highly degenerate with $y$ and $Z_0$ parameters, which we shall fit.

In the simple windless scenario we will set $\lambda = 0$ everywhere.

For our more complex windy model, we consider outflows that are powered by momentum-driven winds from supernovae (SNe). We follow the prescription described in \citet{2009MNRAS.396..141D} to define the mass-loading factor
\begin{equation}
\lambda = \frac {p_{\textsc{sn}} \eta_{\textsc{sn}}} {V_{\mathrm{esc}}(\vec{r})} ,
\end{equation}
where $p_{\textsc{sn}}=3\times10^4\,\textrm{M}_\odot\,\textrm{km}\,\textrm{s}^{-1}$ is the momentum per SN, $\eta_{\textsc{sn}} = 8.3\times10^{-3}$ is the number of SNe per solar mass of stars formed, and $V_{\mathrm{esc}}(\vec{r})$ is the escape velocity at a given point, $\vec{r}$, in the disc.
The escape velocity itself is defined in terms of the gravitational potential
\begin{equation}
V_{\mathrm{esc}}(\vec{r}) = \sqrt{2\left|\Phi_\mathrm{tot}(\vec{r})\right|} ,
\end{equation}
where the gravitational potential,  $\Phi_\mathrm{tot}$, is the sum of contributions from stars, gas and dark matter
\begin{equation}
\Phi_\mathrm{tot}(\vec{r}) = \Phi_\ast(\vec{r}) + \Phi_\mathrm{gas}(\vec{r}) + \Phi_\mathrm{DM}(\vec{r}) .
\end{equation}
We must calculate the gravitational potential at the position in the galaxy where each reservoir/zone is situated.
To achieve this we model the three potentials separately. We shall use thin discs to represent both the stellar and gaseous components. The dark matter component we assume to be distributed in a spherically symmetric halo.

We determine the stellar contribution using the stellar mass maps (Section \ref{subsec:stellar_mass_densities}).
Assuming that the stars lie in a thin plane, we assign a point mass to every map pixel.
The potential at any point is the galaxy is then calculated as a sum of the individual point mass potentials, i.e.
\begin{equation}
\Phi_\ast(\vec{r}) = - \sum\limits_i \frac{G \Delta m_i}{\left| \vec{r}_i - \vec{r} \right|} ,
\end{equation}
where $G$ is the gravitational constant, $\Delta m_i$ is the mass of a pixel, and $\left| \vec{r}_i - \vec{r} \right|$ is the distance in the plane of the galaxy to the centre of the mass pixel.

To estimate the gravitational potential arising from the gas, we adopt the following characteristic surface density profile from \citet{2012ApJ...756..183B}
\begin{equation}
\frac{\Sigma_\textrm{gas} }{14\ \textrm{M}_\odot\,\textrm{pc}^{-2}} =  2.1 \exp\left( -1.65 r / r_{25} \right) ,
\end{equation}
where $r_{25}$ is the optical radius where the surface brightness becomes 25\,mag\,arcsec$^{-2}$.
We use equation~2.164a from \citet{2008gady.book.....B} to calculate the gas contribution to the potential.
It was shown in \citet{2014MNRAS.441.2159W} that the characteristic surface density profile provides a good description of the \ion{H}{i}-rich galaxies in our sample. Admittedly the profile does not provide as good an approximation to the control sample. However, since the contribution of the gas to the total potential is small ($\la10$\%) this will not affect our conclusions.

The dark matter halo provides the dominant contribution to the halo, making up 50--80\% of the total potential, but it is also the most uncertain as we do not have direct constraints on its properties. In view of this we follow common practice to parametrise the dark matter halo mass distribution with the spherically symmetric NFW profile \citep{1997ApJ...490..493N}. In order to do so we the halo mass and concentration. We get the former from halo mass-stellar relation derived by \citet[their equation~3]{2010MNRAS.404.1111G} and the halo concentration from \citet[their fig.~3]{2007MNRAS.378...55M} with the virial radius of the halo using equation~3 from \citet{2009MNRAS.396..141D}. With this best guess dark matter potential, combined with the potentials of the stellar and gas discs, we are now able to estimate the mass-loading factor $\lambda$.

\subsubsection{Fitting $y$ and $Z_0$}\label{subsubsec:fitting_reg_model}

Two components in the model remain unconstrained, namely the yield and the metallicity of the infalling gas.

The stellar yield, $y$, represents the metallicity of the gas returned by short-lived stars.
If we assume there is a universal initial mass function, then we expect $y$ to be constant between galaxies and independent of location within a specific galaxy. The stellar yield can in principle be calculated from stellar evolution models. 
However, the large and poorly understood systematic offsets between the various gas-phase metallicity indicators (see Appendix~\ref{app:comparing_indicators}) mean that we are unable to determine absolute abundances for our galaxies so we have decided to assume that  $y$ is the same for all galaxies, but unknown so we fit it as a global constant.

The metallicity of gas infalling from the halo, $Z_0$, is a poorly known quantity.
For simplicity we therefore assume that the infalling gas has the same metallicity at all radii for each galaxy and that  the halos of all the Bluedisk galaxies have the same metallicity. This may not be a bad assumption since all the galaxies are of similar total stellar mass, and therefore may possess similar mass halos.
As with the stellar yield, our prediction of $Z_0$ also suffers from effects of systematic offsets due to the choice in metallicity indicators.
Thus we also make $Z_0$ a global constant that is to be fit.

In summary given the stellar mass maps and gas mass distributions, the resulting model, equation~\ref{eq:reg_equib_metal}, has two free parameters, $y$ and $Z_0$. These global parameters, namely the stellar yield and the halo metallicity, are fit for all radial bins, across all galaxies, simultaneously. Due to the systematic offsets between metallicity indicators, we caution that inferences should not be made on the fitted values themselves.

\subsection{Bluedisk galaxies as local gas regulators}\label{section:showLGR}

\begin{figure*}
\includegraphics[width=\linewidth]{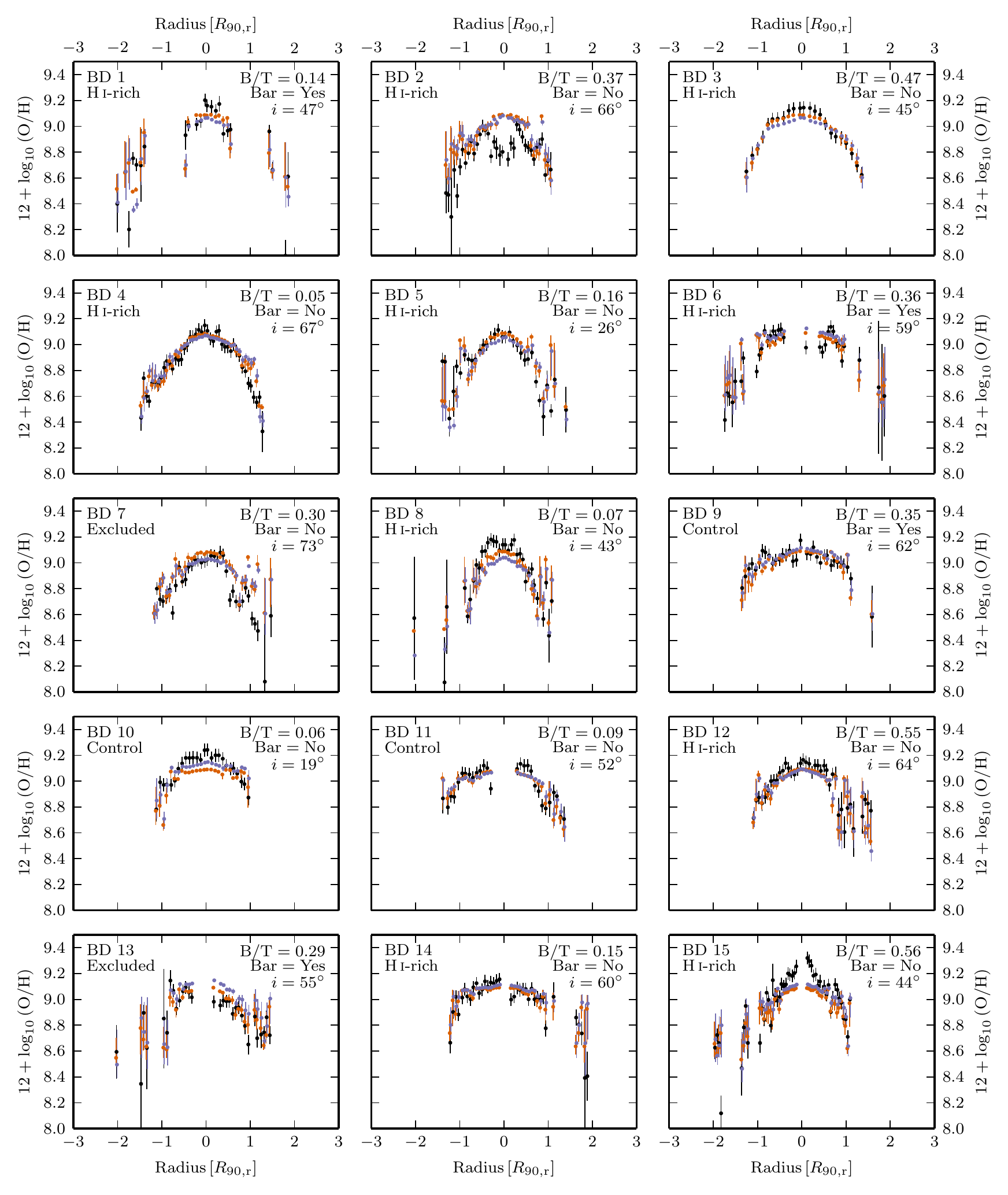}
\caption{The local gas regulator model compared to the metallicity profiles of the Bluedisk galaxies. The observed CL01 metallicity is shown in black. Windless and windy models are shown in orange and blue, respectively. Error bars on the models are not the true full error, but rather they indicate the effects of $\pm$\,1\,$\sigma$ deviations in the gas-to-stellar mass ratio, $r_\mathrm{gas}$. In the top right of each plot we label the bulge-to-total light ratio, $\textrm{B}/\textrm{T}$. We also denote whether a bar is present. In highly inclined systems where we that would not be able to determine the presence of a bar, we denote this with a ``?'' symbol. Since our modelling may be problematic at high inclinations, we also include the measured inclination $i$. The globally fitted parameter values are $y = 1.27\times10^{-3}$, $Z_{0} = 1.91\times10^{-4}$ and $y = 0.55\times10^{-3}$, $Z_{0} = 2.97\times10^{-4}$ in the windy and windless cases, respectively.}
\label{fig:LGR_fits}
\end{figure*}

Having outlined the local gas regulator model, we demonstrate the results for all 50 Bluedisk galaxies in Fig.~\ref{fig:LGR_fits}.
Surprisingly this simplistic equilibrium model appears to match well for many, but by no means all, of the galaxies.
It is strikingly clear that with only two globally-set free parameters we can reproduce a large variety of observed metallicity profile shapes that these galaxies exhibit.
The model also reproduces the observed outer metallicity drops, which is attributed to the transition from a stellar-dominated inner disc to a more gas-dominated outer disc.

The centres of the galaxies appear to be most problematic for the local gas regulator model to reproduce.
One of our key model assumptions is that we assume independence between radial zones.
But the presence of bars and bulges at the centres of galaxies might invalidate this assumption.
For example, a bar could be expected to drive strong radial flows inwards, which if this were the case, we could expect steepened metallicity gradients \citep{1992A&A...262..455G}.
However, we see no obvious connection between deviations from the local gas regulator model and the presence of a strong bars or a prominent bulges.

\addtocounter{figure}{-1}
\begin{figure*}
\includegraphics[width=\linewidth]{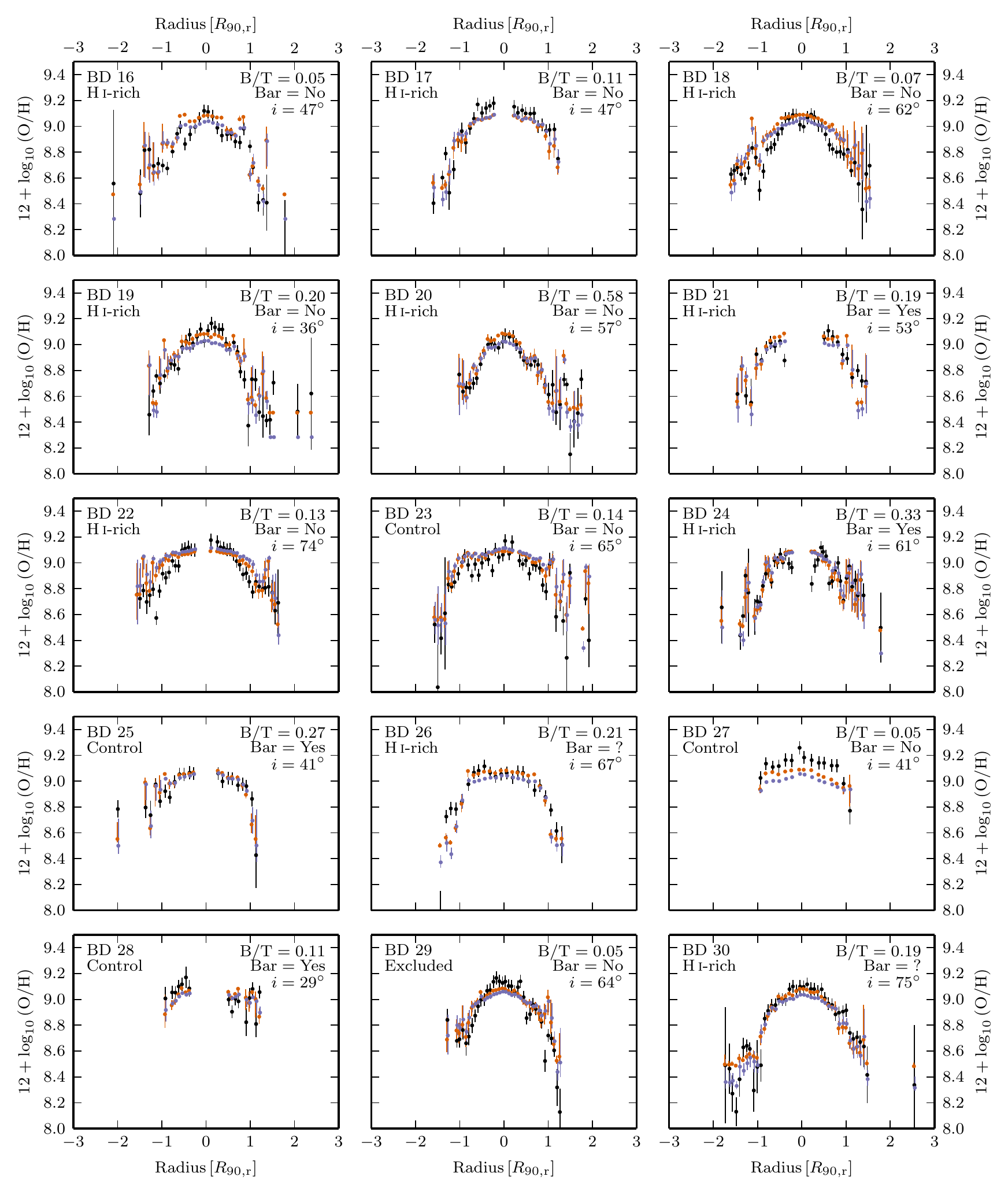}
\caption{The local gas regulator model compared to the metallicity profiles of the Bluedisk galaxies -- \textit{continued}.}
\end{figure*}

\addtocounter{figure}{-1}
\begin{figure*}
\includegraphics[width=\linewidth]{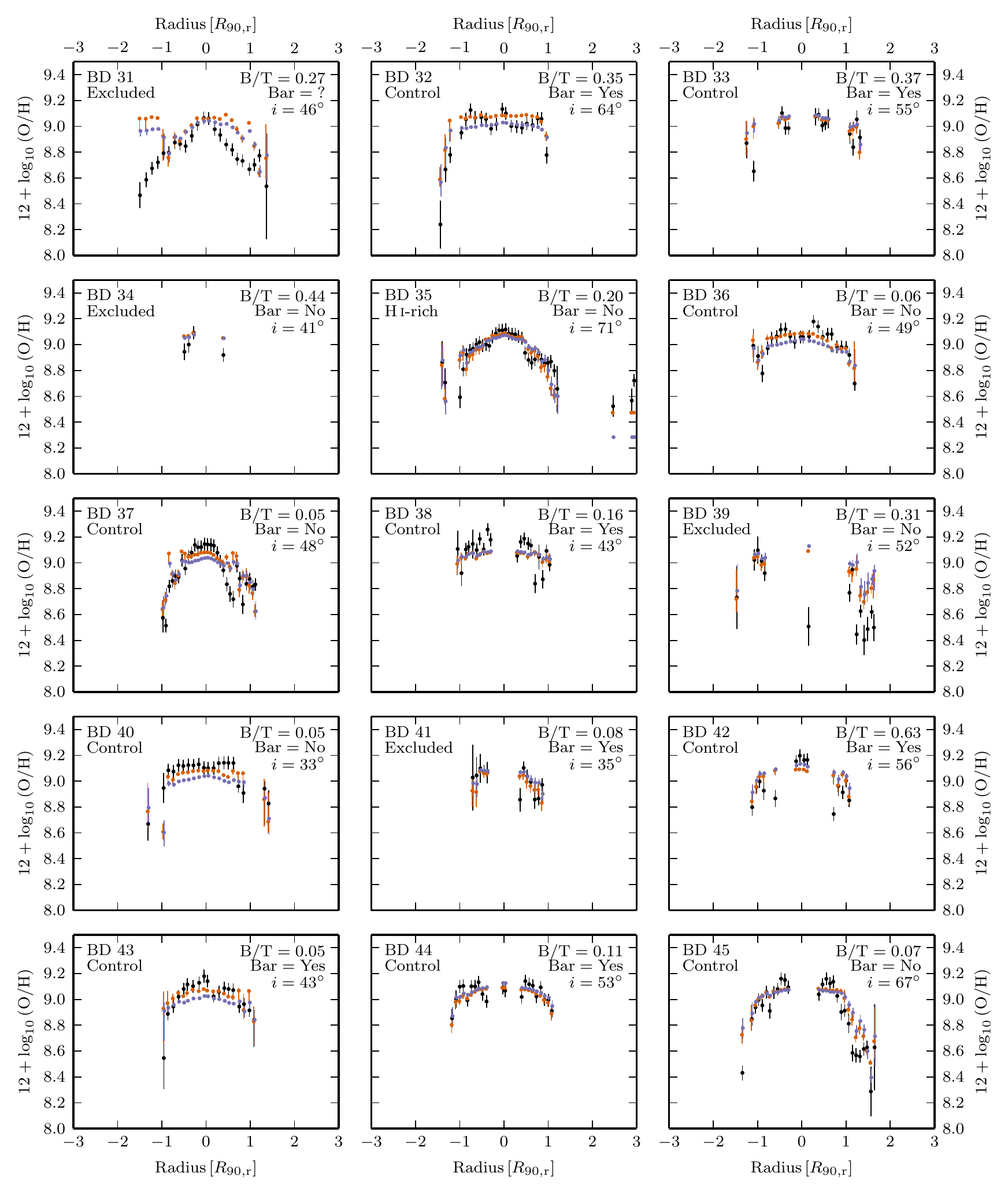}
\caption{The local gas regulator model compared to the metallicity profiles of the Bluedisk galaxies -- \textit{continued}.}
\end{figure*}

\addtocounter{figure}{-1}
\begin{figure*}
\includegraphics[width=\linewidth]{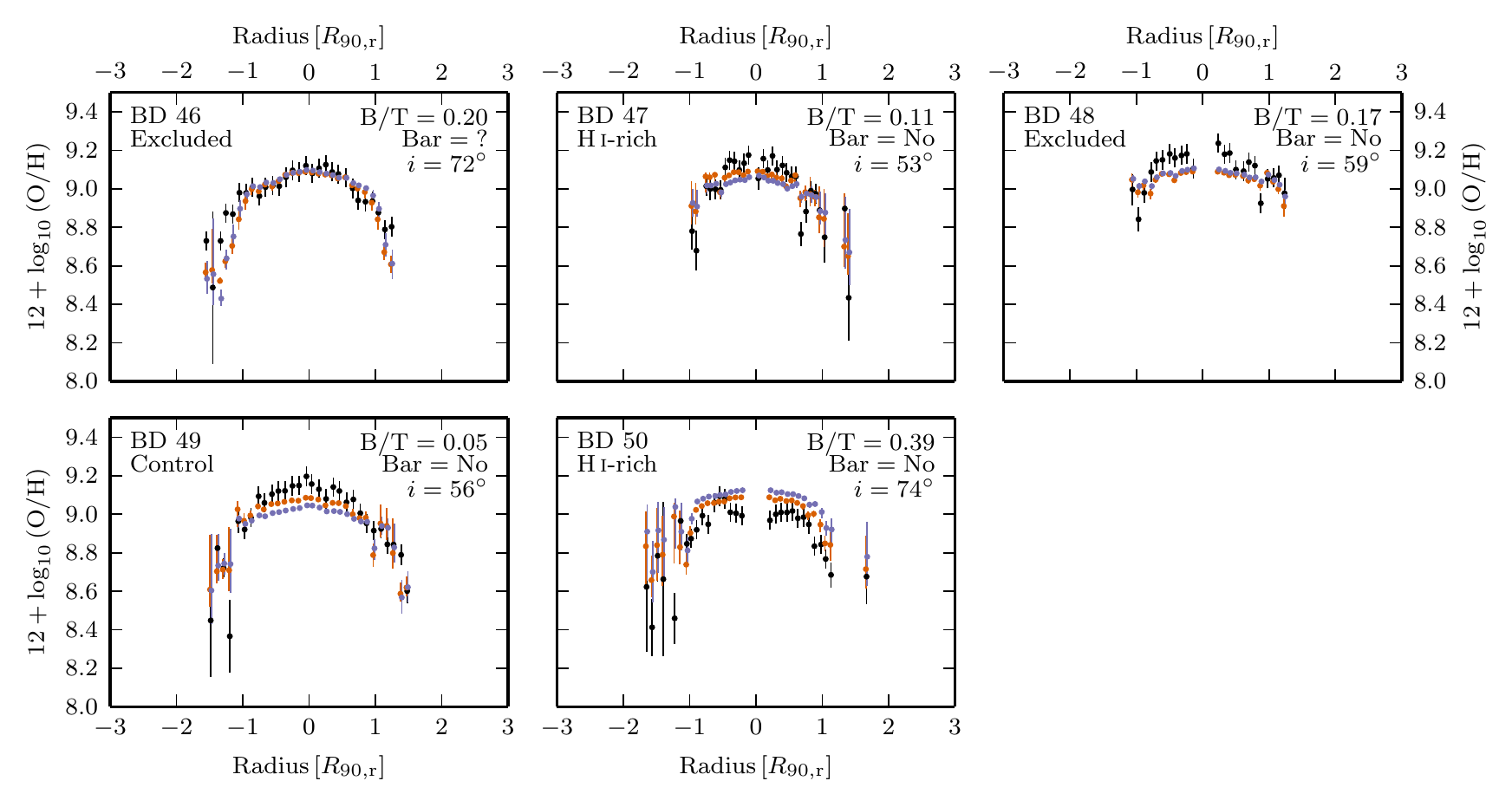}
\caption{The local gas regulator model compared to the metallicity profiles of the Bluedisk galaxies -- \textit{continued}.}
\end{figure*}

Alternatively the failure of the model may indicate that some of these deviant central regions are contaminated by emission whose origin is not photoionization, such as shocks and LINERs, which our selection criteria have failed to exclude.
Emission from non-photoionizing origins can impact different metallicity determination methods in different ways.
Although as shown in Appendix~\ref{app:comparing_indicators} different metallicity indicators yield different absolute and relative abundances, we should expect there to exist a monotonic mapping between the methods.
For example, in Fig.~\ref{fig:profiles_blue} we can see the inner regions of BDs 2,15 that they show contradictory behaviour of the metallicity of the CL01 and KK04 indicators. This primarily indicates contamination from non-photoionizing origins, thus it is not surprising the local gas regulator model appears to fail in these regions.

We find no difference in the quality of fit of the local gas regulator model between the \ion{H}{i}-rich and control galaxies.
Overall both our \ion{H}{i}-rich, control galaxies fitting equally well.
However, our local gas regulator model should not be expected to succeed for galaxies that are interacting, since interactions could also drive strong gas flows.
Indeed our excluded (non-isolated) sample of galaxies exhibit some of the most discrepant fits, e.g. BD~31, which has a very different metallicity profile from the one predicted by the model.

We construct the local gas regulator model with momentum-driven winds and windless cases, both of which appear to represent the data more or less equally well.
This is not because winds have no effect, indeed they do impact strongly on the metallicity, but the effects of the wind is largely degenerate with our fitted parameters: $y$, and $Z_{0}$.
The wind model we employ modifies the metallicity changing the peak metallicity and inner slope (where $r_\mathrm{gas} \sim 0$).
The loss of metals due to winds can be compensated by increasing the values of $y$ and/or $Z_{0}$.
Since we are forbidden from interpreting the values of $y$ and $Z_{0}$, we are unable to conclude anything either for or against the existence of enriched wind-driven outflows.

We note that our results here are compatible with the recent work of \citet{2015MNRAS.448.2030H}.
Using a analytical formalism similar to the L13 gas regulator they are able to reproduce the distribution of metallicity gradients observed.
Therein the metallicity profile is also determined by the current ratio of gas-to-stellar mass.


\section{Conclusions}\label{sec:conclusions}

We present radial gas-phase metallicity profiles of 50 late-type galaxies that form the Bluedisk survey. We explore how the \ion{H}{i} content of these galaxies affects their metallicity profiles.
Although we find a correlation between \ion{H}{i} mass fraction and the metallicity gradient, we observe that the metallicity profiles of our \ion{H}{i}-rich and control samples show remarkable similarity. Furthermore we find that using a simple equilibrium model we are able to approximate the metallicity profiles of both samples with equal success.
We summarize our main results as follows:

\begin{enumerate}

\item We confirm the local mass-metallicity relation for the Bluedisk galaxies. Although we note that at low stellar mass density there appears to be a residual anti-correlation of metallicity with gas mass density.

\item The metallicity gradient of the galaxies is strongly correlated with their \ion{H}{i} mass fraction. Galaxies with higher \ion{H}{i} mass fractions have steeper metallicity gradients.\label{item:b}.

\item We find that in some galaxies the outer disc exhibits steeper metallicity gradients than in the inner disc. However, unlike previous work that has shown this, we find these galaxies to be ubiquitous in both the \ion{H}{i}-rich and control samples.

\item The barred galaxies in our sample bars tend to have flatter metallicity profiles. This is not sufficient, however, to explain observed differences between the metallicity profiles of the \ion{H}{i}-rich and control samples.


\item By applying a simple equilibrium analytical model, we find that we are able to approximate the metallicity profile shapes with the ratio of gas-to-stellar mass, $r_\mathrm{gas} = \Sigma_{\mathrm{gas}} / \Sigma_{\ast}$. In the outer disc where $\Sigma_{\ast}$ is low, a transition to  $r_\mathrm{gas} > 1$ occurs. This naturally gives rise to the steeper outer metallicity gradients which are observed.

\end{enumerate}

If metallicity is truly in equilibrium, this would naturally explain the local mass-metallicity relation.
Also, since the dynamic range of the stellar mass density is much greater than the gas mass density, the overall metallicity profile represents the integrated build up of metals. 


\section*{Acknowledgments}

We would like thank the referee for their detailed comments, which have helped to improve this paper.
We also thank Rob Crain and Simon Lilly who provided useful and stimulating discussions for this work.

This work is based on observations performed at the William Herschel Telescope under the programmes W12BN010 and W13AN010.
The William Herschel Telescope is operated on the island of La Palma by the Isaac Newton Group in the Spanish Observatorio del Roque de los Muchachos of the Instituto de Astrof\'{i}sica de Canarias.

FB acknowledges support from DFG grant BI 1546/1-1.
JMvdH acknowledges support from the European Research Council under the European Union's Seventh Framework Programme (FP/2007-2013) / ERC Grant Agreement nr. 291531.

Funding for the SDSS and SDSS-II has been provided by the Alfred P. Sloan Foundation, the Participating Institutions, the National Science Foundation, the U.S. Department of Energy, the National Aeronautics and Space Administration, the Japanese Monbukagakusho, the Max Planck Society, and the Higher Education Funding Council for England. The SDSS Web Site is http://www.sdss.org/.

The SDSS is managed by the Astrophysical Research Consortium for the Participating Institutions. The Participating Institutions are the American Museum of Natural History, Astrophysical Institute Potsdam, University of Basel, University of Cambridge, Case Western Reserve University, University of Chicago, Drexel University, Fermilab, the Institute for Advanced Study, the Japan Participation Group, Johns Hopkins University, the Joint Institute for Nuclear Astrophysics, the Kavli Institute for Particle Astrophysics and Cosmology, the Korean Scientist Group, the Chinese Academy of Sciences (LAMOST), Los Alamos National Laboratory, the Max-Planck-Institute for Astronomy (MPIA), the Max-Planck-Institute for Astrophysics (MPA), New Mexico State University, Ohio State University, University of Pittsburgh, University of Portsmouth, Princeton University, the United States Naval Observatory, and the University of Washington.

We wish to also acknowledge the \texttt{Python} programming language and the Interactive Data Language (\texttt{IDL}) which were both used throughout this project.

\bibliographystyle{mn2e}
\bibliography{2014_bd_metallicity}

\begin{thebibliography}{85}
\expandafter\ifx\csname natexlab\endcsname\relax\def\natexlab#1{#1}\fi

\bibitem[{{Baldwin}, {Phillips} \& {Terlevich}(1981){Baldwin}, {Phillips}, \&
  {Terlevich}}]{1981PASP...93....5B}
{Baldwin} J.~A., {Phillips} M.~M., {Terlevich} R., 1981, \pasp, 93, 5

\bibitem[{{Bertin} \& {Arnouts}(1996)}]{1996A&AS..117..393B}
{Bertin} E., {Arnouts} S., 1996, \aaps, 117, 393

\bibitem[{{Bigiel} \& {Blitz}(2012)}]{2012ApJ...756..183B}
{Bigiel} F., {Blitz} L., 2012, \apj, 756, 183

\bibitem[{{Binney} \& {Tremaine}(2008)}]{2008gady.book.....B}
{Binney} J., {Tremaine} S., 2008, {Galactic Dynamics: Second Edition}.
  Princeton University Press

\bibitem[{{Blanc} {et~al}\mbox{.}(2015){Blanc}, {Kewley}, {Vogt}, \&
  {Dopita}}]{2015ApJ...798...99B}
{Blanc} G.~A., {Kewley} L., {Vogt} F.~P.~A., {Dopita} M.~A., 2015, \apj, 798,
  99

\bibitem[{{B{\"o}hm} {et~al}\mbox{.}(2004){B{\"o}hm}, {Ziegler}, {Saglia},
  {Bender}, {Fricke}, {Gabasch}, {Heidt}, {Mehlert}, {Noll}, \&
  {Seitz}}]{2004A&A...420...97B}
{B{\"o}hm} A. {et~al.}, 2004, \aap, 420, 97

\bibitem[{{Bouch{\'e}} {et~al}\mbox{.}(2010){Bouch{\'e}}, {Dekel}, {Genzel},
  {Genel}, {Cresci}, {F{\"o}rster Schreiber}, {Shapiro}, {Davies}, \&
  {Tacconi}}]{2010ApJ...718.1001B}
{Bouch{\'e}} N. {et~al.}, 2010, \apj, 718, 1001

\bibitem[{{Brinchmann} {et~al}\mbox{.}(2013){Brinchmann}, {Charlot},
  {Kauffmann}, {Heckman}, {White}, \& {Tremonti}}]{2013MNRAS.432.2112B}
{Brinchmann} J., {Charlot} S., {Kauffmann} G., {Heckman} T., {White} S.~D.~M.,
  {Tremonti} C., 2013, \mnras, 432, 2112

\bibitem[{{Brinchmann} {et~al}\mbox{.}(2004){Brinchmann}, {Charlot}, {White},
  {Tremonti}, {Kauffmann}, {Heckman}, \& {Brinkmann}}]{2004MNRAS.351.1151B}
{Brinchmann} J., {Charlot} S., {White} S.~D.~M., {Tremonti} C., {Kauffmann} G.,
  {Heckman} T., {Brinkmann} J., 2004, \mnras, 351, 1151

\bibitem[{{Bruzual} \& {Charlot}(2003)}]{2003MNRAS.344.1000B}
{Bruzual} G., {Charlot} S., 2003, \mnras, 344, 1000

\bibitem[{{Cappellari} \& {Copin}(2003)}]{2003MNRAS.342..345C}
{Cappellari} M., {Copin} Y., 2003, \mnras, 342, 345

\bibitem[{{Catinella} {et~al}\mbox{.}(2010){Catinella}, {Schiminovich},
  {Kauffmann}, {Fabello}, {Wang}, {Hummels}, {Lemonias}, {Moran}, {Wu},
  {Giovanelli}, {Haynes}, {Heckman}, {Basu-Zych}, {Blanton}, {Brinchmann},
  {Budav{\'a}ri}, {Gon{\c c}alves}, {Johnson}, {Kennicutt}, {Madore}, {Martin},
  {Rich}, {Tacconi}, {Thilker}, {Wild}, \& {Wyder}}]{2010MNRAS.403..683C}
{Catinella} B. {et~al.}, 2010, \mnras, 403, 683

\bibitem[{{Charlot} \& {Fall}(2000)}]{2000ApJ...539..718C}
{Charlot} S., {Fall} S.~M., 2000, \apj, 539, 718

\bibitem[{{Charlot} \& {Longhetti}(2001)}]{2001MNRAS.323..887C}
{Charlot} S., {Longhetti} M., 2001, \mnras, 323, 887

\bibitem[{{Cid Fernandes} {et~al}\mbox{.}(2011){Cid Fernandes},
  {Stasi{\'n}ska}, {Mateus}, \& {Vale Asari}}]{2011MNRAS.413.1687C}
{Cid Fernandes} R., {Stasi{\'n}ska} G., {Mateus} A., {Vale Asari} N., 2011,
  \mnras, 413, 1687

\bibitem[{{Dav{\'e}}, {Finlator} \& {Oppenheimer}(2012){Dav{\'e}}, {Finlator},
  \& {Oppenheimer}}]{2012MNRAS.421...98D}
{Dav{\'e}} R., {Finlator} K., {Oppenheimer} B.~D., 2012, \mnras, 421, 98

\bibitem[{{Davoust} \& {Contini}(2004)}]{2004A&A...416..515D}
{Davoust} E., {Contini} T., 2004, \aap, 416, 515

\bibitem[{{Di Matteo} {et~al}\mbox{.}(2013){Di Matteo}, {Haywood}, {Combes},
  {Semelin}, \& {Snaith}}]{2013A&A...553A.102D}
{Di Matteo} P., {Haywood} M., {Combes} F., {Semelin} B., {Snaith} O.~N., 2013,
  \aap, 553, A102

\bibitem[{{Diehl} \& {Statler}(2006)}]{2006MNRAS.368..497D}
{Diehl} S., {Statler} T.~S., 2006, \mnras, 368, 497

\bibitem[{{Dopita} {et~al}\mbox{.}(2013){Dopita}, {Sutherland}, {Nicholls},
  {Kewley}, \& {Vogt}}]{2013ApJS..208...10D}
{Dopita} M.~A., {Sutherland} R.~S., {Nicholls} D.~C., {Kewley} L.~J., {Vogt}
  F.~P.~A., 2013, \apjs, 208, 10

\bibitem[{{Driver} {et~al}\mbox{.}(2011){Driver}, {Hill}, {Kelvin}, {Robotham},
  {Liske}, {Norberg}, {Baldry}, {Bamford}, {Hopkins}, {Loveday}, {Peacock},
  {Andrae}, {Bland-Hawthorn}, {Brough}, {Brown}, {Cameron}, {Ching}, {Colless},
  {Conselice}, {Croom}, {Cross}, {de Propris}, {Dye}, {Drinkwater}, {Ellis},
  {Graham}, {Grootes}, {Gunawardhana}, {Jones}, {van Kampen}, {Maraston},
  {Nichol}, {Parkinson}, {Phillipps}, {Pimbblet}, {Popescu}, {Prescott},
  {Roseboom}, {Sadler}, {Sansom}, {Sharp}, {Smith}, {Taylor}, {Thomas},
  {Tuffs}, {Wijesinghe}, {Dunne}, {Frenk}, {Jarvis}, {Madore}, {Meyer},
  {Seibert}, {Staveley-Smith}, {Sutherland}, \& {Warren}}]{2011MNRAS.413..971D}
{Driver} S.~P. {et~al.}, 2011, \mnras, 413, 971

\bibitem[{{Dutil} \& {Roy}(1999)}]{1999ApJ...516...62D}
{Dutil} Y., {Roy} J.-R., 1999, \apj, 516, 62

\bibitem[{{Dutton} \& {van den Bosch}(2009)}]{2009MNRAS.396..141D}
{Dutton} A.~A., {van den Bosch} F.~C., 2009, \mnras, 396, 141

\bibitem[{{Ellison} {et~al}\mbox{.}(2008){Ellison}, {Patton}, {Simard}, \&
  {McConnachie}}]{2008ApJ...672L.107E}
{Ellison} S.~L., {Patton} D.~R., {Simard} L., {McConnachie} A.~W., 2008, \apjl,
  672, L107

\bibitem[{{Fall} \& {Efstathiou}(1980)}]{1980MNRAS.193..189F}
{Fall} S.~M., {Efstathiou} G., 1980, \mnras, 193, 189

\bibitem[{{Finlator} \& {Dav{\'e}}(2008)}]{2008MNRAS.385.2181F}
{Finlator} K., {Dav{\'e}} R., 2008, \mnras, 385, 2181

\bibitem[{{Foster} {et~al}\mbox{.}(2012){Foster}, {Hopkins}, {Gunawardhana},
  {Lara-L{\'o}pez}, {Sharp}, {Steele}, {Taylor}, {Driver}, {Baldry}, {Bamford},
  {Liske}, {Loveday}, {Norberg}, {Peacock}, {Alpaslan}, {Bauer},
  {Bland-Hawthorn}, {Brough}, {Cameron}, {Colless}, {Conselice}, {Croom},
  {Frenk}, {Hill}, {Jones}, {Kelvin}, {Kuijken}, {Nichol}, {Owers},
  {Parkinson}, {Pimbblet}, {Popescu}, {Prescott}, {Robotham}, {Lopez-Sanchez},
  {Sutherland}, {Thomas}, {Tuffs}, {van Kampen}, \&
  {Wijesinghe}}]{2012A&A...547A..79F}
{Foster} C. {et~al.}, 2012, \aap, 547, A79

\bibitem[{{Friedli}, {Benz} \& {Kennicutt}(1994){Friedli}, {Benz}, \&
  {Kennicutt}}]{1994ApJ...430L.105F}
{Friedli} D., {Benz} W., {Kennicutt} R., 1994, \apjl, 430, L105

\bibitem[{{Fu} {et~al}\mbox{.}(2013){Fu}, {Kauffmann}, {Huang}, {Yates},
  {Moran}, {Heckman}, {Dav{\'e}}, {Guo}, \& {Henriques}}]{2013MNRAS.434.1531F}
{Fu} J. {et~al.}, 2013, \mnras, 434, 1531

\bibitem[{{Gadotti} \& {Kauffmann}(2009)}]{2009MNRAS.399..621G}
{Gadotti} D.~A., {Kauffmann} G., 2009, \mnras, 399, 621

\bibitem[{{Gallazzi} {et~al}\mbox{.}(2005){Gallazzi}, {Charlot}, {Brinchmann},
  {White}, \& {Tremonti}}]{2005MNRAS.362...41G}
{Gallazzi} A., {Charlot} S., {Brinchmann} J., {White} S.~D.~M., {Tremonti}
  C.~A., 2005, \mnras, 362, 41

\bibitem[{{Goetz} \& {Koeppen}(1992)}]{1992A&A...262..455G}
{Goetz} M., {Koeppen} J., 1992, \aap, 262, 455

\bibitem[{{Guo} {et~al}\mbox{.}(2010){Guo}, {White}, {Li}, \&
  {Boylan-Kolchin}}]{2010MNRAS.404.1111G}
{Guo} Q., {White} S., {Li} C., {Boylan-Kolchin} M., 2010, \mnras, 404, 1111

\bibitem[{{Ho} {et~al}\mbox{.}(2015){Ho}, {Kudritzki}, {Kewley}, {Zahid},
  {Dopita}, {Bresolin}, \& {Rupke}}]{2015MNRAS.448.2030H}
{Ho} I.-T., {Kudritzki} R.-P., {Kewley} L.~J., {Zahid} H.~J., {Dopita} M.~A.,
  {Bresolin} F., {Rupke} D.~S.~N., 2015, \mnras, 448, 2030

\bibitem[{{Hoopes} \& {Walterbos}(2003)}]{2003ApJ...586..902H}
{Hoopes} C.~G., {Walterbos} R.~A.~M., 2003, \apj, 586, 902

\bibitem[{{Hoopes}, {Walterbos} \& {Bothun}(2001){Hoopes}, {Walterbos}, \&
  {Bothun}}]{2001ApJ...559..878H}
{Hoopes} C.~G., {Walterbos} R.~A.~M., {Bothun} G.~D., 2001, \apj, 559, 878

\bibitem[{{Kauffmann} {et~al}\mbox{.}(2003{\natexlab{a}}){Kauffmann},
  {Heckman}, {Tremonti}, {Brinchmann}, {Charlot}, {White}, {Ridgway},
  {Brinkmann}, {Fukugita}, {Hall}, {Ivezi{\'c}}, {Richards}, \&
  {Schneider}}]{2003MNRAS.346.1055K}
{Kauffmann} G. {et~al.}, 2003{\natexlab{a}}, \mnras, 346, 1055

\bibitem[{{Kauffmann} {et~al}\mbox{.}(2003{\natexlab{b}}){Kauffmann},
  {Heckman}, {White}, {Charlot}, {Tremonti}, {Brinchmann}, {Bruzual}, {Peng},
  {Seibert}, {Bernardi}, {Blanton}, {Brinkmann}, {Castander}, {Cs{\'a}bai},
  {Fukugita}, {Ivezic}, {Munn}, {Nichol}, {Padmanabhan}, {Thakar}, {Weinberg},
  \& {York}}]{2003MNRAS.341...33K}
{Kauffmann} G. {et~al.}, 2003{\natexlab{b}}, \mnras, 341, 33

\bibitem[{{Kennicutt}(1998)}]{1998ARA&A..36..189K}
{Kennicutt}, Jr. R.~C., 1998, \araa, 36, 189

\bibitem[{{Kewley} {et~al}\mbox{.}(2001){Kewley}, {Dopita}, {Sutherland},
  {Heisler}, \& {Trevena}}]{2001ApJ...556..121K}
{Kewley} L.~J., {Dopita} M.~A., {Sutherland} R.~S., {Heisler} C.~A., {Trevena}
  J., 2001, \apj, 556, 121

\bibitem[{{Kewley} \& {Ellison}(2008)}]{2008ApJ...681.1183K}
{Kewley} L.~J., {Ellison} S.~L., 2008, \apj, 681, 1183

\bibitem[{{Kim} {et~al}\mbox{.}(2013){Kim}, {Krumholz}, {Wise}, {Turk},
  {Goldbaum}, \& {Abel}}]{2013ApJ...775..109K}
{Kim} J.-h., {Krumholz} M.~R., {Wise} J.~H., {Turk} M.~J., {Goldbaum} N.~J.,
  {Abel} T., 2013, \apj, 775, 109

\bibitem[{{Kobulnicky} \& {Kewley}(2004)}]{2004ApJ...617..240K}
{Kobulnicky} H.~A., {Kewley} L.~J., 2004, \apj, 617, 240

\bibitem[{{Leroy} {et~al}\mbox{.}(2009){Leroy}, {Walter}, {Bigiel}, {Usero},
  {Weiss}, {Brinks}, {de Blok}, {Kennicutt}, {Schuster}, {Kramer},
  {Wiesemeyer}, \& {Roussel}}]{2009AJ....137.4670L}
{Leroy} A.~K. {et~al.}, 2009, \aj, 137, 4670

\bibitem[{{Lilly} {et~al}\mbox{.}(2013){Lilly}, {Carollo}, {Pipino}, {Renzini},
  \& {Peng}}]{2013ApJ...772..119L}
{Lilly} S.~J., {Carollo} C.~M., {Pipino} A., {Renzini} A., {Peng} Y., 2013,
  \apj, 772, 119

\bibitem[{{Macci{\`o}} {et~al}\mbox{.}(2007){Macci{\`o}}, {Dutton}, {van den
  Bosch}, {Moore}, {Potter}, \& {Stadel}}]{2007MNRAS.378...55M}
{Macci{\`o}} A.~V., {Dutton} A.~A., {van den Bosch} F.~C., {Moore} B., {Potter}
  D., {Stadel} J., 2007, \mnras, 378, 55

\bibitem[{{Mannucci} {et~al}\mbox{.}(2010){Mannucci}, {Cresci}, {Maiolino},
  {Marconi}, \& {Gnerucci}}]{2010MNRAS.408.2115M}
{Mannucci} F., {Cresci} G., {Maiolino} R., {Marconi} A., {Gnerucci} A., 2010,
  \mnras, 408, 2115

\bibitem[{{Martin} \& {Roy}(1994)}]{1994ApJ...424..599M}
{Martin} P., {Roy} J.-R., 1994, \apj, 424, 599

\bibitem[{{Masters} {et~al}\mbox{.}(2012){Masters}, {Nichol}, {Haynes}, {Keel},
  {Lintott}, {Simmons}, {Skibba}, {Bamford}, {Giovanelli}, \&
  {Schawinski}}]{2012MNRAS.424.2180M}
{Masters} K.~L. {et~al.}, 2012, \mnras, 424, 2180

\bibitem[{{Minchev} {et~al}\mbox{.}(2011){Minchev}, {Famaey}, {Combes}, {Di
  Matteo}, {Mouhcine}, \& {Wozniak}}]{2011A&A...527A.147M}
{Minchev} I., {Famaey} B., {Combes} F., {Di Matteo} P., {Mouhcine} M.,
  {Wozniak} H., 2011, \aap, 527, A147

\bibitem[{{Mo}, {Mao} \& {White}(1998){Mo}, {Mao}, \&
  {White}}]{1998MNRAS.295..319M}
{Mo} H.~J., {Mao} S., {White} S.~D.~M., 1998, \mnras, 295, 319

\bibitem[{{Moran} {et~al}\mbox{.}(2012){Moran}, {Heckman}, {Kauffmann},
  {Dav{\'e}}, {Catinella}, {Brinchmann}, {Wang}, {Schiminovich}, {Saintonge},
  {Gracia-Carpio}, {Tacconi}, {Giovanelli}, {Haynes}, {Fabello}, {Hummels},
  {Lemonias}, \& {Wu}}]{2012ApJ...745...66M}
{Moran} S.~M. {et~al.}, 2012, \apj, 745, 66

\bibitem[{{Moran} {et~al}\mbox{.}(2010){Moran}, {Kauffmann}, {Heckman},
  {Gracia-Carpio}, {Saintonge}, {Catinella}, {Wang}, {Chen}, {Tacconi},
  {Schiminovich}, {Cox}, {Giovanelli}, {Haynes}, \&
  {Kramer}}]{2010ApJ...720.1126M}
{Moran} S.~M. {et~al.}, 2010, \apj, 720, 1126

\bibitem[{{Moustakas} {et~al}\mbox{.}(2010){Moustakas}, {Kennicutt},
  {Tremonti}, {Dale}, {Smith}, \& {Calzetti}}]{2010ApJS..190..233M}
{Moustakas} J., {Kennicutt}, Jr. R.~C., {Tremonti} C.~A., {Dale} D.~A., {Smith}
  J.-D.~T., {Calzetti} D., 2010, \apjs, 190, 233

\bibitem[{{Navarro}, {Frenk} \& {White}(1997){Navarro}, {Frenk}, \&
  {White}}]{1997ApJ...490..493N}
{Navarro} J.~F., {Frenk} C.~S., {White} S.~D.~M., 1997, \apj, 490, 493

\bibitem[{{Osterbrock} \& {Ferland}(2006)}]{2006agna.book.....O}
{Osterbrock} D.~E., {Ferland} G.~J., 2006, {Astrophysics of gaseous nebulae and
  active galactic nuclei}

\bibitem[{{Pagel}(1997)}]{1997nceg.book.....P}
{Pagel} B.~E.~J., 1997, {Nucleosynthesis and Chemical Evolution of Galaxies}

\bibitem[{{Peng} {et~al}\mbox{.}(2002){Peng}, {Ho}, {Impey}, \&
  {Rix}}]{2002AJ....124..266P}
{Peng} C.~Y., {Ho} L.~C., {Impey} C.~D., {Rix} H.-W., 2002, \aj, 124, 266

\bibitem[{{P{\'e}rez-Montero}(2014)}]{2014MNRAS.441.2663P}
{P{\'e}rez-Montero} E., 2014, \mnras, 441, 2663

\bibitem[{{Petit} {et~al}\mbox{.}(2014){Petit}, {Krumholz}, {Goldbaum}, \&
  {Forbes}}]{2014arXiv1411.7585P}
{Petit} A.~C., {Krumholz} M.~R., {Goldbaum} N.~J., {Forbes} J.~C., 2014, ArXiv
  e-prints

\bibitem[{{Pichon} {et~al}\mbox{.}(2011){Pichon}, {Pogosyan}, {Kimm}, {Slyz},
  {Devriendt}, \& {Dubois}}]{2011MNRAS.418.2493P}
{Pichon} C., {Pogosyan} D., {Kimm} T., {Slyz} A., {Devriendt} J., {Dubois} Y.,
  2011, \mnras, 418, 2493

\bibitem[{{Pilyugin} {et~al}\mbox{.}(2014){Pilyugin}, {Grebel}, {Zinchenko}, \&
  {Kniazev}}]{2014AJ....148..134P}
{Pilyugin} L.~S., {Grebel} E.~K., {Zinchenko} I.~A., {Kniazev} A.~Y., 2014,
  \aj, 148, 134

\bibitem[{{Pilyugin} \& {Mattsson}(2011)}]{2011MNRAS.412.1145P}
{Pilyugin} L.~S., {Mattsson} L., 2011, \mnras, 412, 1145

\bibitem[{{Rich} {et~al}\mbox{.}(2012){Rich}, {Torrey}, {Kewley}, {Dopita}, \&
  {Rupke}}]{2012ApJ...753....5R}
{Rich} J.~A., {Torrey} P., {Kewley} L.~J., {Dopita} M.~A., {Rupke} D.~S.~N.,
  2012, \apj, 753, 5

\bibitem[{{Rosales-Ortega} {et~al}\mbox{.}(2012){Rosales-Ortega},
  {S{\'a}nchez}, {Iglesias-P{\'a}ramo}, {D{\'{\i}}az}, {V{\'{\i}}lchez},
  {Bland-Hawthorn}, {Husemann}, \& {Mast}}]{2012ApJ...756L..31R}
{Rosales-Ortega} F.~F., {S{\'a}nchez} S.~F., {Iglesias-P{\'a}ramo} J.,
  {D{\'{\i}}az} A.~I., {V{\'{\i}}lchez} J.~M., {Bland-Hawthorn} J., {Husemann}
  B., {Mast} D., 2012, \apjl, 756, L31

\bibitem[{{Ro{\v s}kar} {et~al}\mbox{.}(2008){Ro{\v s}kar}, {Debattista},
  {Quinn}, {Stinson}, \& {Wadsley}}]{2008ApJ...684L..79R}
{Ro{\v s}kar} R., {Debattista} V.~P., {Quinn} T.~R., {Stinson} G.~S., {Wadsley}
  J., 2008, \apjl, 684, L79

\bibitem[{{Rupke}, {Kewley} \& {Barnes}(2010){Rupke}, {Kewley}, \&
  {Barnes}}]{2010ApJ...710L.156R}
{Rupke} D.~S.~N., {Kewley} L.~J., {Barnes} J.~E., 2010, \apjl, 710, L156

\bibitem[{{S{\'a}nchez} {et~al}\mbox{.}(2015){S{\'a}nchez}, {Galbany},
  {P{\'e}rez}, {S{\'a}nchez-Bl{\'a}zquez}, {Falc{\'o}n-Barroso},
  {Rosales-Ortega}, {S{\'a}nchez-Menguiano}, {Marino}, {Kuncarayakti},
  {Anderson}, {Kruehler}, {Cano-D{\'{\i}}az}, {Barrera-Ballesteros}, \&
  {Gonz{\'a}lez-Gonz{\'a}lez}}]{2015A&A...573A.105S}
{S{\'a}nchez} S.~F. {et~al.}, 2015, \aap, 573, A105

\bibitem[{{S{\'a}nchez} {et~al}\mbox{.}(2014){S{\'a}nchez}, {Rosales-Ortega},
  {Iglesias-P{\'a}ramo}, {Moll{\'a}}, {Barrera-Ballesteros}, {Marino},
  {P{\'e}rez}, {S{\'a}nchez-Blazquez}, {Gonz{\'a}lez Delgado}, {Cid Fernandes},
  {de Lorenzo-C{\'a}ceres}, {Mendez-Abreu}, {Galbany}, {Falcon-Barroso},
  {Miralles-Caballero}, {Husemann}, {Garc{\'{\i}}a-Benito}, {Mast}, {Walcher},
  {Gil de Paz}, {Garc{\'{\i}}a-Lorenzo}, {Jungwiert}, {V{\'{\i}}lchez},
  {J{\'{\i}}lkov{\'a}}, {Lyubenova}, {Cortijo-Ferrero}, {D{\'{\i}}az},
  {Wisotzki}, {M{\'a}rquez}, {Bland-Hawthorn}, {Ellis}, {van de Ven}, {Jahnke},
  {Papaderos}, {Gomes}, {Mendoza}, \&
  {L{\'o}pez-S{\'a}nchez}}]{2014A&A...563A..49S}
{S{\'a}nchez} S.~F. {et~al.}, 2014, \aap, 563, A49

\bibitem[{{S{\'a}nchez} {et~al}\mbox{.}(2013){S{\'a}nchez}, {Rosales-Ortega},
  {Jungwiert}, {Iglesias-P{\'a}ramo}, {V{\'{\i}}lchez}, {Marino}, {Walcher},
  {Husemann}, {Mast}, {Monreal-Ibero}, {Cid Fernandes}, {P{\'e}rez},
  {Gonz{\'a}lez Delgado}, {Garc{\'{\i}}a-Benito}, {Galbany}, {van de Ven},
  {Jahnke}, {Flores}, {Bland-Hawthorn}, {L{\'o}pez-S{\'a}nchez}, {Stanishev},
  {Miralles-Caballero}, {D{\'{\i}}az}, {S{\'a}nchez-Blazquez}, {Moll{\'a}},
  {Gallazzi}, {Papaderos}, {Gomes}, {Gruel}, {P{\'e}rez}, {Ruiz-Lara},
  {Florido}, {de Lorenzo-C{\'a}ceres}, {Mendez-Abreu}, {Kehrig}, {Roth},
  {Ziegler}, {Alves}, {Wisotzki}, {Kupko}, {Quirrenbach}, {Bomans}, \& {Califa
  Collaboration}}]{2013A&A...554A..58S}
{S{\'a}nchez} S.~F. {et~al.}, 2013, \aap, 554, A58

\bibitem[{{Skibba} {et~al}\mbox{.}(2012){Skibba}, {Masters}, {Nichol},
  {Zehavi}, {Hoyle}, {Edmondson}, {Bamford}, {Cardamone}, {Keel}, {Lintott}, \&
  {Schawinski}}]{2012MNRAS.423.1485S}
{Skibba} R.~A. {et~al.}, 2012, \mnras, 423, 1485

\bibitem[{{Stewart} {et~al}\mbox{.}(2013){Stewart}, {Brooks}, {Bullock},
  {Maller}, {Diemand}, {Wadsley}, \& {Moustakas}}]{2013ApJ...769...74S}
{Stewart} K.~R., {Brooks} A.~M., {Bullock} J.~S., {Maller} A.~H., {Diemand} J.,
  {Wadsley} J., {Moustakas} L.~A., 2013, \apj, 769, 74

\bibitem[{{Torrey} {et~al}\mbox{.}(2012){Torrey}, {Cox}, {Kewley}, \&
  {Hernquist}}]{2012ApJ...746..108T}
{Torrey} P., {Cox} T.~J., {Kewley} L., {Hernquist} L., 2012, \apj, 746, 108

\bibitem[{{Tremonti} {et~al}\mbox{.}(2004){Tremonti}, {Heckman}, {Kauffmann},
  {Brinchmann}, {Charlot}, {White}, {Seibert}, {Peng}, {Schlegel}, {Uomoto},
  {Fukugita}, \& {Brinkmann}}]{2004ApJ...613..898T}
{Tremonti} C.~A. {et~al.}, 2004, \apj, 613, 898

\bibitem[{{Vila-Costas} \& {Edmunds}(1992)}]{1992MNRAS.259..121V}
{Vila-Costas} M.~B., {Edmunds} M.~G., 1992, \mnras, 259, 121

\bibitem[{{Walter} {et~al}\mbox{.}(2008){Walter}, {Brinks}, {de Blok},
  {Bigiel}, {Kennicutt}, {Thornley}, \& {Leroy}}]{2008AJ....136.2563W}
{Walter} F., {Brinks} E., {de Blok} W.~J.~G., {Bigiel} F., {Kennicutt}, Jr.
  R.~C., {Thornley} M.~D., {Leroy} A., 2008, \aj, 136, 2563

\bibitem[{{Wang} {et~al}\mbox{.}(2014){Wang}, {Fu}, {Aumer}, {Kauffmann},
  {J{\'o}zsa}, {Serra}, {Huang}, {Brinchmann}, {van der Hulst}, \&
  {Bigiel}}]{2014MNRAS.441.2159W}
{Wang} J. {et~al.}, 2014, \mnras, 441, 2159

\bibitem[{{Wang} {et~al}\mbox{.}(2013){Wang}, {Kauffmann}, {J{\'o}zsa},
  {Serra}, {van der Hulst}, {Bigiel}, {Brinchmann}, {Verheijen}, {Oosterloo},
  {Wang}, {Li}, {den Heijer}, \& {Kerp}}]{2013MNRAS.433..270W}
{Wang} J. {et~al.}, 2013, \mnras, 433, 270

\bibitem[{{Wang} {et~al}\mbox{.}(2011){Wang}, {Kauffmann}, {Overzier},
  {Catinella}, {Schiminovich}, {Heckman}, {Moran}, {Haynes}, {Giovanelli}, \&
  {Kong}}]{2011MNRAS.412.1081W}
{Wang} J. {et~al.}, 2011, \mnras, 412, 1081

\bibitem[{{Wang} {et~al}\mbox{.}(2012){Wang}, {Kauffmann}, {Overzier},
  {Tacconi}, {Kong}, {Saintonge}, {Catinella}, {Schiminovich}, {Moran}, \&
  {Johnson}}]{2012MNRAS.423.3486W}
{Wang} J. {et~al.}, 2012, \mnras, 423, 3486

\bibitem[{{Weinzirl} {et~al}\mbox{.}(2009){Weinzirl}, {Jogee}, {Khochfar},
  {Burkert}, \& {Kormendy}}]{2009ApJ...696..411W}
{Weinzirl} T., {Jogee} S., {Khochfar} S., {Burkert} A., {Kormendy} J., 2009,
  \apj, 696, 411

\bibitem[{{White} \& {Rees}(1978)}]{1978MNRAS.183..341W}
{White} S.~D.~M., {Rees} M.~J., 1978, \mnras, 183, 341

\bibitem[{{Wuyts} {et~al}\mbox{.}(2014){Wuyts}, {Kurk}, {F{\"o}rster
  Schreiber}, {Genzel}, {Wisnioski}, {Bandara}, {Wuyts}, {Beifiori}, {Bender},
  {Brammer}, {Burkert}, {Buschkamp}, {Carollo}, {Chan}, {Davies}, {Eisenhauer},
  {Fossati}, {Kulkarni}, {Lang}, {Lilly}, {Lutz}, {Mancini}, {Mendel},
  {Momcheva}, {Naab}, {Nelson}, {Renzini}, {Rosario}, {Saglia}, {Seitz},
  {Sharples}, {Sternberg}, {Tacchella}, {Tacconi}, {van Dokkum}, \&
  {Wilman}}]{2014ApJ...789L..40W}
{Wuyts} E. {et~al.}, 2014, \apjl, 789, L40

\bibitem[{{York} {et~al}\mbox{.}(2000){York}, {Adelman}, {Anderson},
  {Anderson}, {Annis}, {Bahcall}, {Bakken}, {Barkhouser}, {Bastian}, {Berman},
  {Boroski}, {Bracker}, {Briegel}, {Briggs}, {Brinkmann}, {Brunner}, {Burles},
  {Carey}, {Carr}, {Castander}, {Chen}, {Colestock}, {Connolly}, {Crocker},
  {Csabai}, {Czarapata}, {Davis}, {Doi}, {Dombeck}, {Eisenstein}, {Ellman},
  {Elms}, {Evans}, {Fan}, {Federwitz}, {Fiscelli}, {Friedman}, {Frieman},
  {Fukugita}, {Gillespie}, {Gunn}, {Gurbani}, {de Haas}, {Haldeman}, {Harris},
  {Hayes}, {Heckman}, {Hennessy}, {Hindsley}, {Holm}, {Holmgren}, {Huang},
  {Hull}, {Husby}, {Ichikawa}, {Ichikawa}, {Ivezi{\'c}}, {Kent}, {Kim},
  {Kinney}, {Klaene}, {Kleinman}, {Kleinman}, {Knapp}, {Korienek}, {Kron},
  {Kunszt}, {Lamb}, {Lee}, {Leger}, {Limmongkol}, {Lindenmeyer}, {Long},
  {Loomis}, {Loveday}, {Lucinio}, {Lupton}, {MacKinnon}, {Mannery}, {Mantsch},
  {Margon}, {McGehee}, {McKay}, {Meiksin}, {Merelli}, {Monet}, {Munn},
  {Narayanan}, {Nash}, {Neilsen}, {Neswold}, {Newberg}, {Nichol}, {Nicinski},
  {Nonino}, {Okada}, {Okamura}, {Ostriker}, {Owen}, {Pauls}, {Peoples},
  {Peterson}, {Petravick}, {Pier}, {Pope}, {Pordes}, {Prosapio},
  {Rechenmacher}, {Quinn}, {Richards}, {Richmond}, {Rivetta}, {Rockosi},
  {Ruthmansdorfer}, {Sandford}, {Schlegel}, {Schneider}, {Sekiguchi}, {Sergey},
  {Shimasaku}, {Siegmund}, {Smee}, {Smith}, {Snedden}, {Stone}, {Stoughton},
  {Strauss}, {Stubbs}, {SubbaRao}, {Szalay}, {Szapudi}, {Szokoly}, {Thakar},
  {Tremonti}, {Tucker}, {Uomoto}, {Vanden Berk}, {Vogeley}, {Waddell}, {Wang},
  {Watanabe}, {Weinberg}, {Yanny}, {Yasuda}, \& {SDSS
  Collaboration}}]{2000AJ....120.1579Y}
{York} D.~G. {et~al.}, 2000, \aj, 120, 1579

\bibitem[{{Zaritsky}, {Kennicutt} \& {Huchra}(1994){Zaritsky}, {Kennicutt}, \&
  {Huchra}}]{1994ApJ...420...87Z}
{Zaritsky} D., {Kennicutt}, Jr. R.~C., {Huchra} J.~P., 1994, \apj, 420, 87

\end{thebibliography}

\appendix

\section{Comparison of metallicity indicators}\label{app:comparing_indicators}

In Section~\ref{subsec:inferring_metallicities} we discuss a variety of methods for determining metallicity.
We shall now discuss the similarities and differences between these methods and justify the use of our primary method that uses the CL01 models.

In Figs.~\ref{fig:profiles_blue}, \ref{fig:profiles_cont} and \ref{fig:profiles_excl} we present metallicities derived from the CL01, KK04 and NS methods.
It should be immediately apparent that large (0.6\,dex) systematic offsets in metallicity exist between these methods.
The CL01 and KK04 methods, both derived from theory, produce largely consistent metallicities. However, they report much higher metallicities than those from the NS method, which is derived empirically.
It is also noticeable that the NS method produces shallower profiles than the CL01 models.
Of course disparities between metallicity indicators are not unique to this work and they have been well documented by \citet{2008ApJ...681.1183K} and \citet{2010ApJS..190..233M}.

\begin{figure*}
\includegraphics[width=\linewidth]{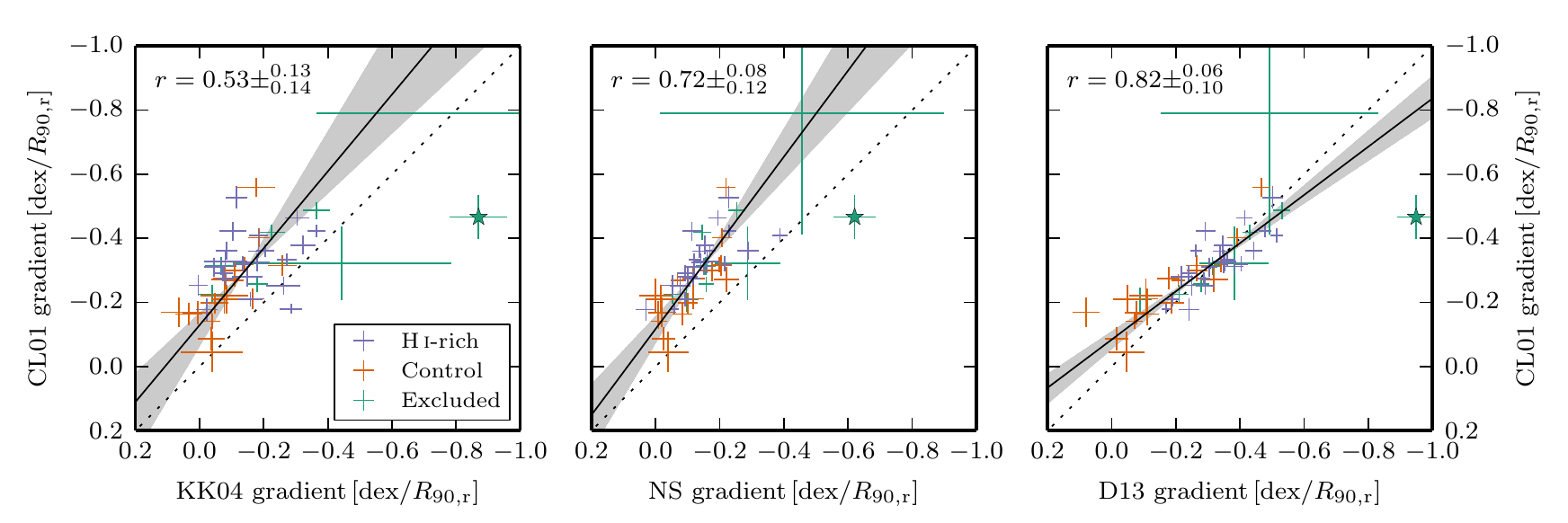}
\caption{Comparison of straight-line gradients derived for CL01 against three difference metallicity indicators (KK04, NS and D13), for all 50 Bluedisk galaxies, \ion{H}{i}-rich (blue), control (orange) and excluded (green). Linear fit, using orthogonal-distance regression, is shown (solid line) and the shaded region indicates the associated error in the slope. The dotted line indicates equal $x$\,=\,$y$ mapping. The Spearman's rank correlation coefficient is given in the top-left corner of each figure. A star indicates BD 39, which is excluded from the regression and r-statistic computation, due to its companion galaxy. Median values and $\pm$1\,$\sigma$ errors on statistics and gradients are computed by bootstrapping Monte-Carlo-scattered resampled data. For reference, $r = 0.34$ is the one-tailed Spearman's $r$-value at a $\alpha = 0.01$ significance level.}
\label{fig:compare_grad}
\end{figure*}

Although it is not desirable to have absolute and relative differences between metallicity methods, we assert that this is of no major significance provided the metallicity indicators are consistent. In other words, a galaxy with a steeper profile in one method should yield a steeper profile in all methods.
We test this assumption by fitting simple straight-line models to metallicity profiles derived from different methods, and compare the inferred gradients.
In Fig.~\ref{fig:compare_grad} we show the gradients inferred from the CL01 method versus KK04, NS and D13 methods.
We note the strongest correlation exists between the CL01 and D13 models, indicating our results are not strongly dependant on the photoionization models employed.
However, particularly remarkable is the similarity between CL01 and NS gradients, since the NS method is empirically calibrated.
We note that the KK04 method often produces shallower gradients than CL01.
Indeed it appears that the KK04 methods can produce much higher metallicities than CL01 in the outskirts of the galaxies (e.g. BD 50).
This is often associated with significant differences in inferred dust attenuation strengths, with the standard Case B method, producing erratic results between adjacent bins.
Nevertheless we derive comfort in using our CL01 metallicities as our default method for metallicity determination. 

\section{Gas surface density estimates}\label{app:gas_surface_density_estimates}

In Section~\ref{gas_mass_densitites} we use a technique developed by B13 to estimate gas surface mass densities from optical lines. It was shown there that when most of the lines in the optical spectrum are available it is possible to use photoionization models with a flexible treatment of metal depletion to place constraints on the gas surface mass density of galaxies. 

This works because emission line ratios are sensitive to temperature and since metals are very important coolants, changing their depletion factor at fixed metallicity, $Z$, changes the temperature in the gas noticeably. Exploiting this fact, B13 showed that it is possible to place constraints on the dust-to-metal ratio, $\xi$, of ionised gas using only optical emission lines. When combined with an estimate of metallicity and the dust optical depth, primarily from Balmer lines, and a simple model for the interstellar medium of a galaxy they show that total gas surface mass densities can be estimated through
\begin{equation}
  \label{eq:sigma_gas_2}
  \Sigma_{\mathrm{gas}} = 0.2\frac{\tau_V}{\xi Z}\ \textrm{M}_\odot\,\textrm{pc}^{-2}.
\end{equation}

\begin{figure*}
  \includegraphics[width=\linewidth]{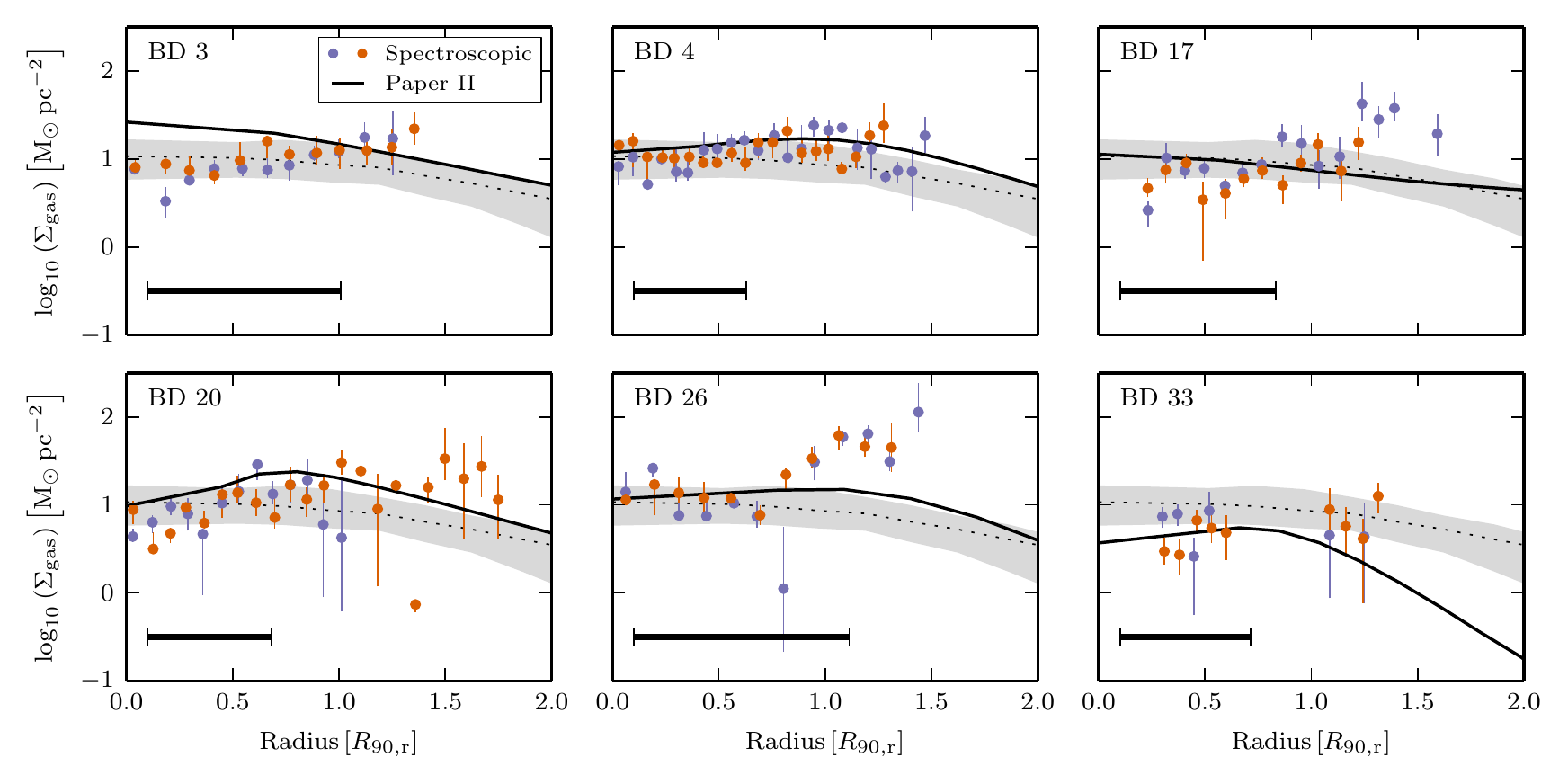}
  \caption{Comparison of gas density profiles of six galaxies using two different estimators. The spectroscopic $\Sigma_\mathrm{gas}$ estimates are represented by blue and orange data points, with the two colours distinguishing either side of the spectroscopic slit. The azimuthally averaged $\Sigma_\mathrm{gas}$ estimates from Paper II are shown as a solid black line. We also show the $\Sigma_\mathrm{gas}$ profile averaged over all galaxies presented in Paper II. These are plotted as a dotted black line and a grey shaded area indicating the median and $\pm$1\,$\sigma$ quantile range, respectively. In the bottom-left corner of each plot, a thick black bar indicates the scale of 13\,arcsec which roughly approximates the resolution limit of the Paper II estimates.}
  \label{fig:CL01vsHI_gasmass}
\end{figure*}

We have calculated this quantity for each spatial bin in the spectra discussed in this paper.

We shall now provide an additional check of these spectroscopic $\Sigma_\mathrm{gas}$ estimates.
In \citet{2014MNRAS.441.2159W}, herein referred to as Paper II, azimuthally averaged $\Sigma_\mathrm{gas}$ were calculated.
These were computed by combining the observed \ion{H}{i} surface density with an estimated contribution from $\textrm{H}_2$. The $\textrm{H}_2$ component was estimated using a SF scaling relation applied to the observed SFRs.
In Fig.~\ref{fig:CL01vsHI_gasmass} we show a few select examples of our spectroscopic gas profile against those from Paper II.
These galaxies were selected to span a range from very poor to very good agreement.
Overall the match between the two estimators is reasonable given the differences in analysis and that the profiles from Paper II are azimuthally averaged while the spectrally determined gas surface densities originate from long-slit spectra.

Nevertheless, it is noticeable that the spectroscopically determined gas surface densities do not show a strong drop in the outer regions of the galaxies.
This is possibly due to a characteristic of the B13 method that is not discussed in detail by B13 (but see their fig.~15), namely that it might give a biased estimate of the \emph{average} gas surface density in the outer regions of galaxies.
The reason for this is that the method only works reliably when there is a clear emission line source, in practice an \ion{H}{ii} region in the spectral aperture.
In the outskirts of galaxies these regions are fewer and tend to coincide with peaks in the local gas density.
But these peaks provide biased estimates of the azimuthally averaged gas surface density at those radii so the spectroscopic method will also provide biased estimates.

\begin{figure*}
  \includegraphics[width=\linewidth]{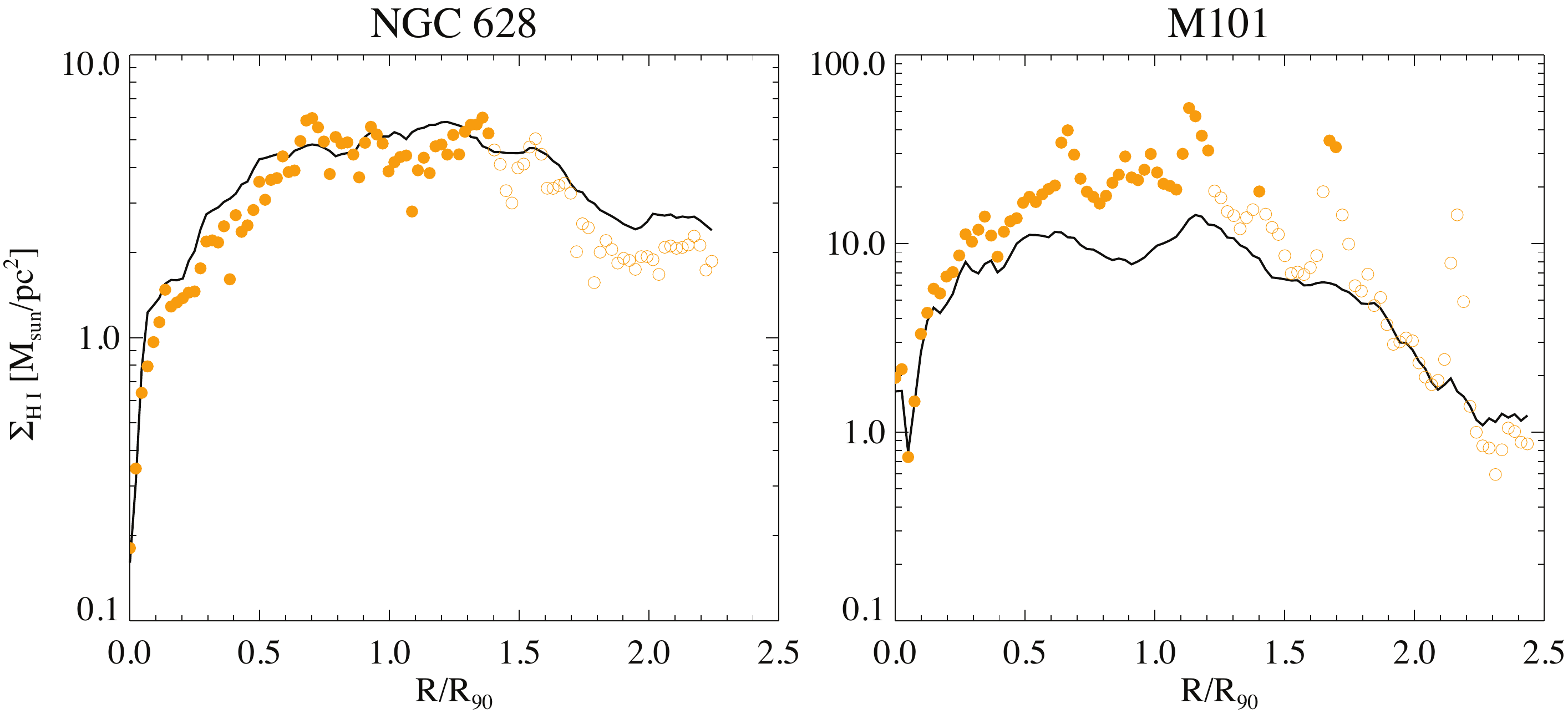}
  \caption{The azimuthally averaged \ion{H}{i} gas density in galaxy disks (solid black line), compared to the \Halpha{} flux weighted density (orange symbols). The left panel shows the results for NGC 628 with the right-hand panel that of M 101. The open orange symbols indicate annuli where the average SFR is $<10^{-3}\ \textrm{M}_\odot\mathrm{yr}^{-1}\mathrm{kpc}^{-2}$.}
  \label{fig:hi_density_ha_comp}
\end{figure*}

To illustrate this fact, Fig.~\ref{fig:hi_density_ha_comp} shows the azimuthally averaged \ion{H}{i} gas profiles for two large nearby spiral galaxies, NGC 628 and M 101 as black solid lines.
The \ion{H}{i} maps were taken from \citet{2008AJ....136.2563W}.
To illustrate the effect of probing the gas density at the location of \ion{H}{ii} regions in the outer disks we overplot the \Halpha{} weighted \ion{H}{i} profiles as orange symbols on top.
The open symbols are for annuli where the mean SFR is $<10^{-3}\ \textrm{M}_\odot\mathrm{yr}^{-1}\mathrm{kpc}^{-2}$, assuming a Salpeter initial mass function and the L(\Halpha{}) to SFR conversion factor of \citep{1998ARA&A..36..189K}. We took the \Halpha{} maps from \citet{2001ApJ...559..878H}. 

What is noticeable is that the two galaxies are rather different with the \Halpha{} weighted profile in NGC 628 being very close to the straight mean profile.
In contrast the \Halpha{}-weighted profile in M101 is noticeably higher than the mean profile and it is also clear that star formation at large radii is connected to relatively high gas densities. 

\begin{figure*}
  \includegraphics[width=\linewidth]{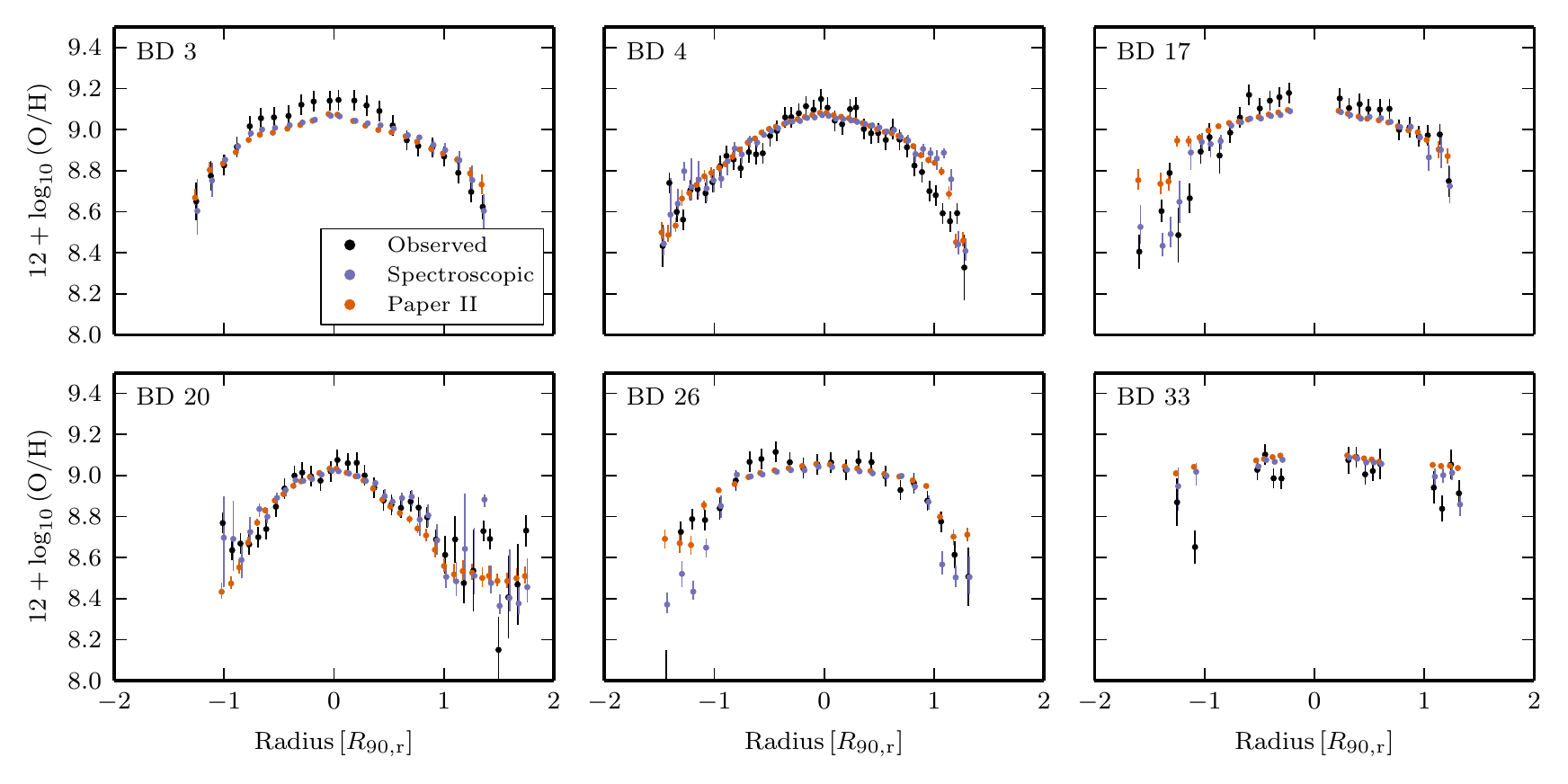}
  \caption{Comparing the effects of two gas density estimators on the local gas regulator model. The observed CL01
metallicity is shown in black. The model using spectroscopic $\Sigma_{\textrm{gas}}$ estimates is shown in blue. The equivalent model using Paper II $\Sigma_{\textrm{gas}}$ estimates is shown in orange. Both models assume the modelling scenario with winds.}
  \label{fig:LGR_CL01vsHI_gasmass}
\end{figure*}

Nevertheless, the differences between two estimators of $\Sigma_\mathrm{gas}$ do not strongly impact on our analysis using the local gas regulator model (Section \ref{section:showLGR}).
This is demonstrated in Fig.~\ref{fig:LGR_CL01vsHI_gasmass}, which highlights the relatively minor effect of the choice of gas-density estimator on our conclusions.
The reason for the lack of significant difference is that the local gas regulator models depends on the ratio of gas-to-stellar mass. Across a galaxy the dynamic range of the stellar-mass density is much greater than that of the gas-mass density. Therefore the overall shape of the metallicity profile is primarily set by the stellar-mass density profile.

\section{Metallicity profiles equally weighed by galaxy}\label{app:weighted_profile_stack}
In Section~\ref{subsec:profile_stack} we produce stacked metallicity profiles.
As each data point receives an equal weight within a bin.
Galaxies with many data points may, however, dominate a bin.
If these galaxies have atypical metallicity profiles, this would be mimicked in the stacked profile.
Such problems are only likely to arise in the outermost radial bin of a stack.
To qualify and counteract this effect we also produce stacked profiles we use a weighted median, where data points are each weighted inversely to the number of data points from the same galaxy per bin. 

\begin{figure*}
\includegraphics[width=\linewidth]{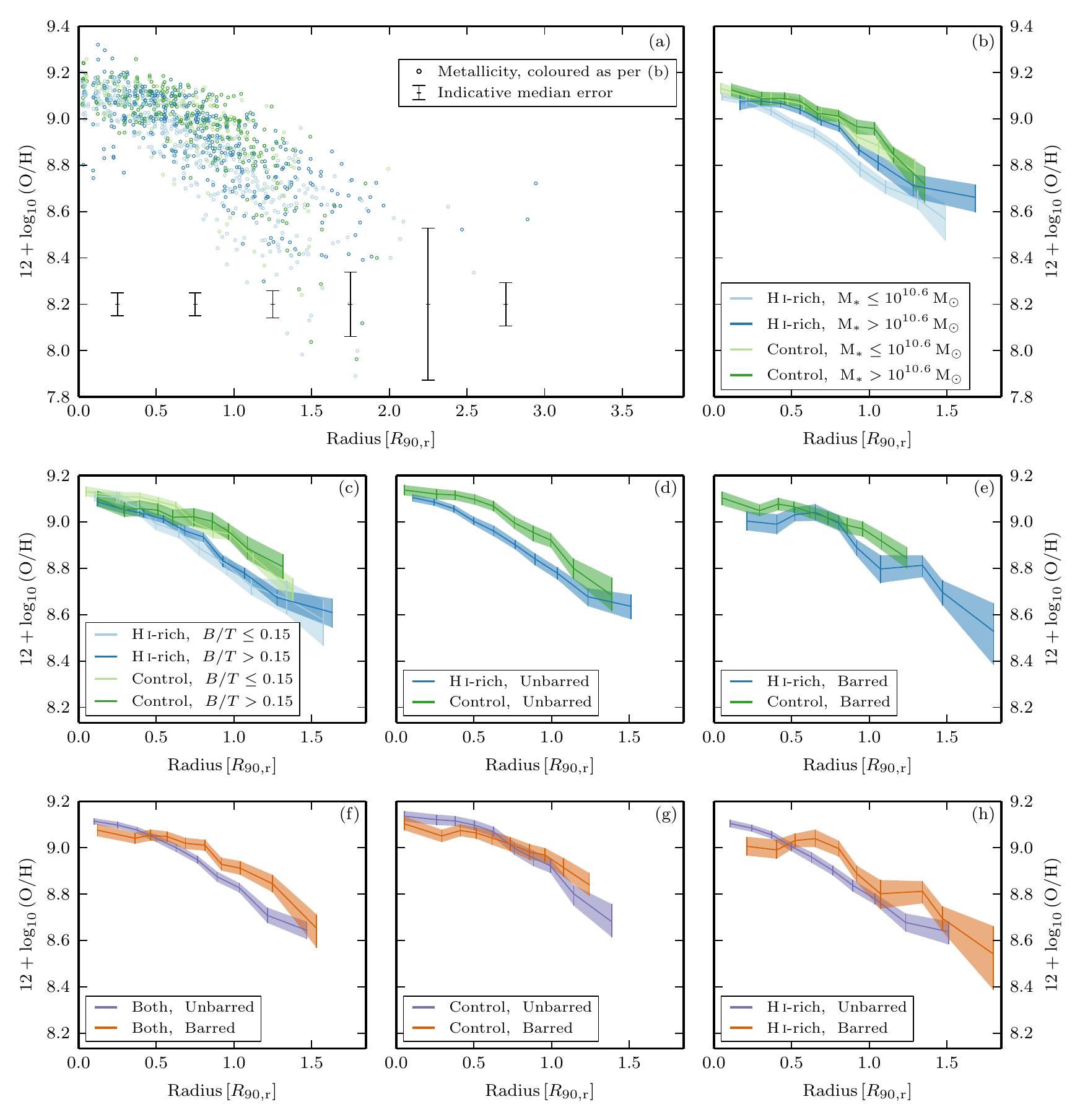}
\caption{Equivalent of Fig.~\ref{fig:profile_stack} with weighted such that galaxies receive equal weight per bin. See Fig.~\ref{fig:profile_stack} for description.}
\label{fig:profile_stack_weighted}
\end{figure*}

In Fig.~\ref{fig:profile_stack_weighted} we reproduce Fig.~\ref{fig:profile_stack} with this new weighting.
The only appreciable difference occurs in the for some control sample stacks, where the outermost metallicity is never lower than the metallicity of the \ion{H}{i}-rich galaxies.
Since these bins are dominated by a few galaxies, they may not be representative of the whole sample.

\section{The Atlas}\label{app:atlas}

In Table \ref{tab:slit_descr} we provide the positions and orientations of each spectroscopic slit.

\begin{figure*}
\includegraphics[width=\linewidth]{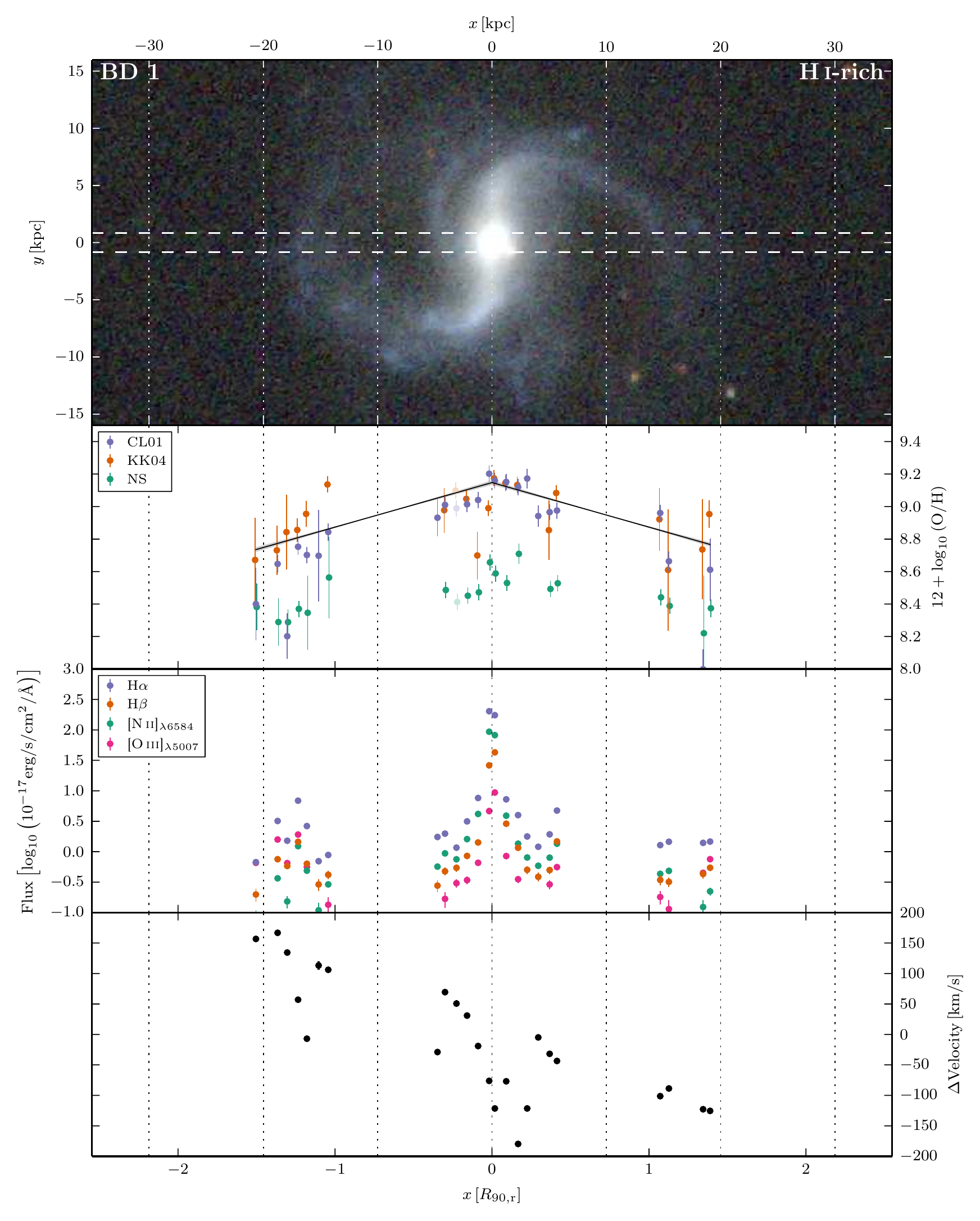}
\caption{Atlas of data for BD 1. See text for details.}
\label{fig:atlas_1}
\end{figure*}

\begin{figure*}
\includegraphics[width=\linewidth]{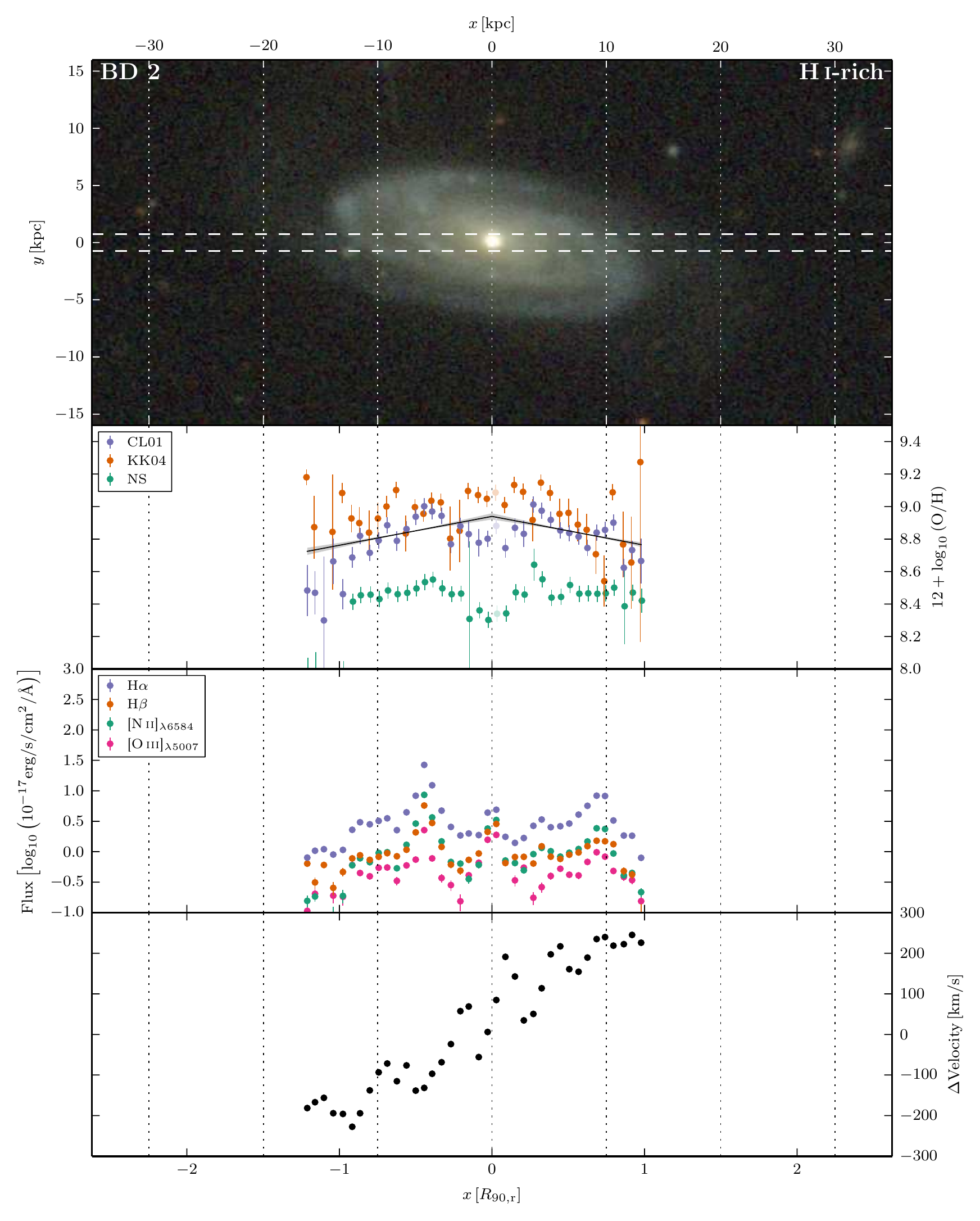}
\caption{Atlas of data for BD 2. See text for details.}
\label{fig:atlas_2}
\end{figure*}

\begin{figure*}
\includegraphics[width=\linewidth]{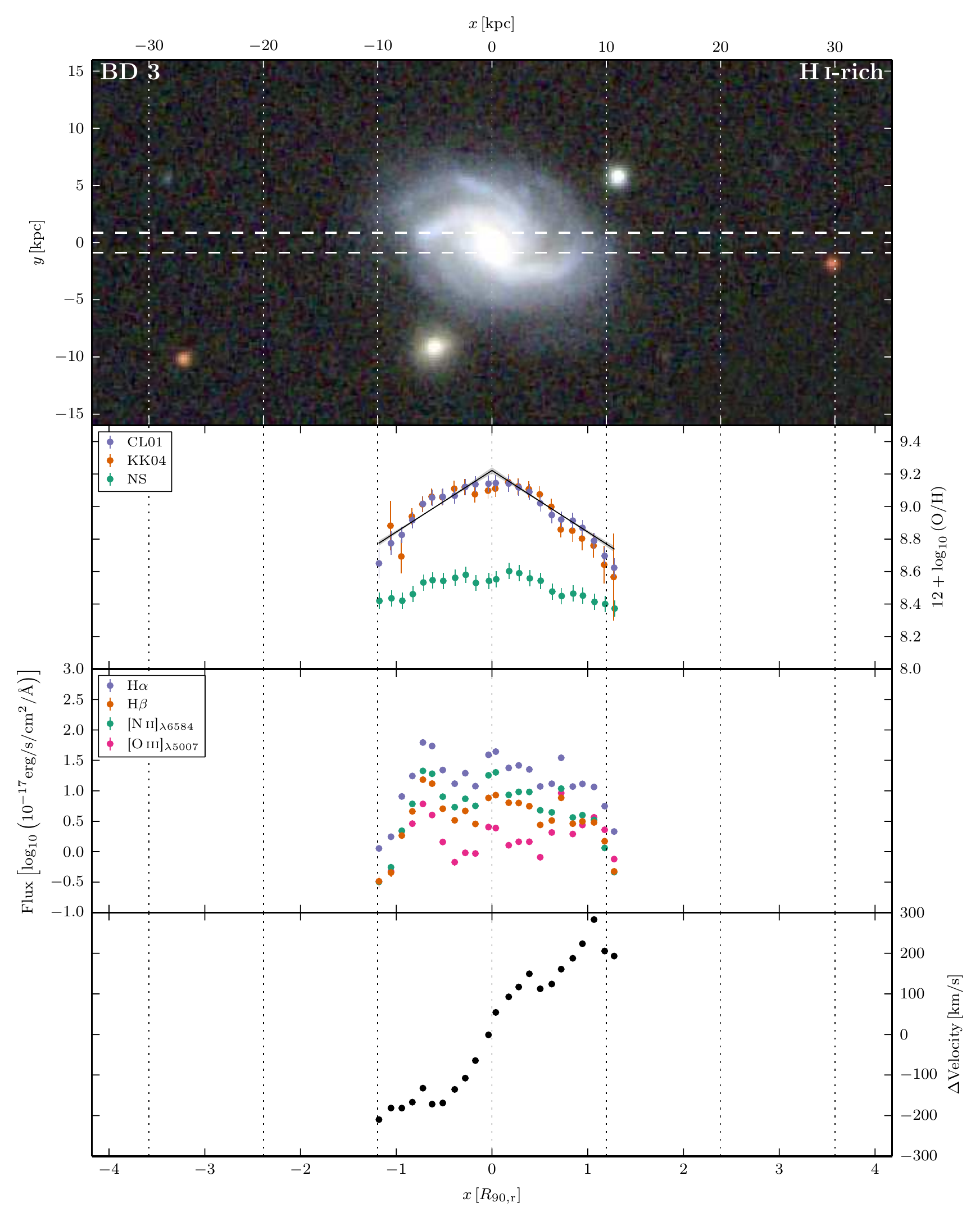}
\caption{Atlas of data for BD 3. See text for details.}
\label{fig:atlas_3}
\end{figure*}

\begin{figure*}
\includegraphics[width=\linewidth]{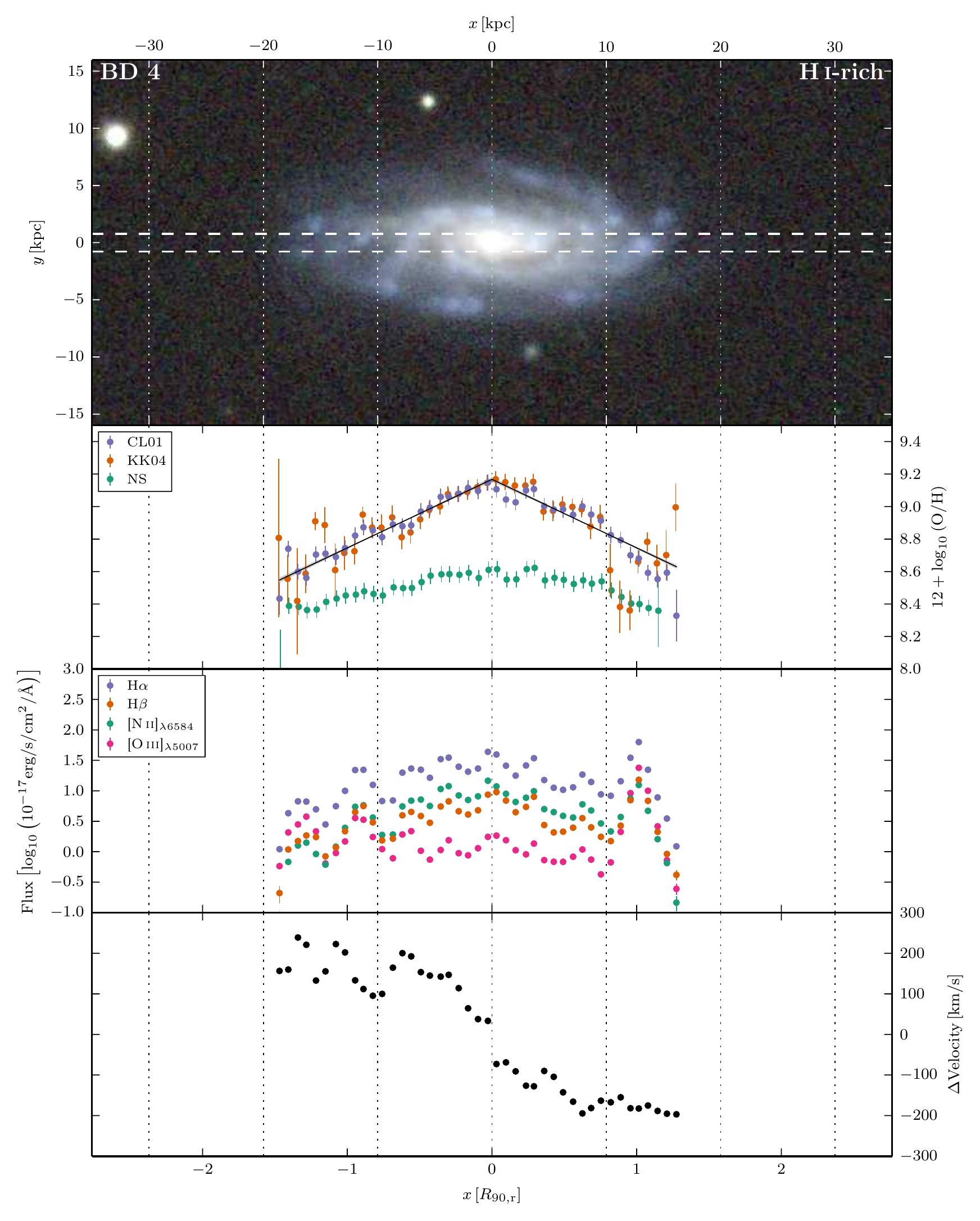}
\caption{Atlas of data for BD 4. See text for details.}
\label{fig:atlas_4}
\end{figure*}

\begin{figure*}
\includegraphics[width=\linewidth]{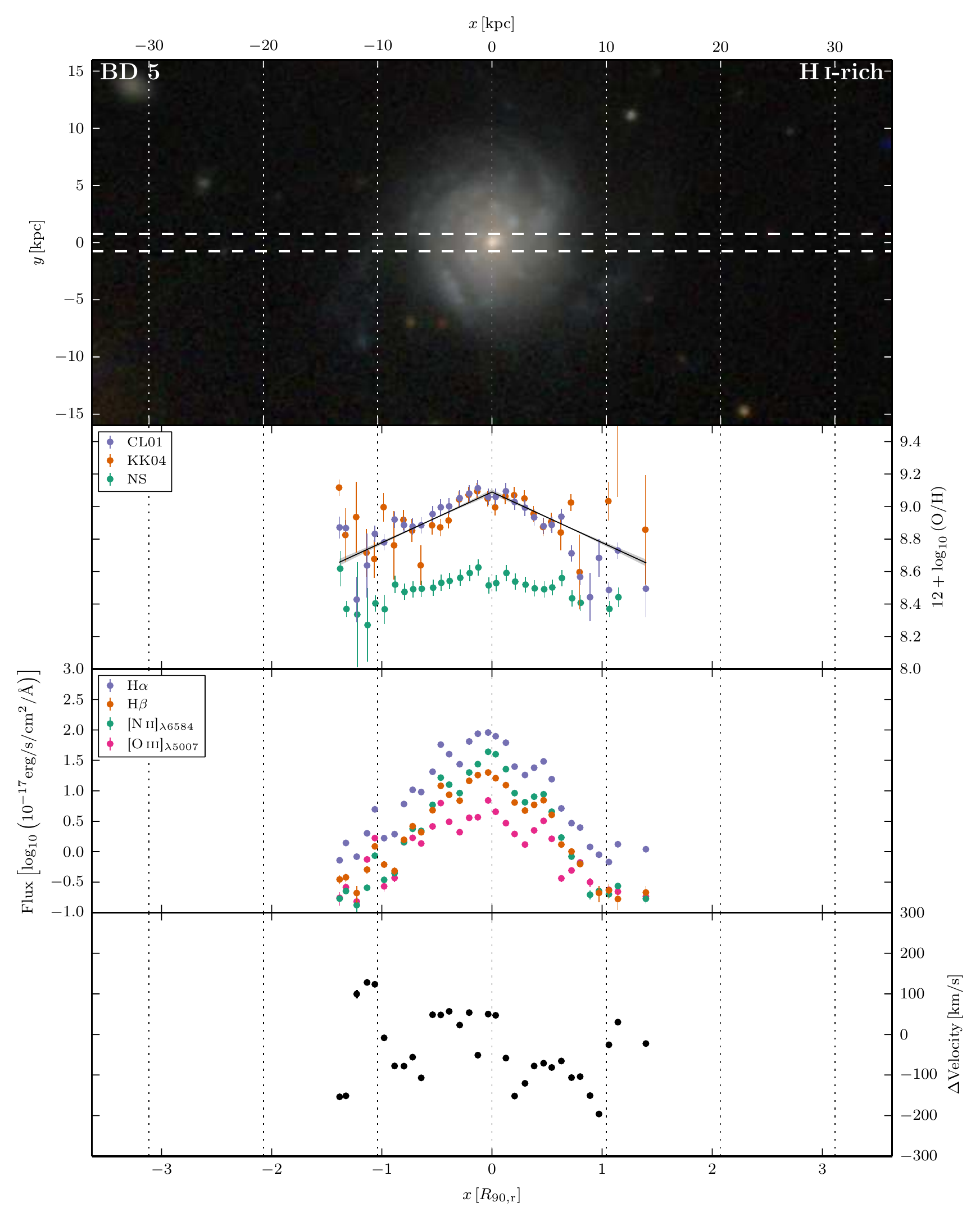}
\caption{Atlas of data for BD 5. See text for details.}
\label{fig:atlas_5}
\end{figure*}

\begin{figure*}
\includegraphics[width=\linewidth]{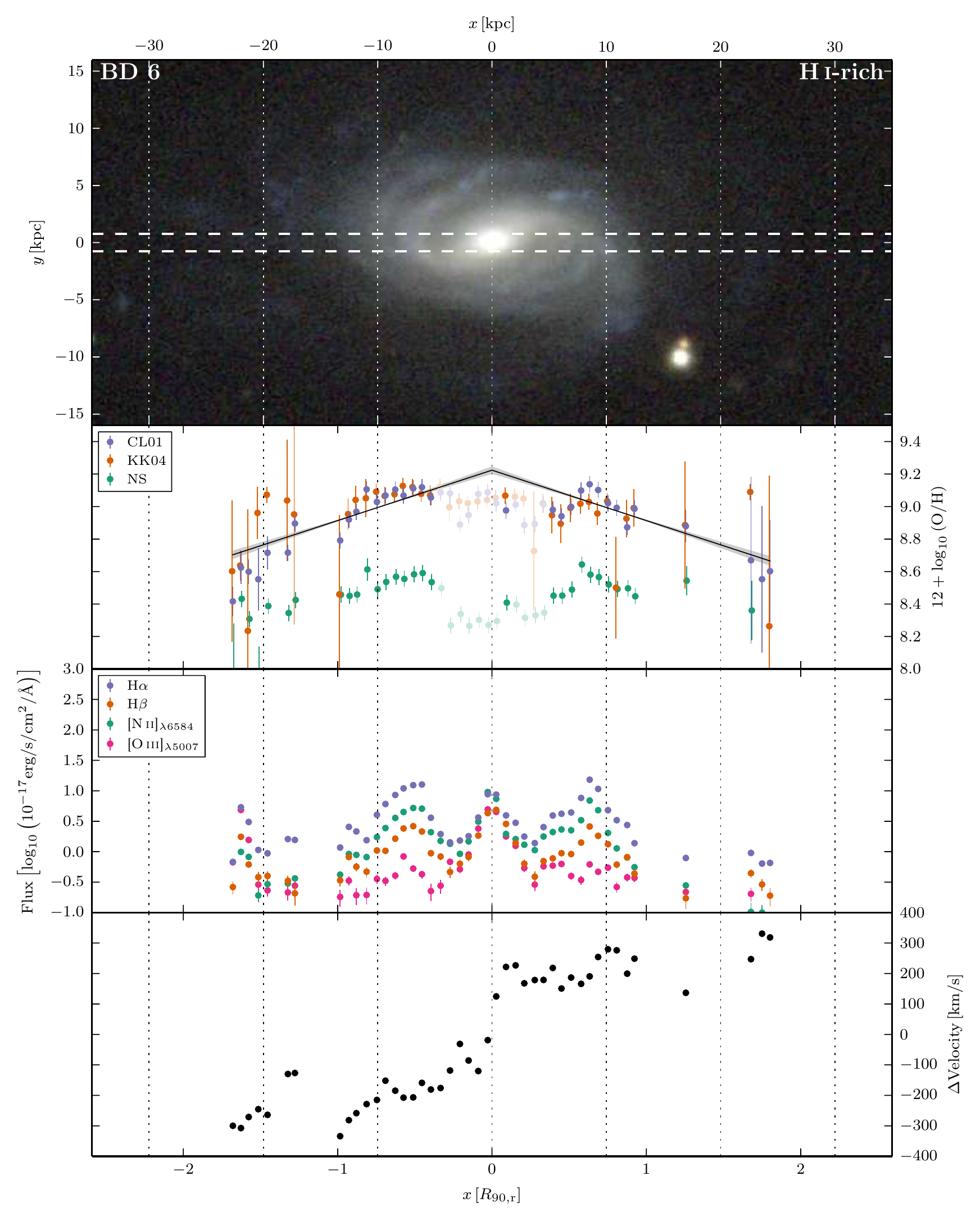}
\caption{Atlas of data for BD 6. See text for details.}
\label{fig:atlas_6}
\end{figure*}

\begin{figure*}
\includegraphics[width=\linewidth]{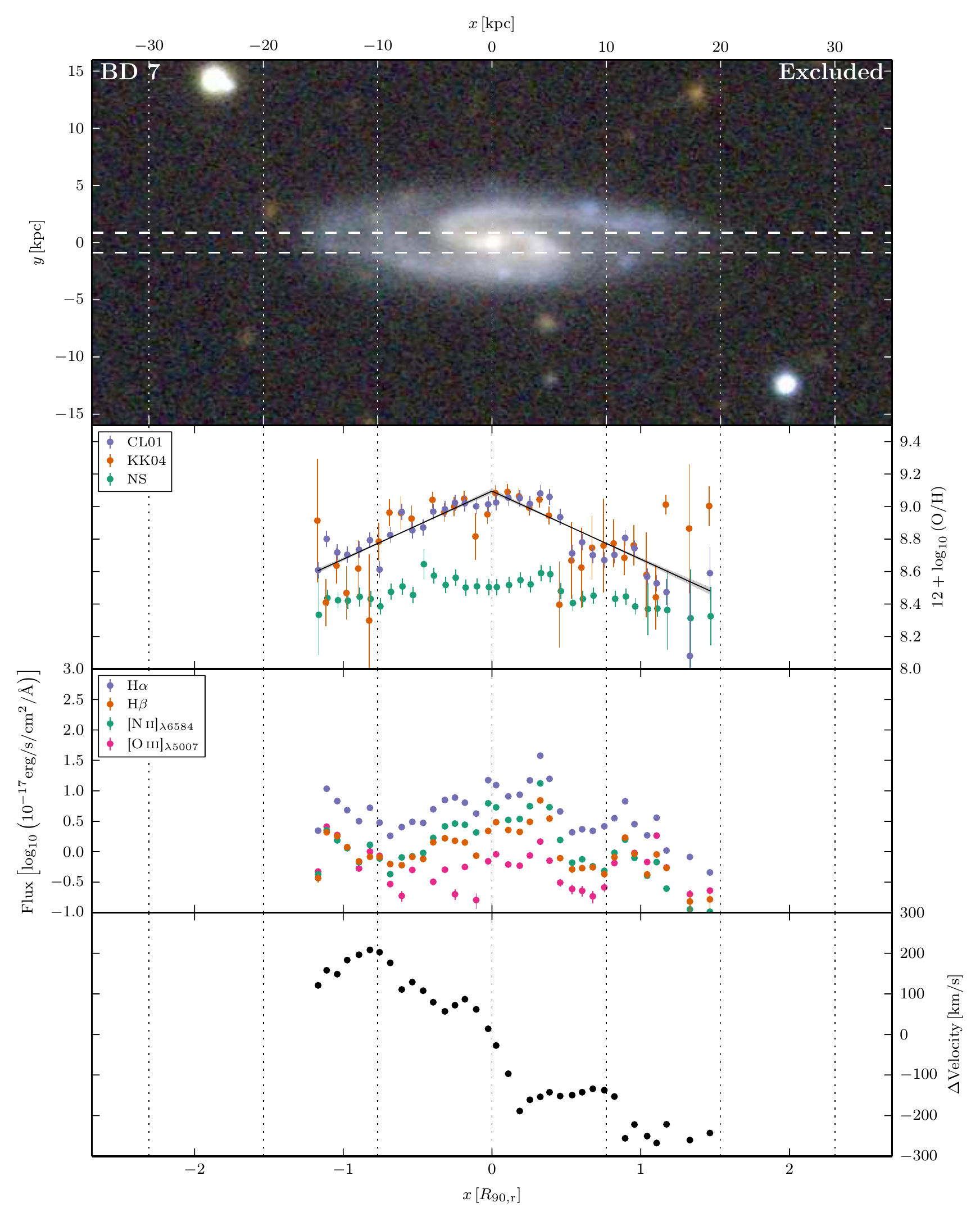}
\caption{Atlas of data for BD 7. See text for details.}
\label{fig:atlas_7}
\end{figure*}

\begin{figure*}
\includegraphics[width=\linewidth]{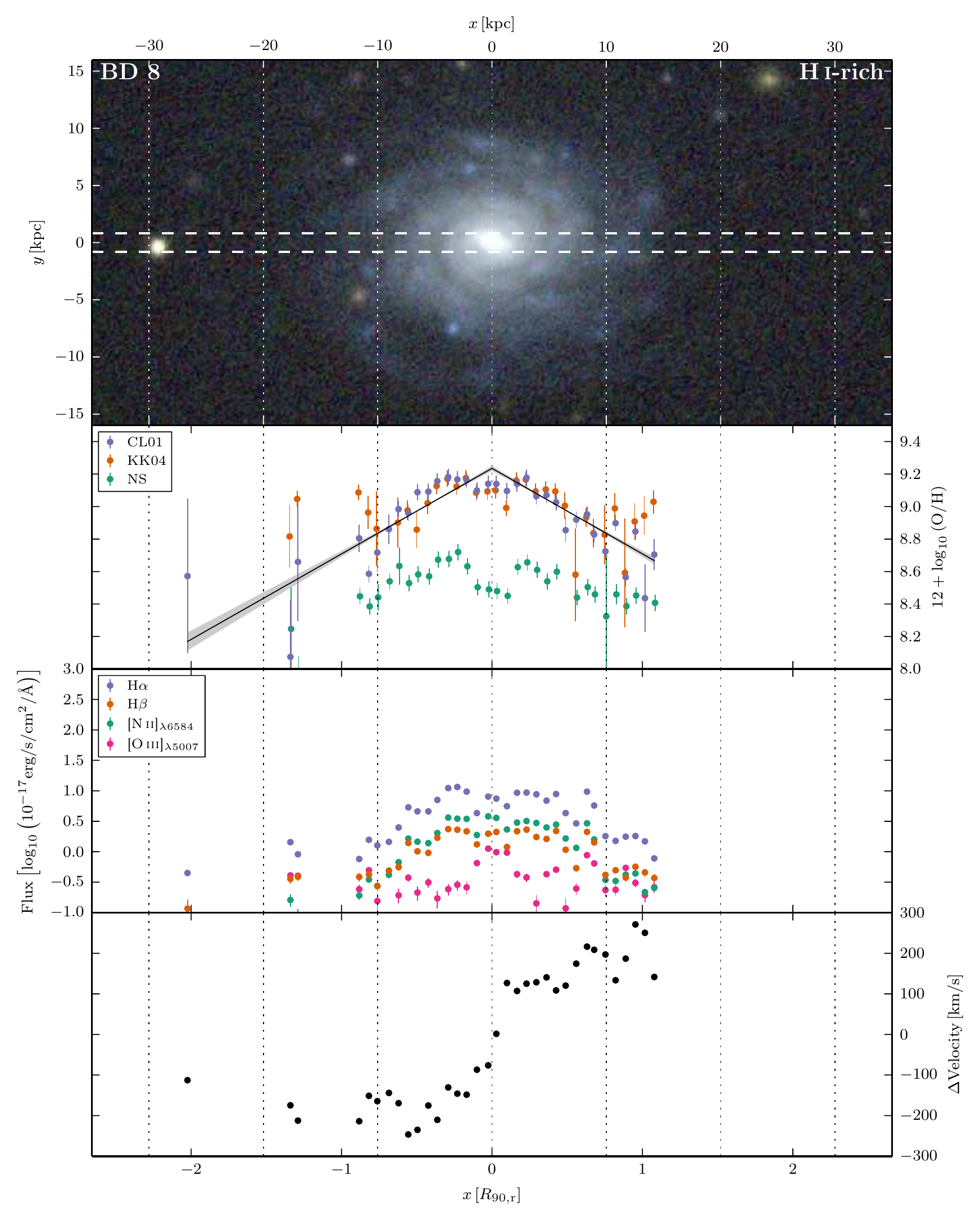}
\caption{Atlas of data for BD 8. See text for details.}
\label{fig:atlas_8}
\end{figure*}

\begin{figure*}
\includegraphics[width=\linewidth]{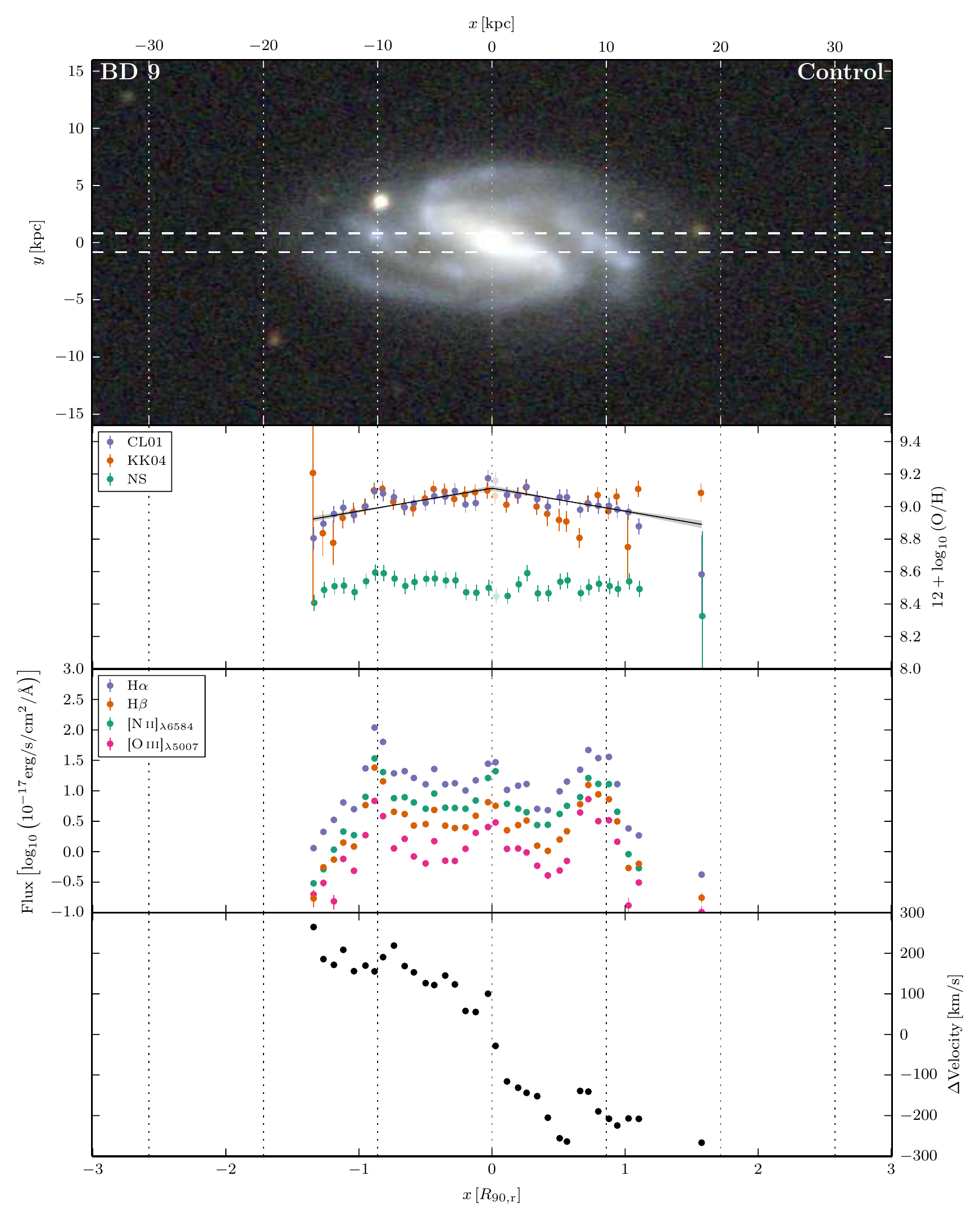}
\caption{Atlas of data for BD 9. See text for details.}
\label{fig:atlas_9}
\end{figure*}

\begin{figure*}
\includegraphics[width=\linewidth]{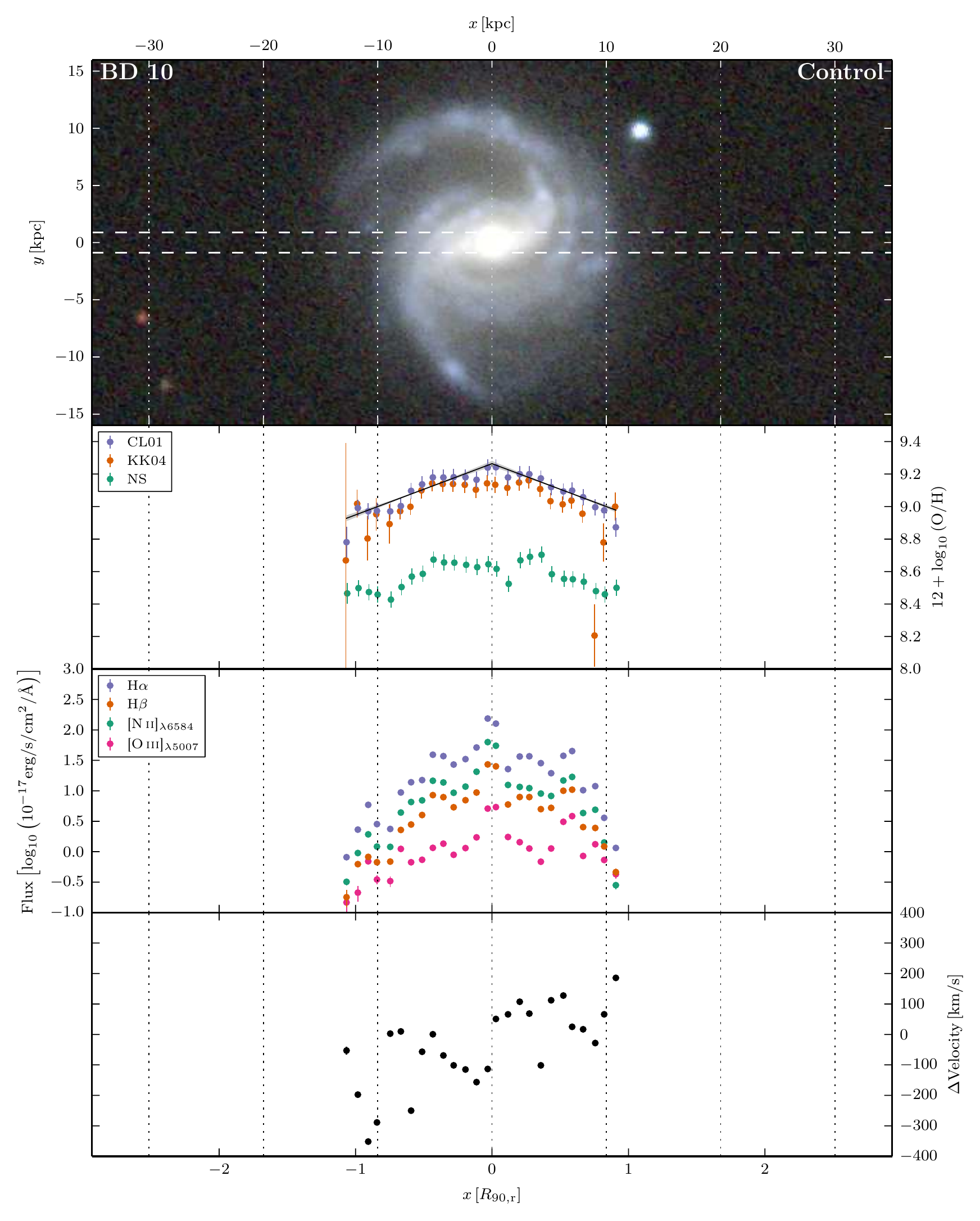}
\caption{Atlas of data for BD 10. See text for details.}
\label{fig:atlas_10}
\end{figure*}

\begin{figure*}
\includegraphics[width=\linewidth]{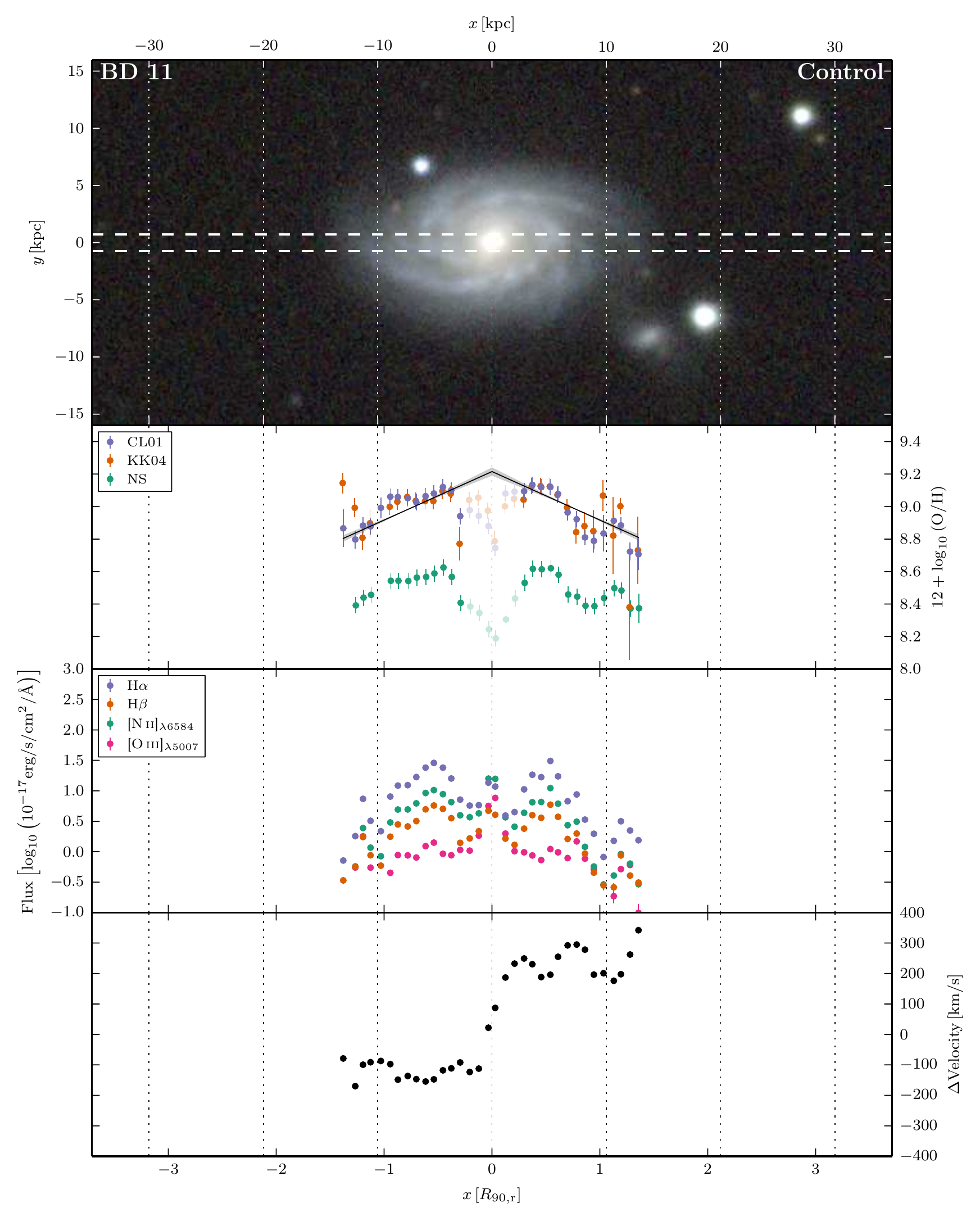}
\caption{Atlas of data for BD 11. See text for details.}
\label{fig:atlas_11}
\end{figure*}

\begin{figure*}
\includegraphics[width=\linewidth]{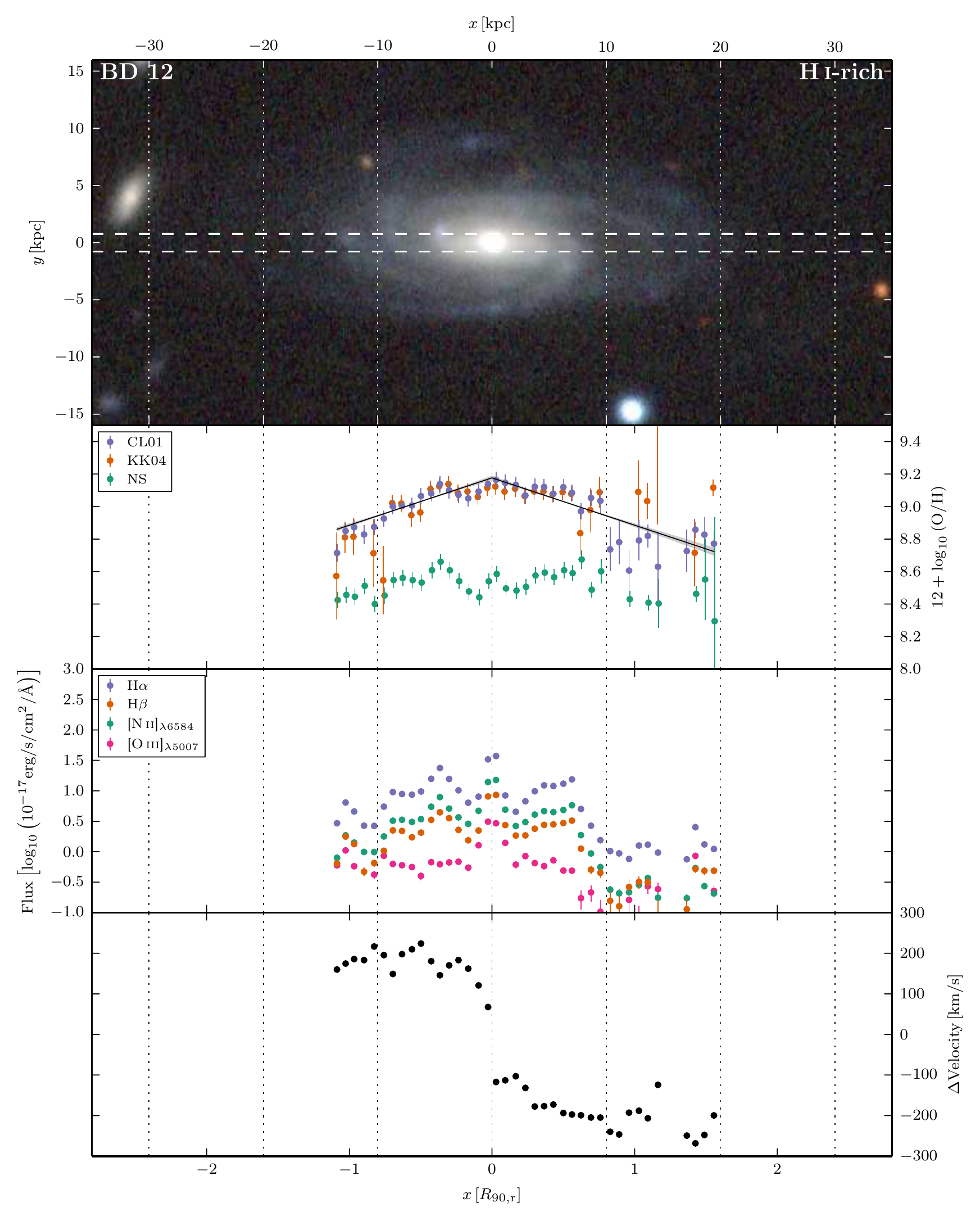}
\caption{Atlas of data for BD 12. See text for details.}
\label{fig:atlas_12}
\end{figure*}

\begin{figure*}
\includegraphics[width=\linewidth]{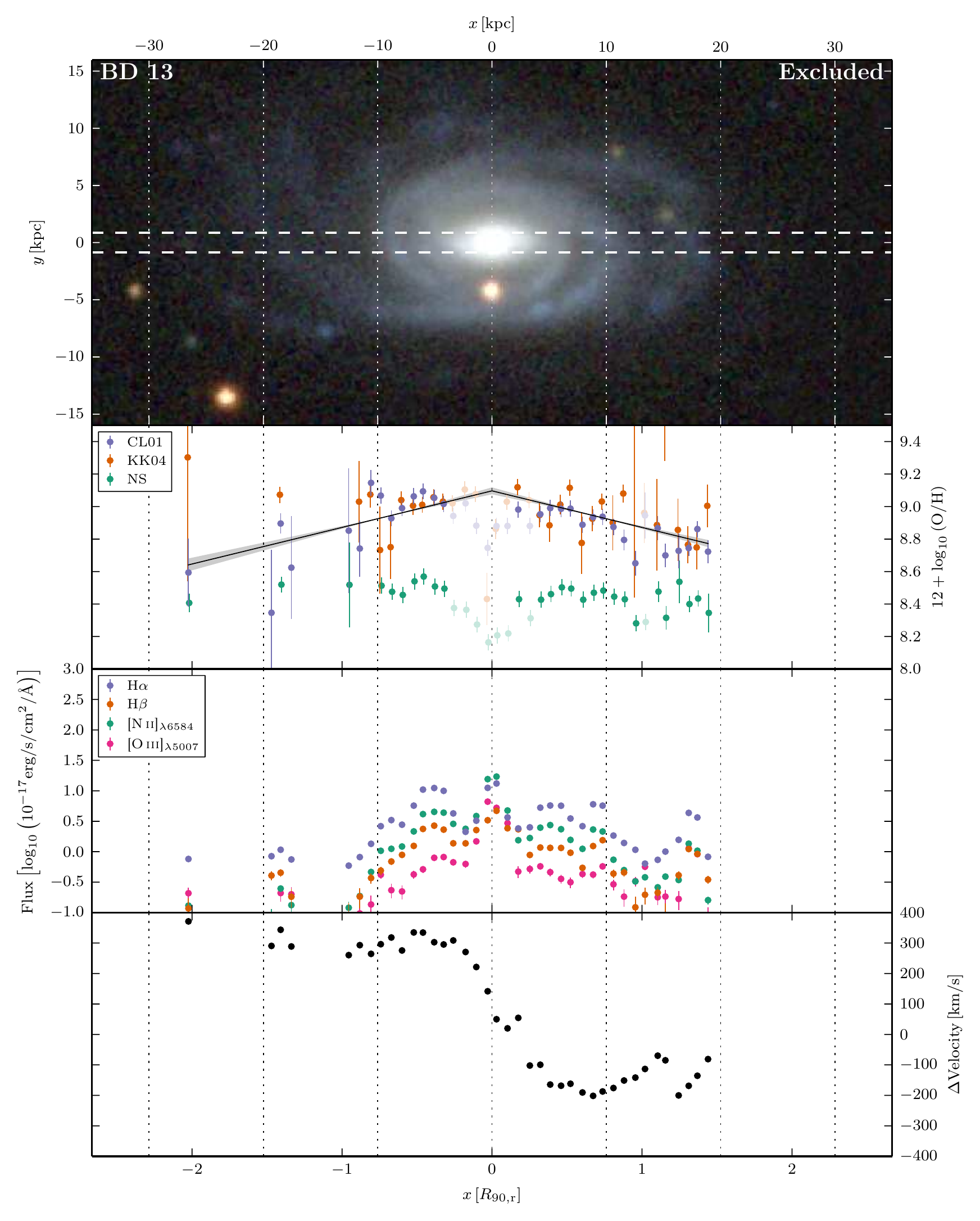}
\caption{Atlas of data for BD 13. See text for details.}
\label{fig:atlas_13}
\end{figure*}

\begin{figure*}
\includegraphics[width=\linewidth]{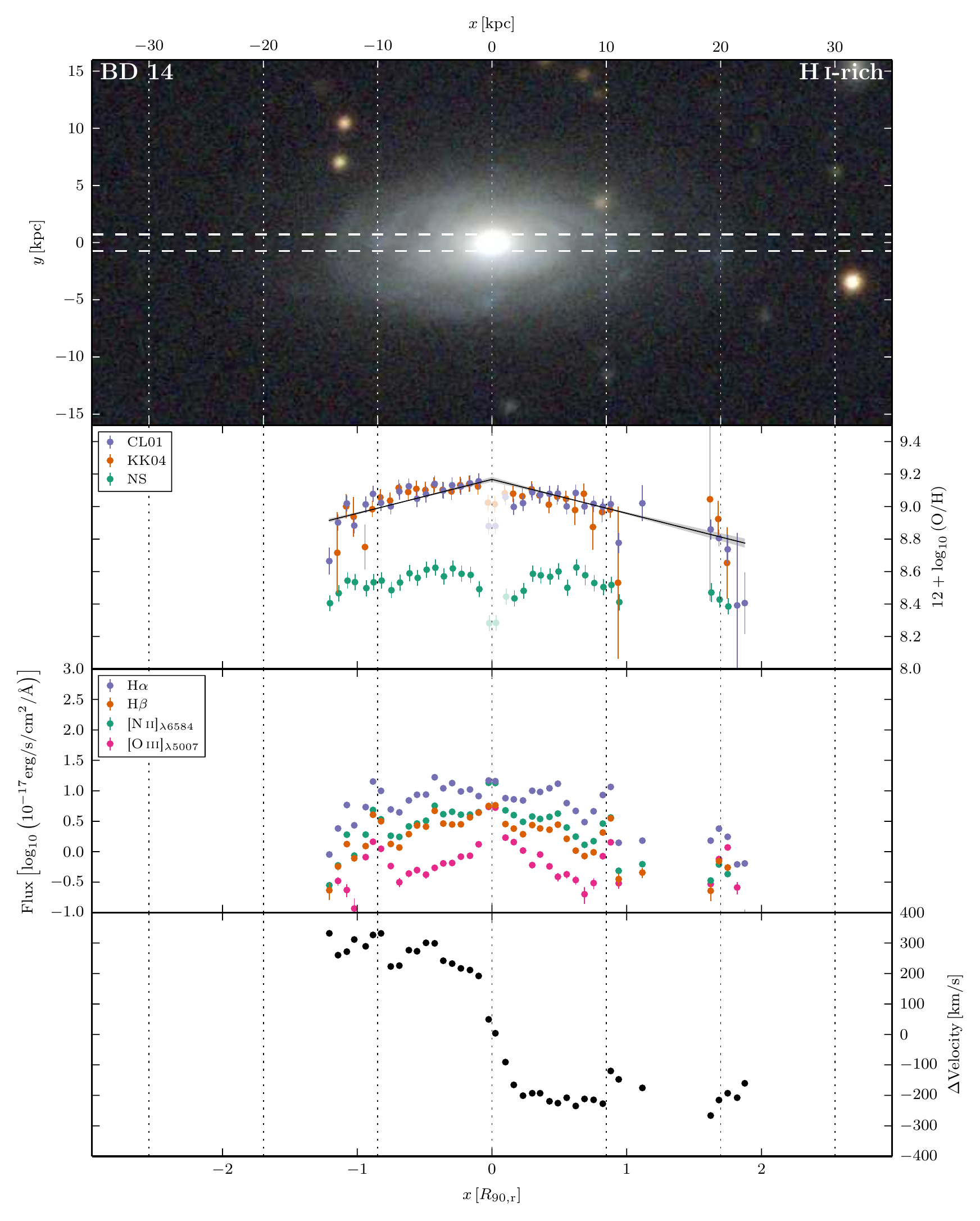}
\caption{Atlas of data for BD 14. See text for details.}
\label{fig:atlas_14}
\end{figure*}

\begin{figure*}
\includegraphics[width=\linewidth]{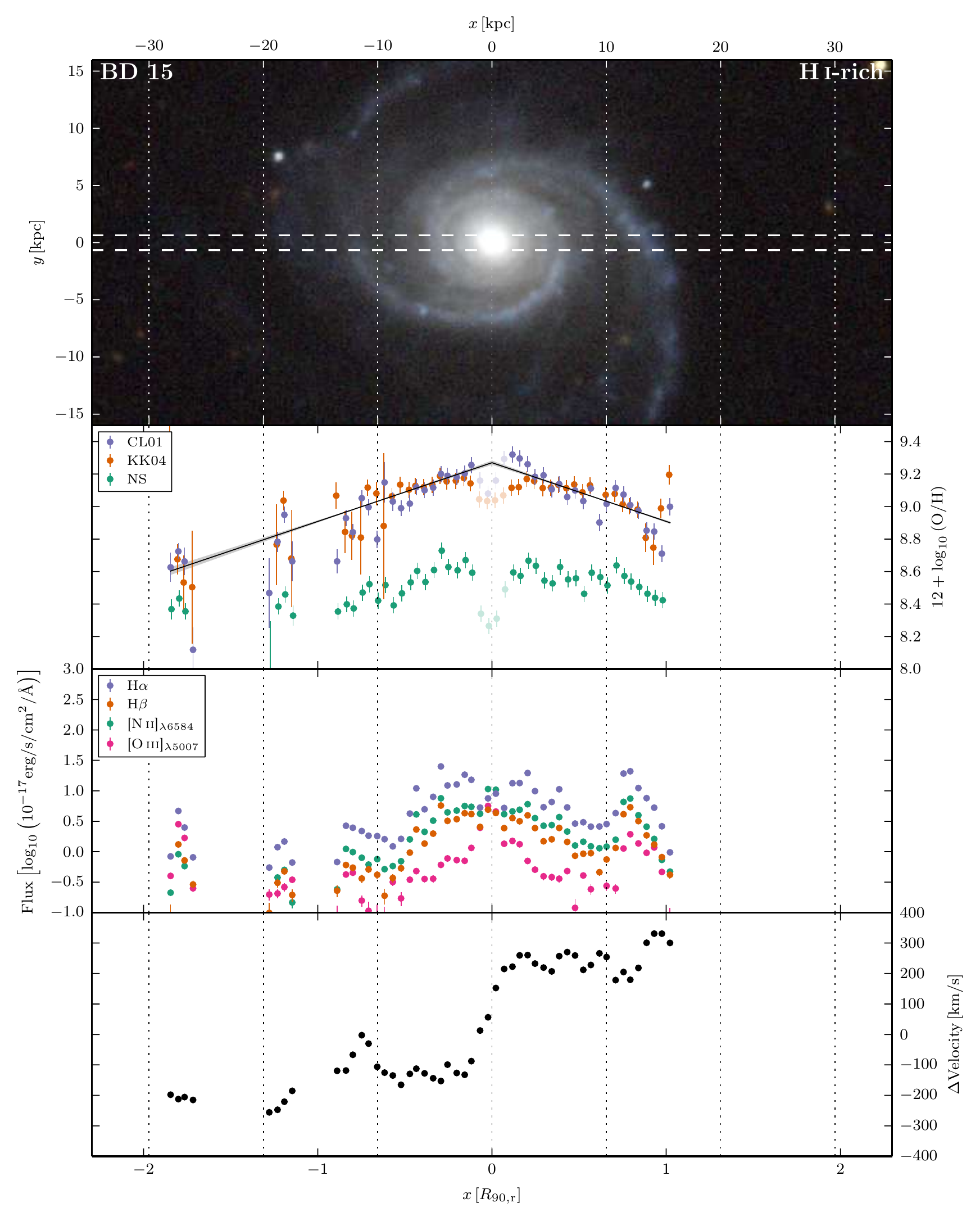}
\caption{Atlas of data for BD 15. See text for details.}
\label{fig:atlas_15}
\end{figure*}

\begin{figure*}
\includegraphics[width=\linewidth]{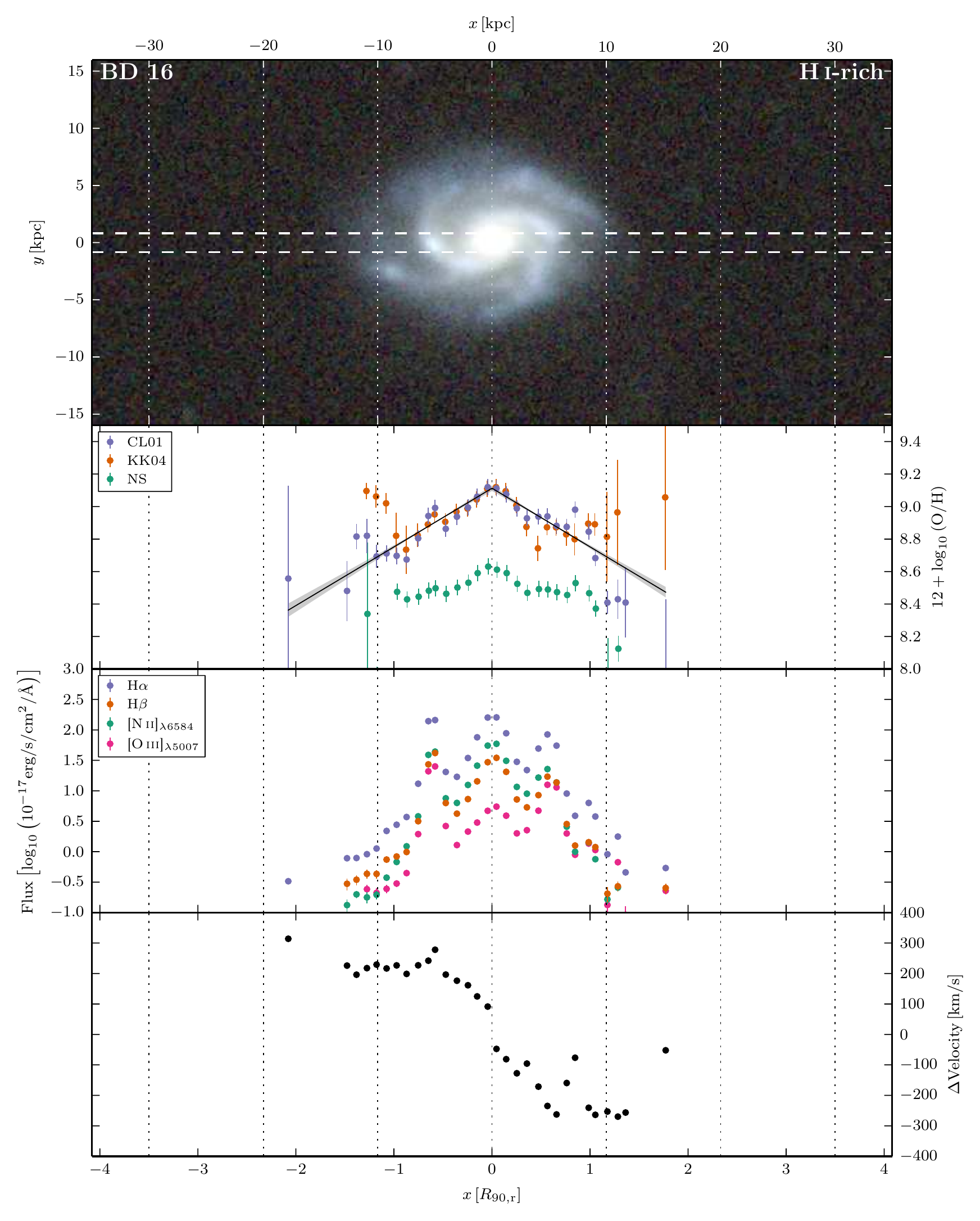}
\caption{Atlas of data for BD 16. See text for details.}
\label{fig:atlas_16}
\end{figure*}

\begin{figure*}
\includegraphics[width=\linewidth]{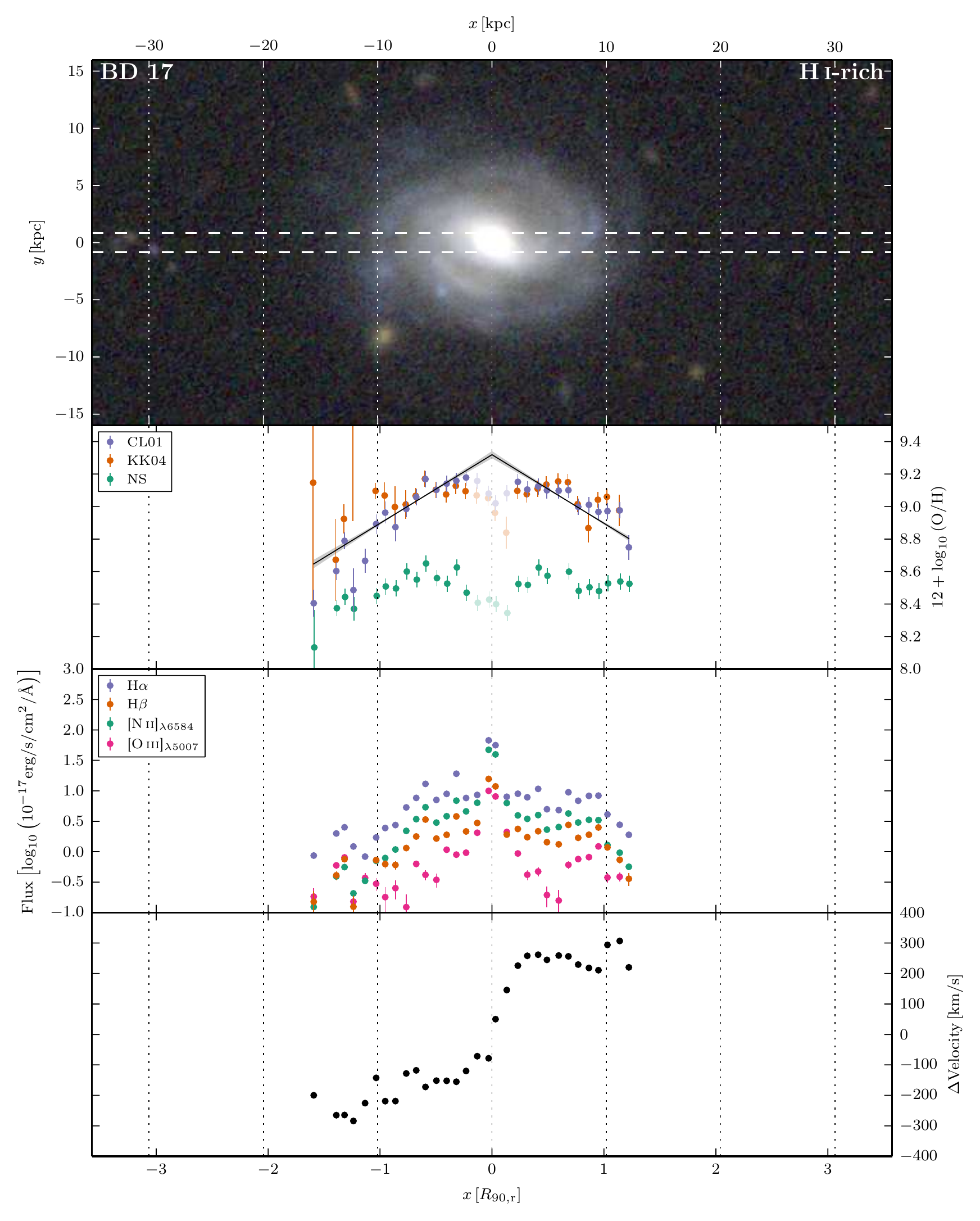}
\caption{Atlas of data for BD 17. See text for details.}
\label{fig:atlas_17}
\end{figure*}

\begin{figure*}
\includegraphics[width=\linewidth]{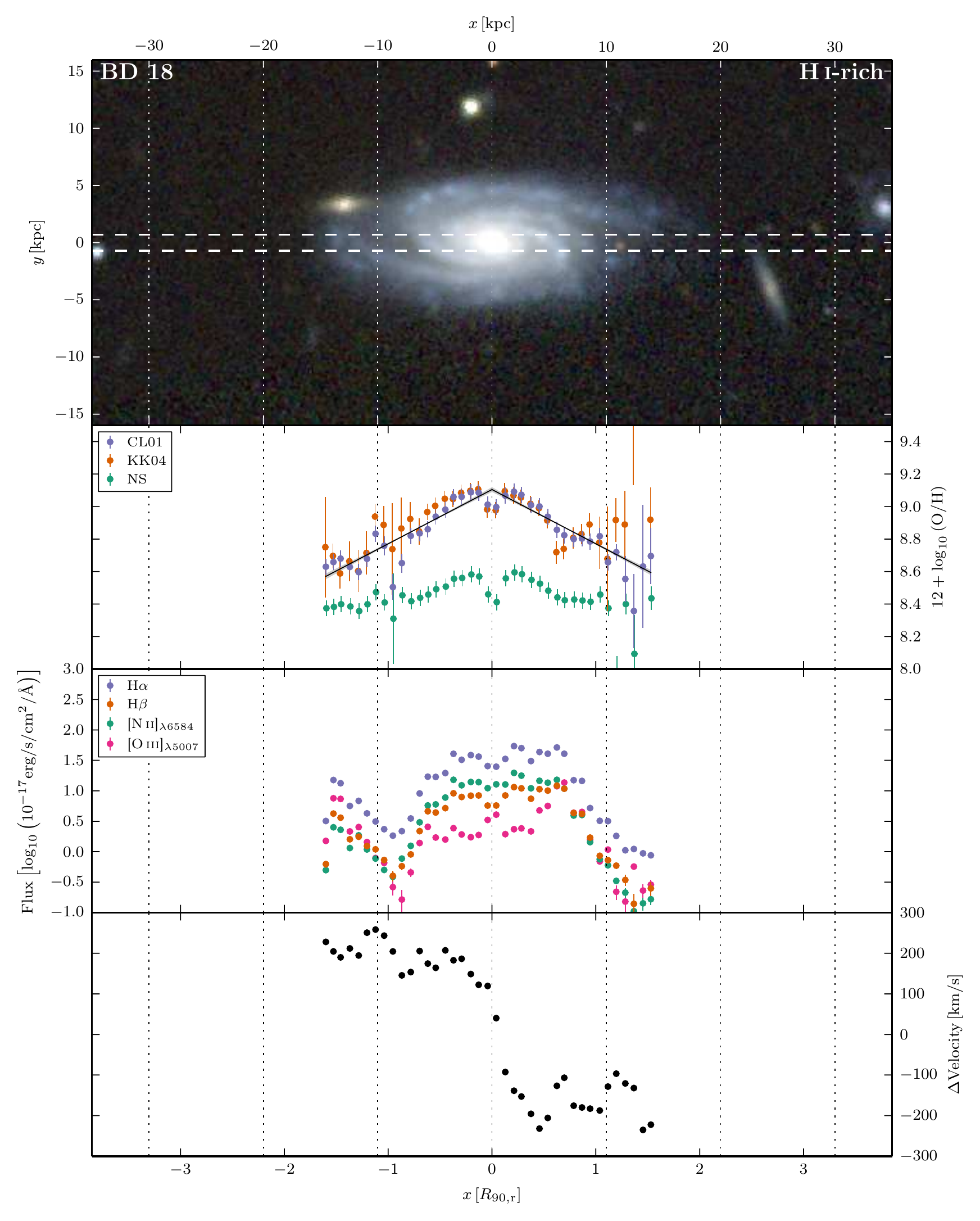}
\caption{Atlas of data for BD 18. See text for details.}
\label{fig:atlas_18}
\end{figure*}

\begin{figure*}
\includegraphics[width=\linewidth]{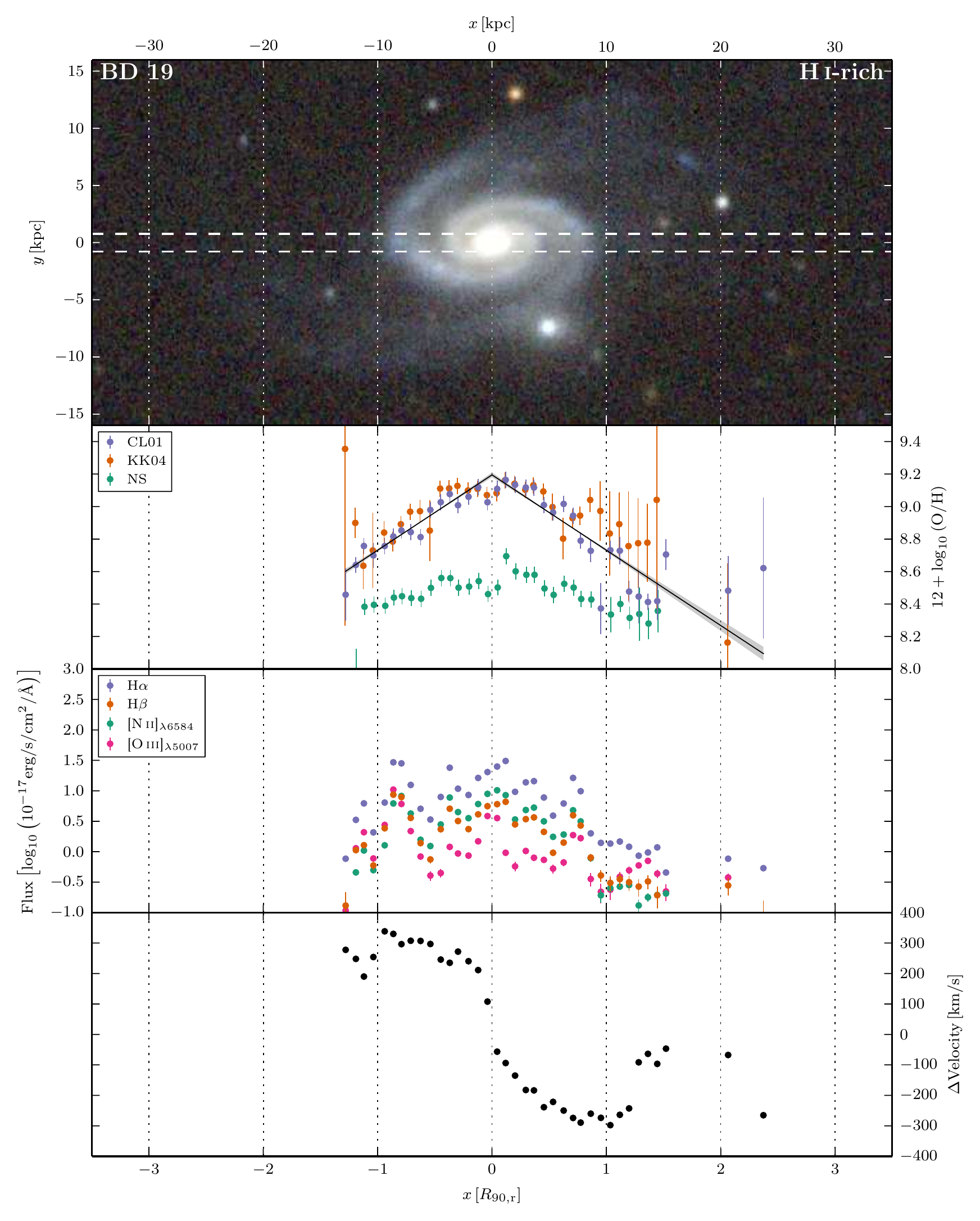}
\caption{Atlas of data for BD 19. See text for details.}
\label{fig:atlas_19}
\end{figure*}

\begin{figure*}
\includegraphics[width=\linewidth]{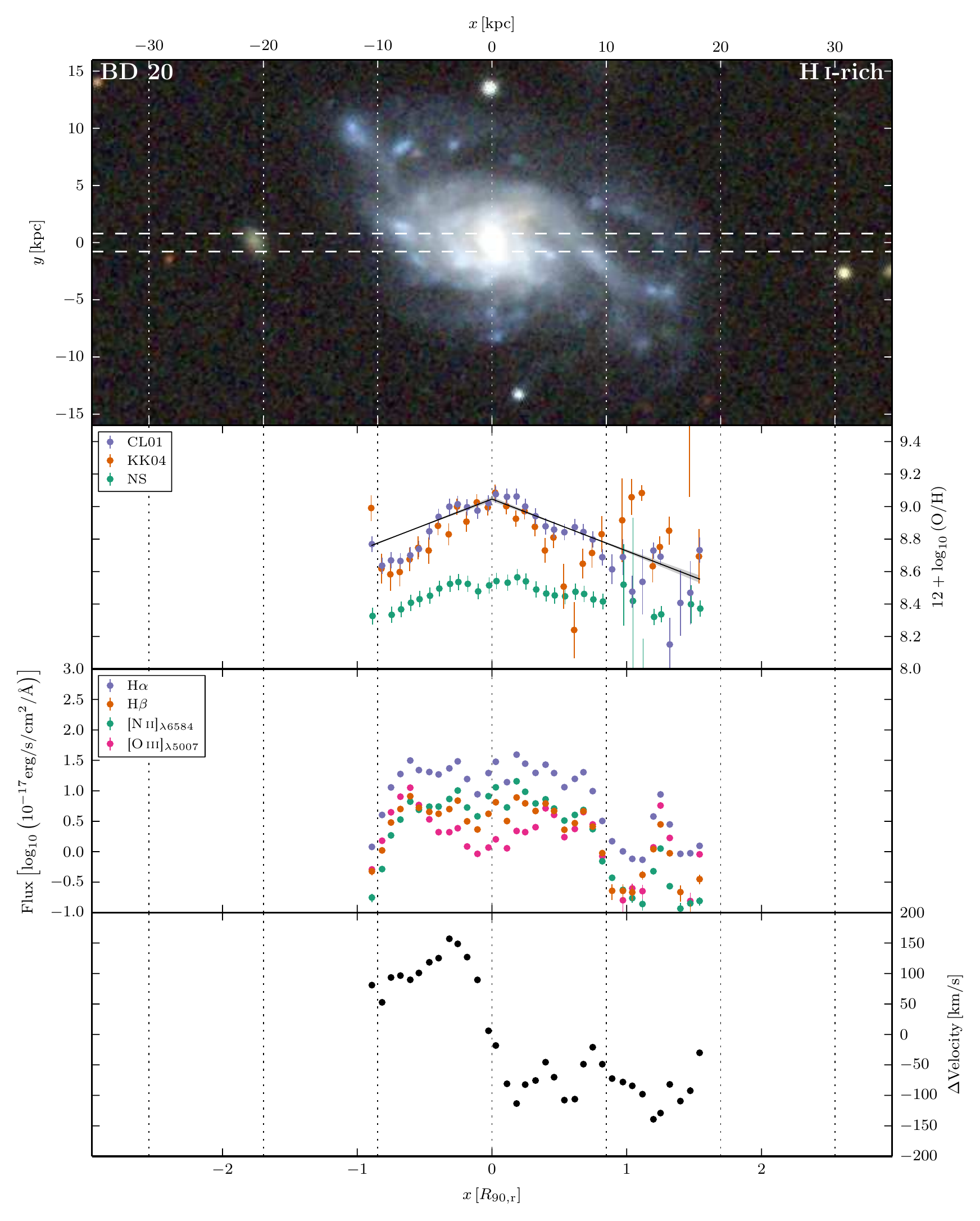}
\caption{Atlas of data for BD 20. See text for details.}
\label{fig:atlas_20}
\end{figure*}

\begin{figure*}
\includegraphics[width=\linewidth]{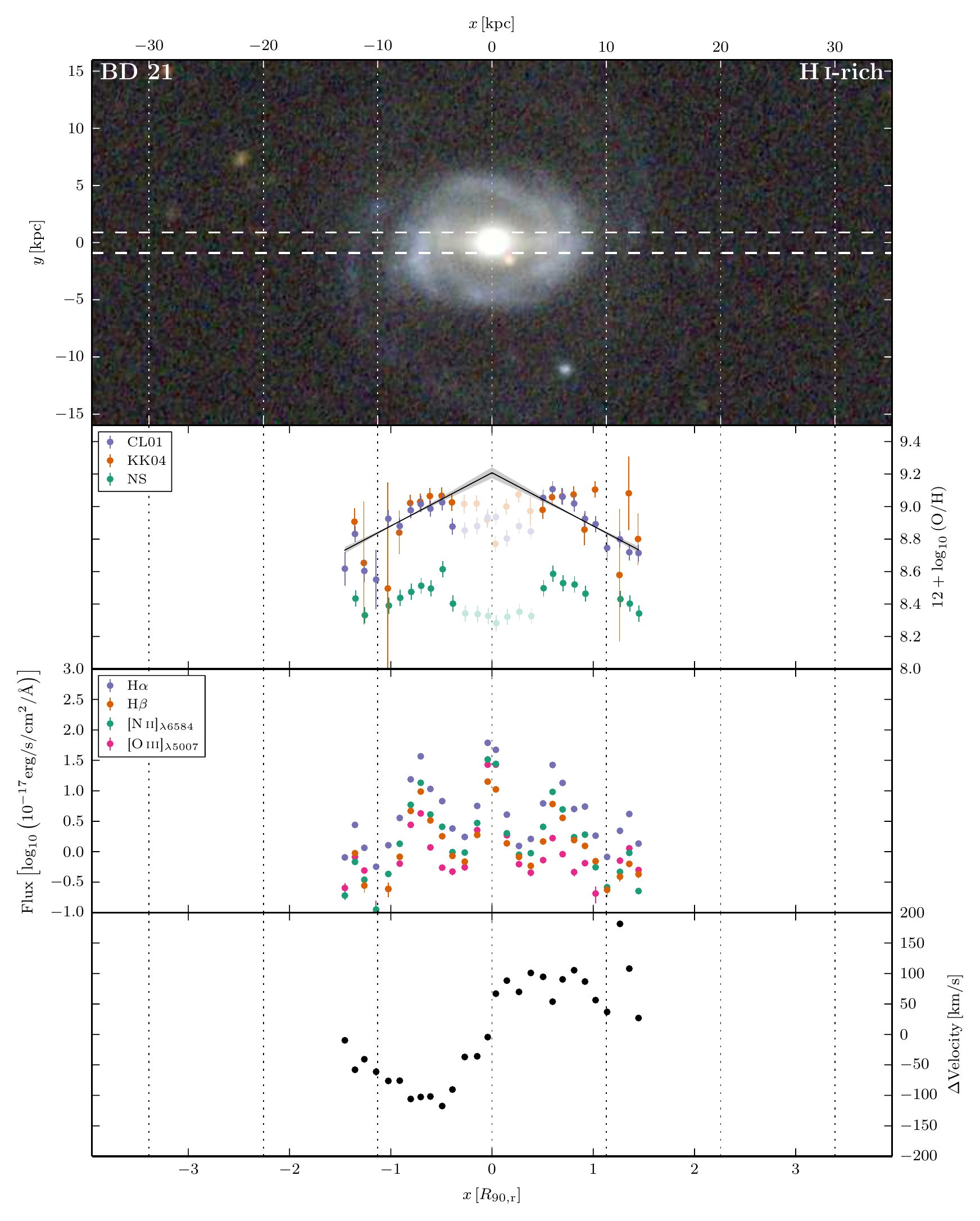}
\caption{Atlas of data for BD 21. See text for details.}
\label{fig:atlas_21}
\end{figure*}

\begin{figure*}
\includegraphics[width=\linewidth]{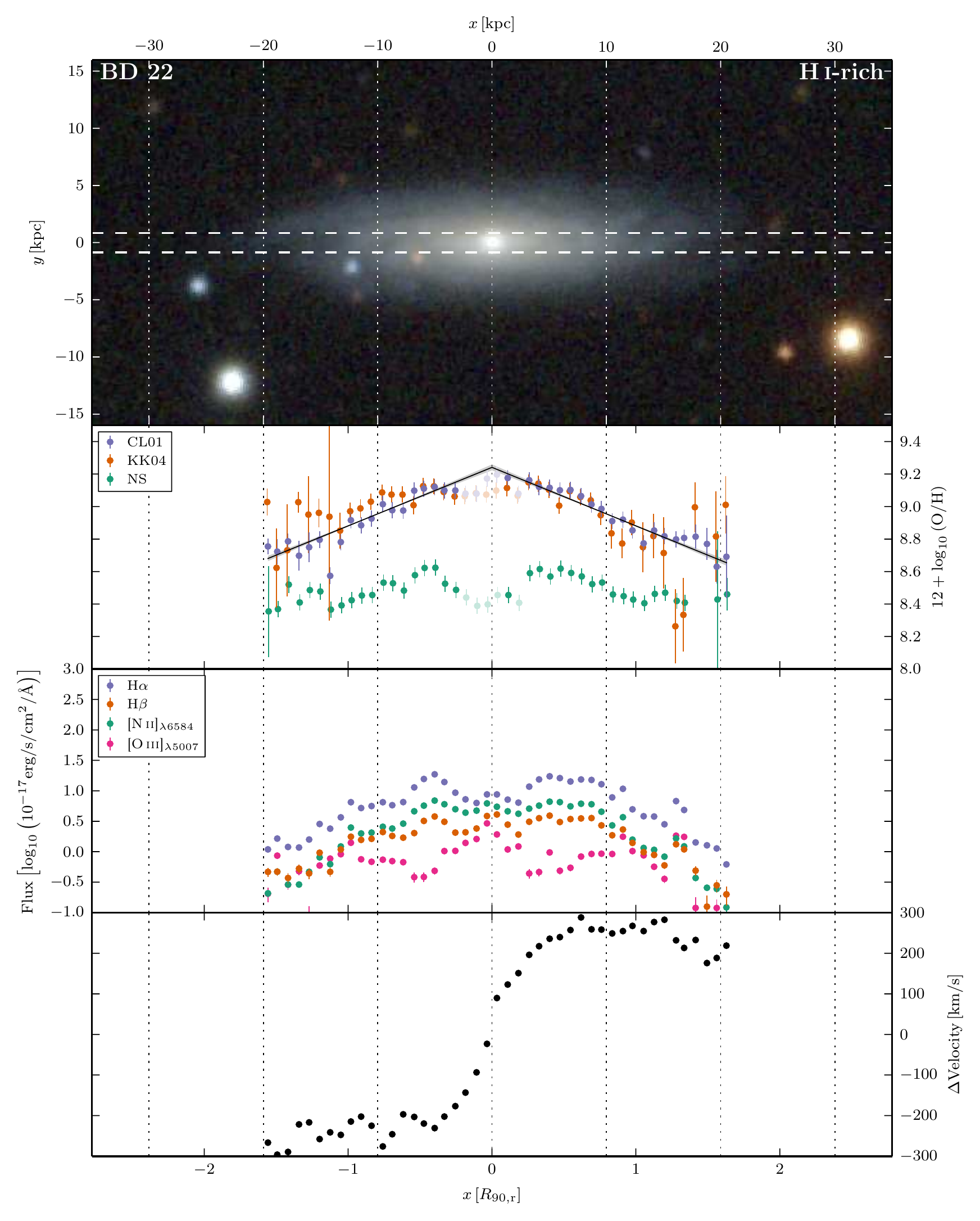}
\caption{Atlas of data for BD 22. See text for details.}
\label{fig:atlas_22}
\end{figure*}

\begin{figure*}
\includegraphics[width=\linewidth]{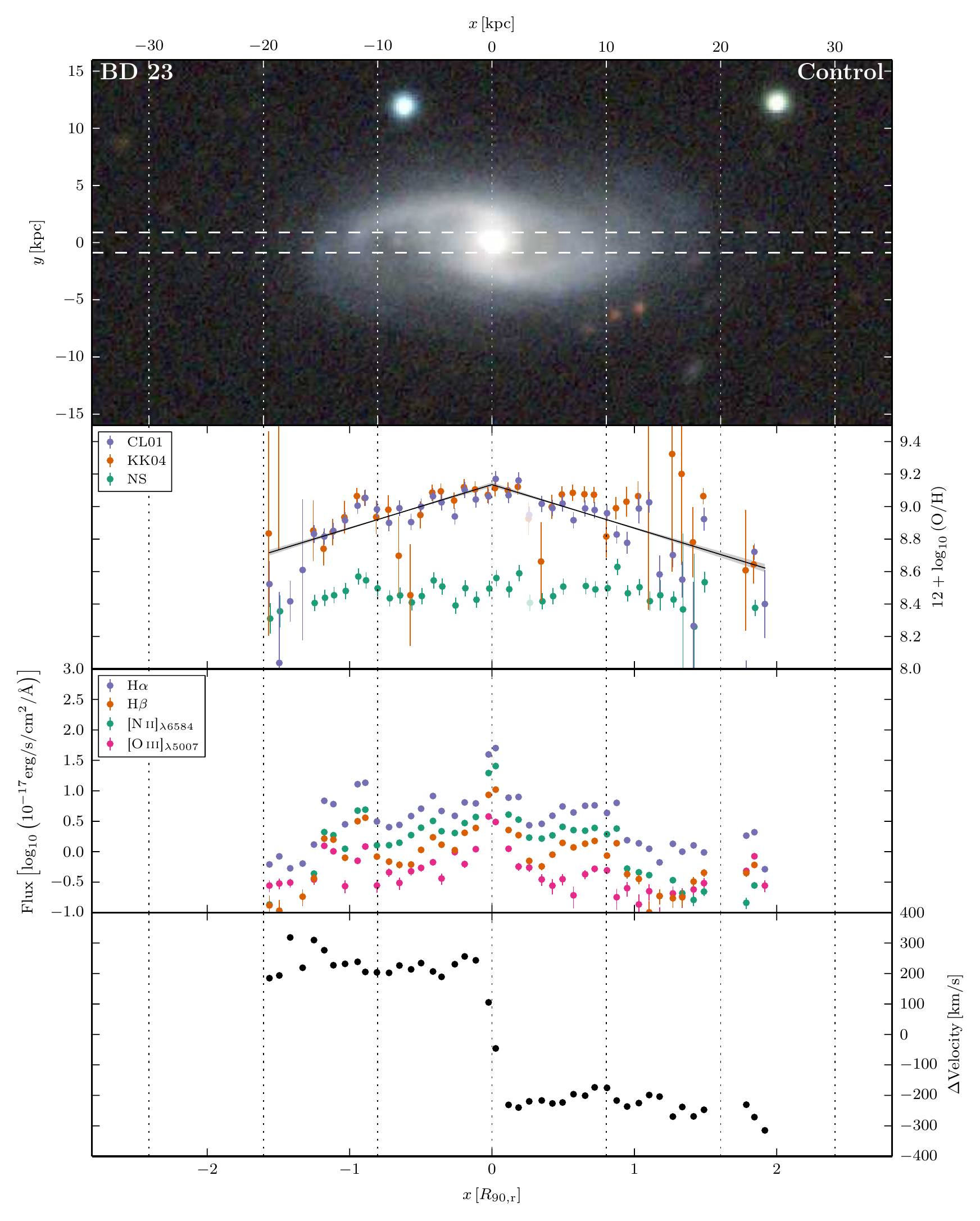}
\caption{Atlas of data for BD 23. See text for details.}
\label{fig:atlas_23}
\end{figure*}

\begin{figure*}
\includegraphics[width=\linewidth]{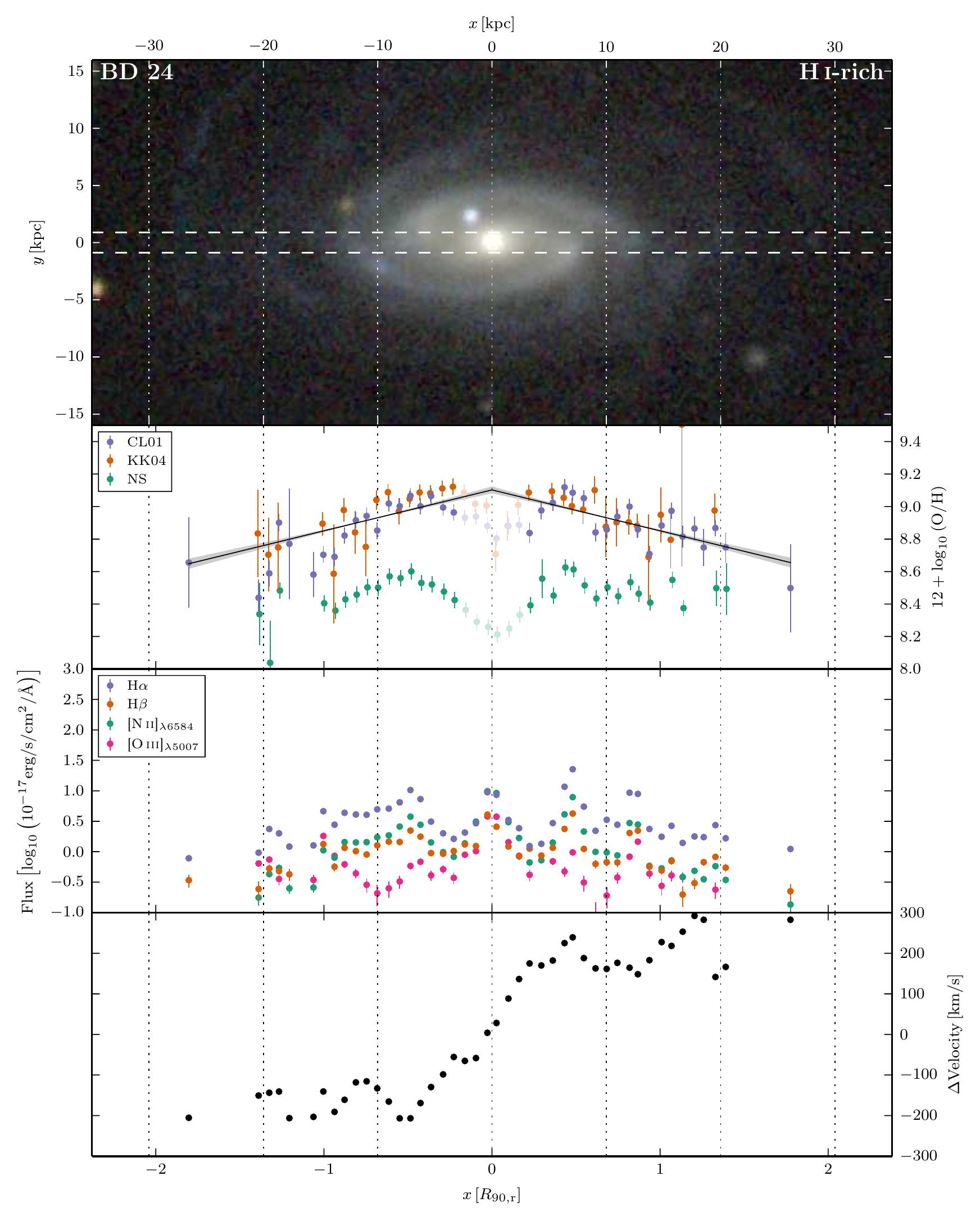}
\caption{Atlas of data for BD 24. See text for details.}
\label{fig:atlas_24}
\end{figure*}

\begin{figure*}
\includegraphics[width=\linewidth]{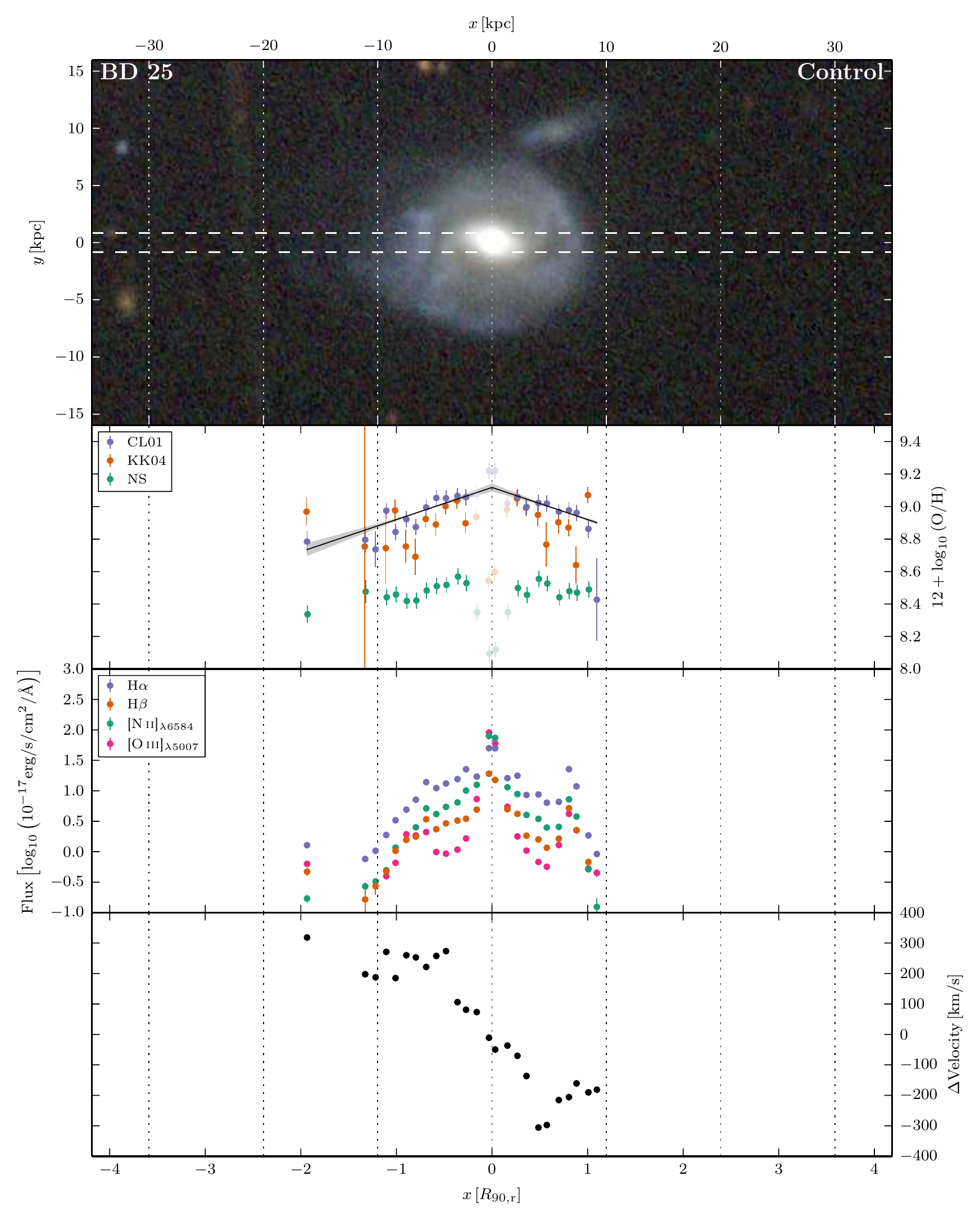}
\caption{Atlas of data for BD 25. See text for details.}
\label{fig:atlas_25}
\end{figure*}

\begin{figure*}
\includegraphics[width=\linewidth]{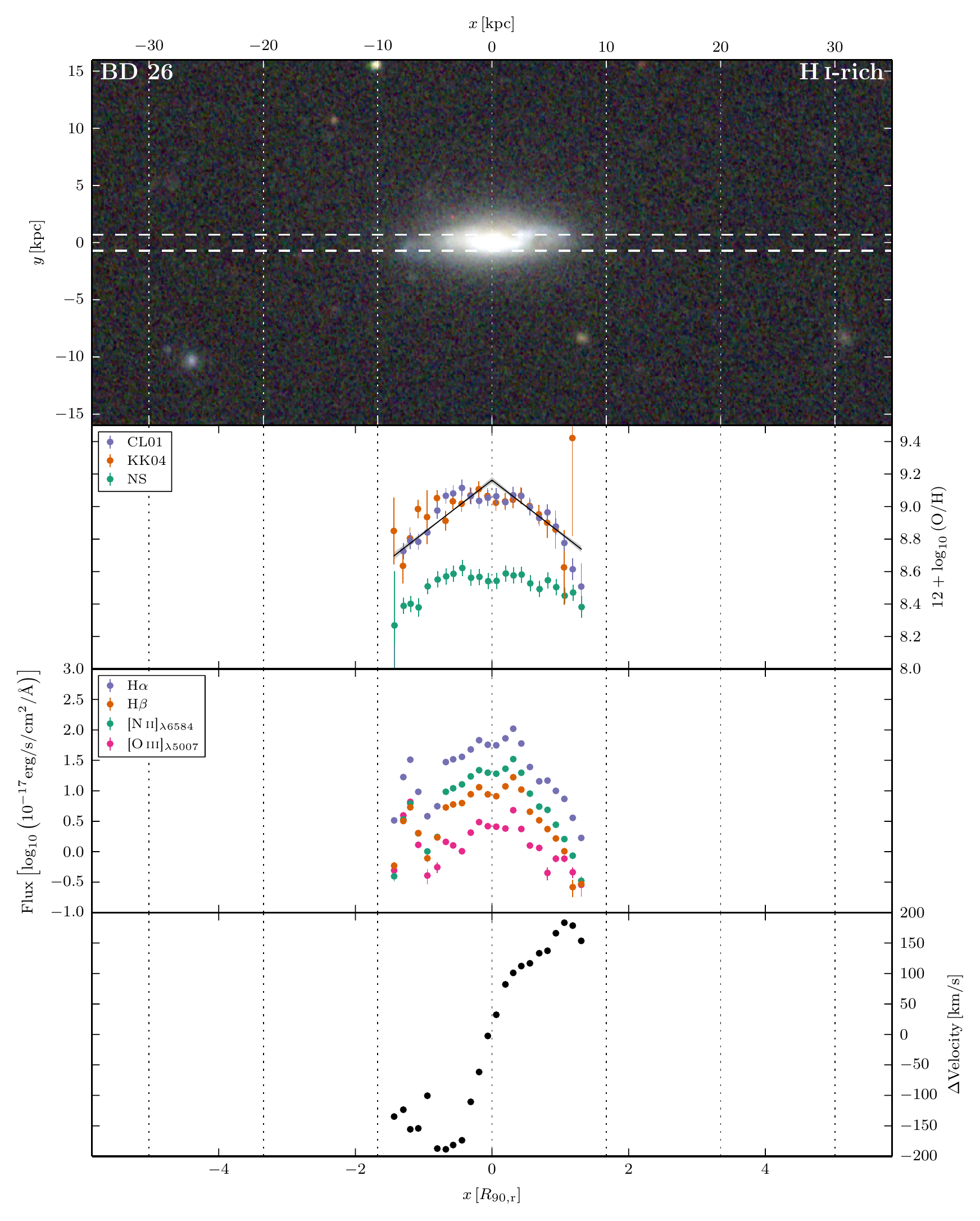}
\caption{Atlas of data for BD 26. See text for details.}
\label{fig:atlas_26}
\end{figure*}

\begin{figure*}
\includegraphics[width=\linewidth]{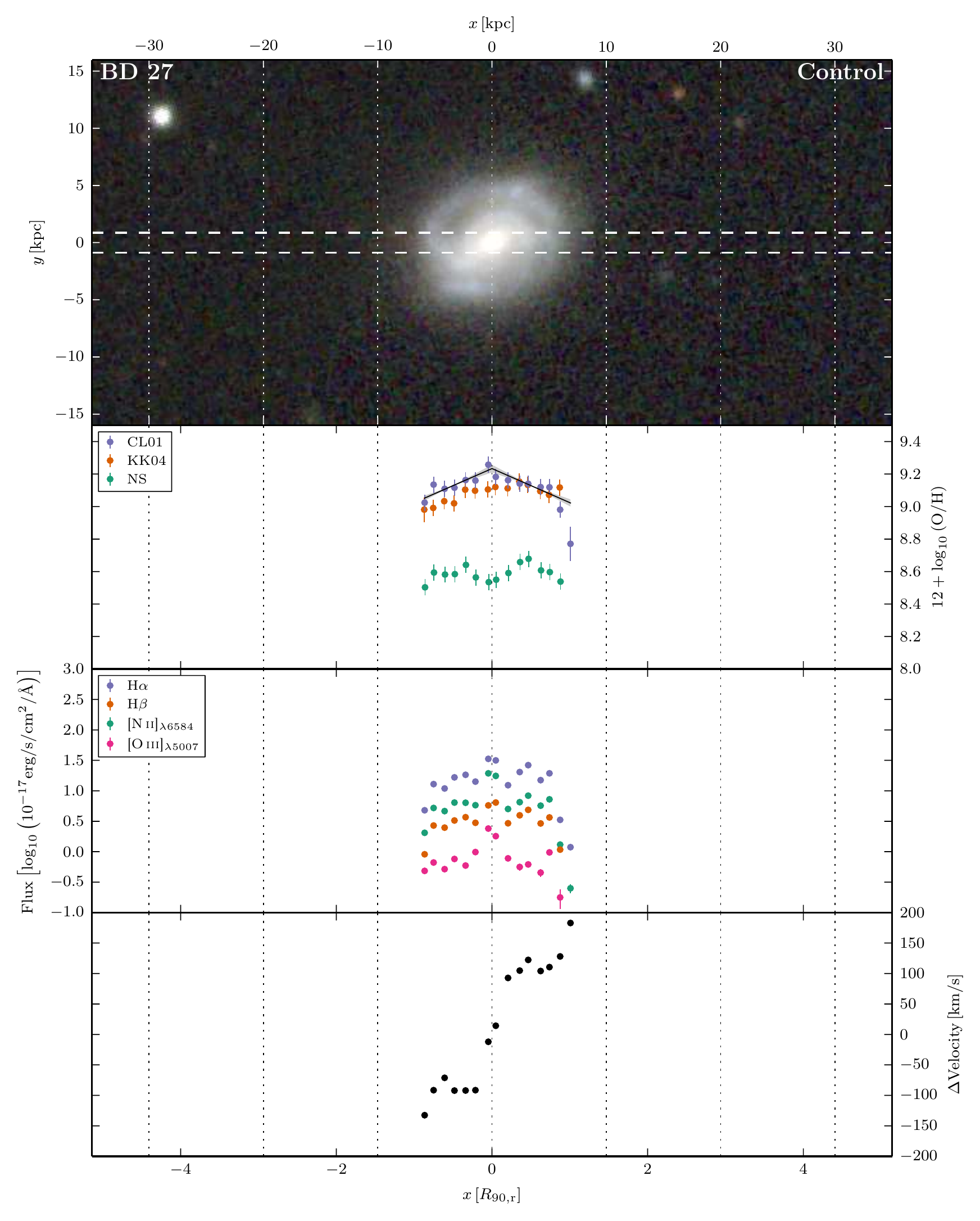}
\caption{Atlas of data for BD 27. See text for details.}
\label{fig:atlas_27}
\end{figure*}

\begin{figure*}
\includegraphics[width=\linewidth]{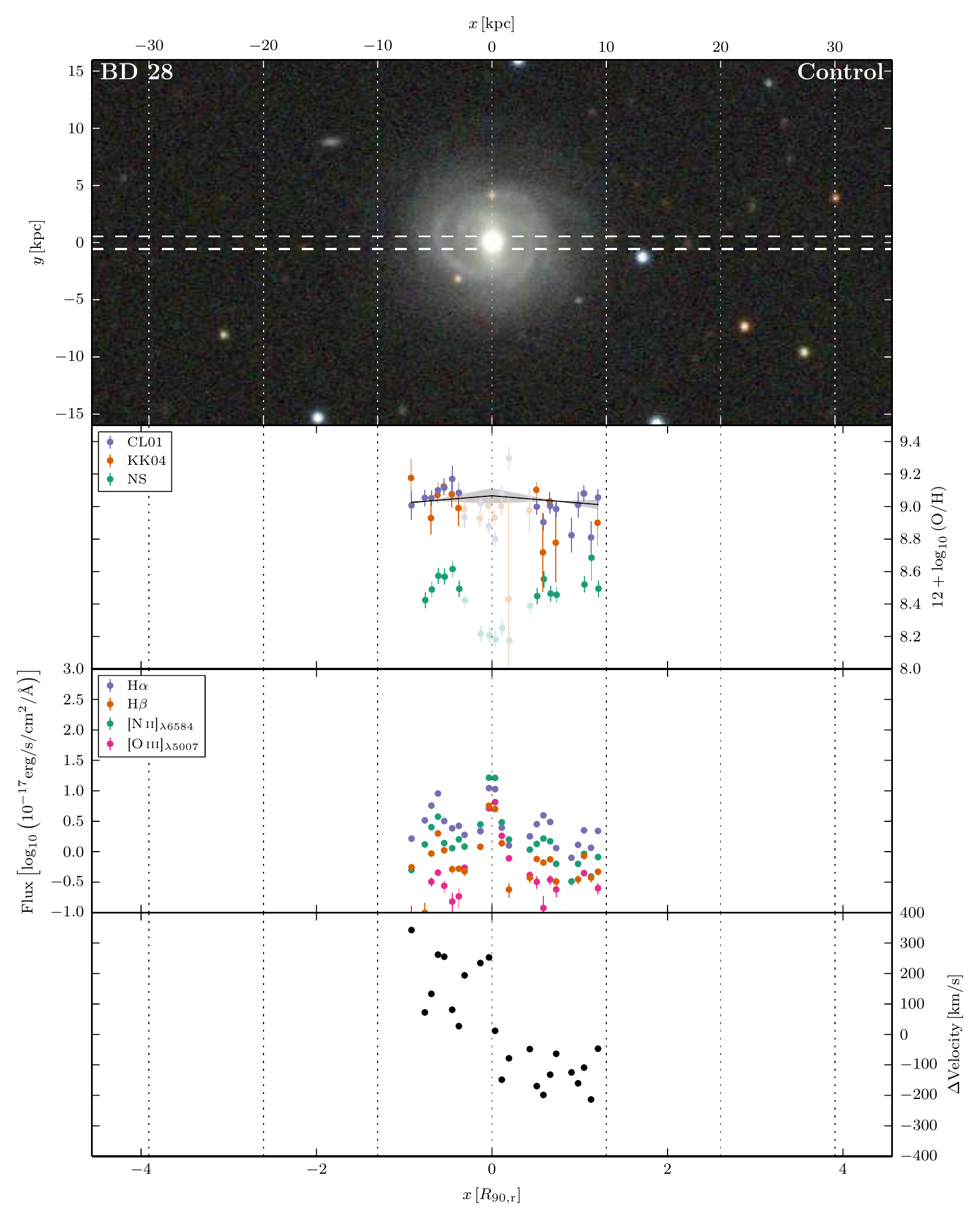}
\caption{Atlas of data for BD 28. See text for details.}
\label{fig:atlas_28}
\end{figure*}

\begin{figure*}
\includegraphics[width=\linewidth]{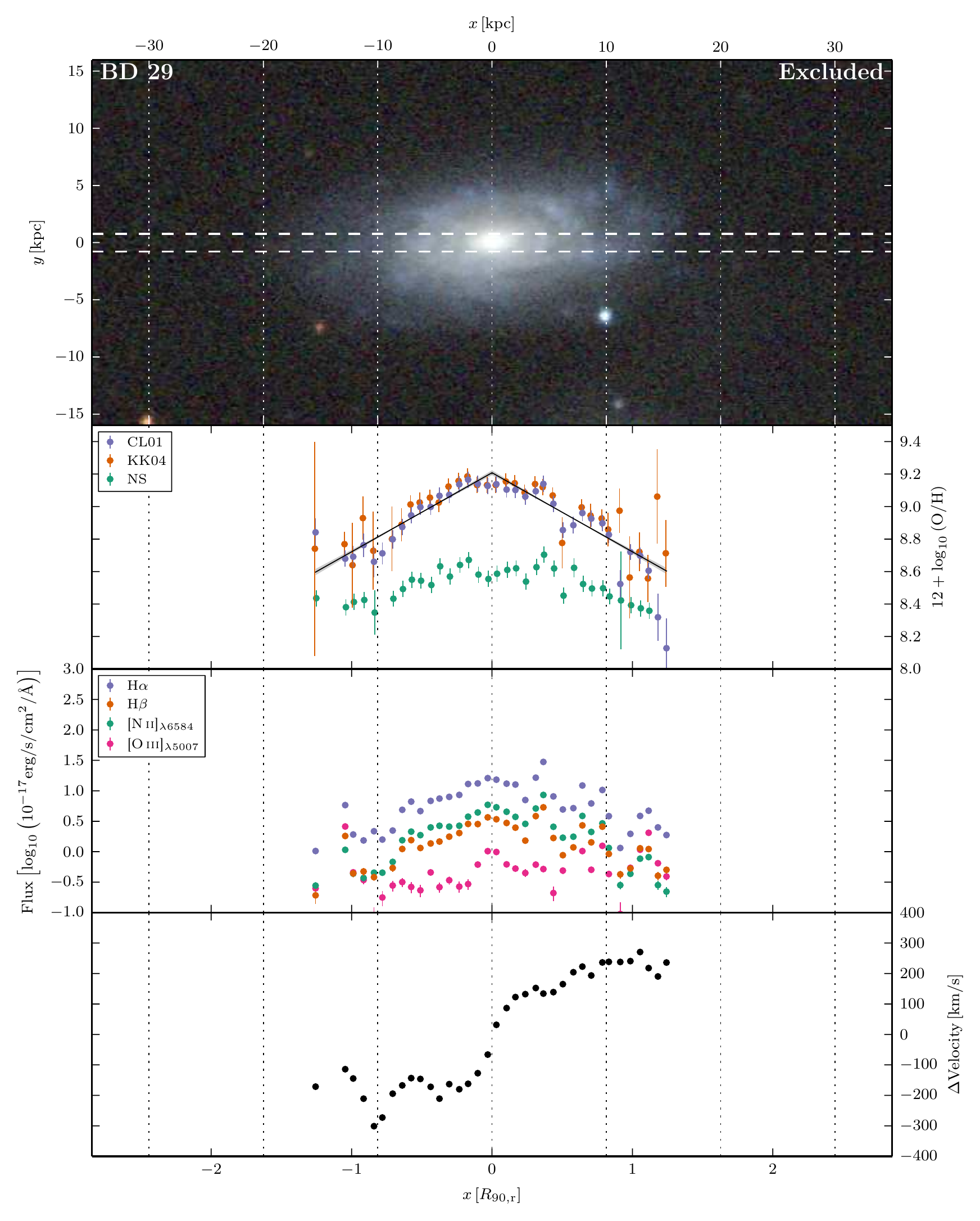}
\caption{Atlas of data for BD 29. See text for details.}
\label{fig:atlas_29}
\end{figure*}

\begin{figure*}
\includegraphics[width=\linewidth]{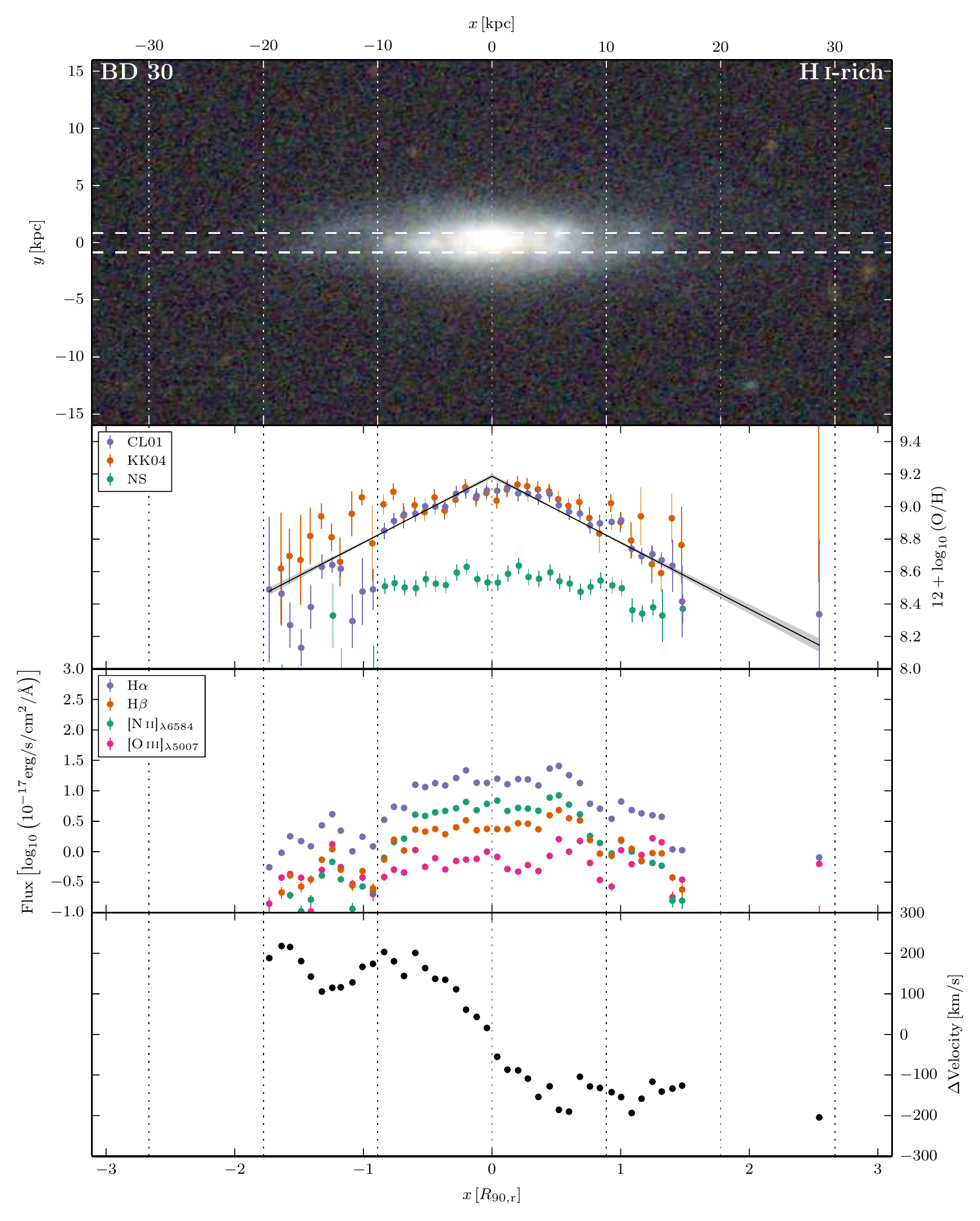}
\caption{Atlas of data for BD 30. See text for details.}
\label{fig:atlas_30}
\end{figure*}

\begin{figure*}
\includegraphics[width=\linewidth]{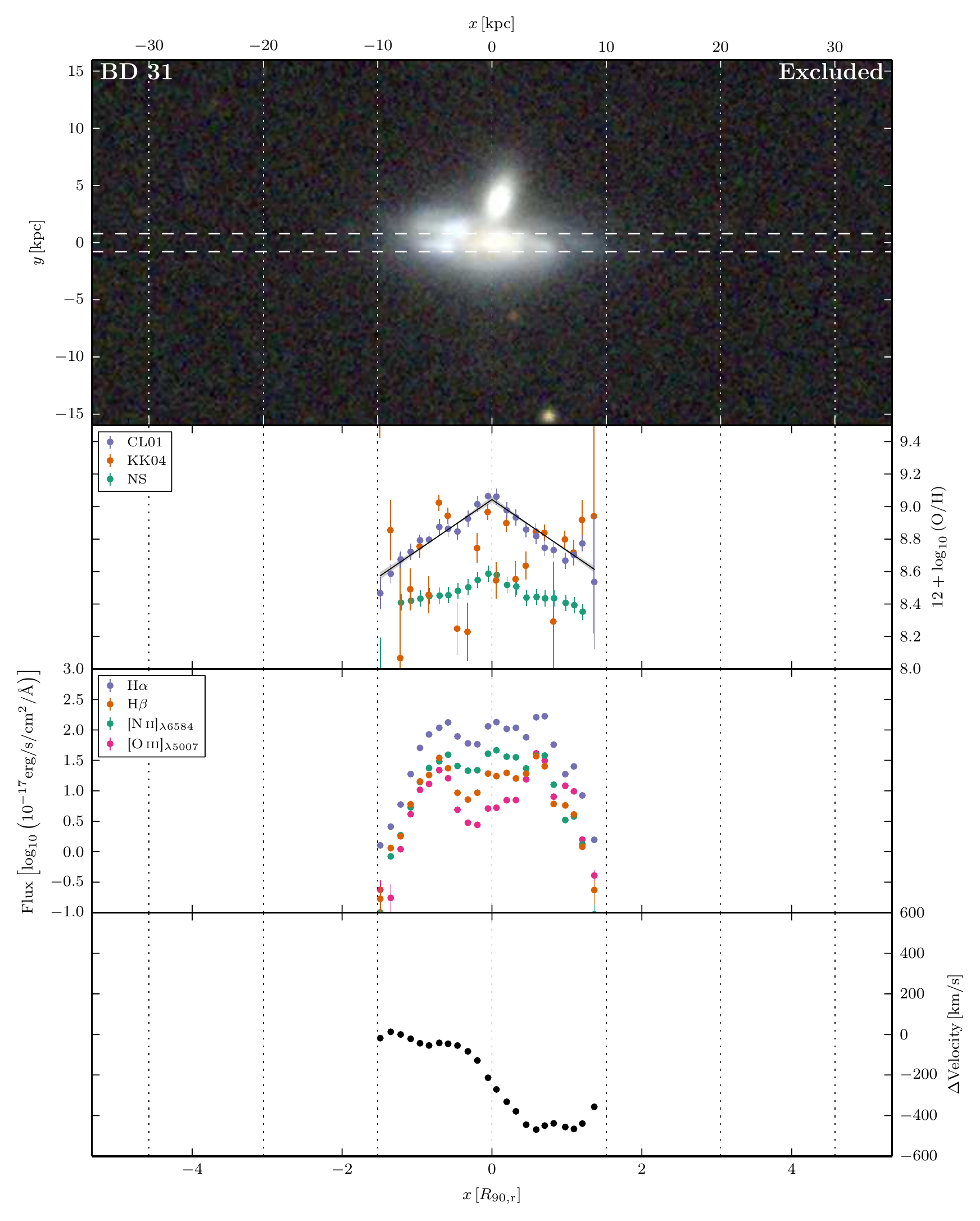}
\caption{Atlas of data for BD 31. See text for details.}
\label{fig:atlas_31}
\end{figure*}

\begin{figure*}
\includegraphics[width=\linewidth]{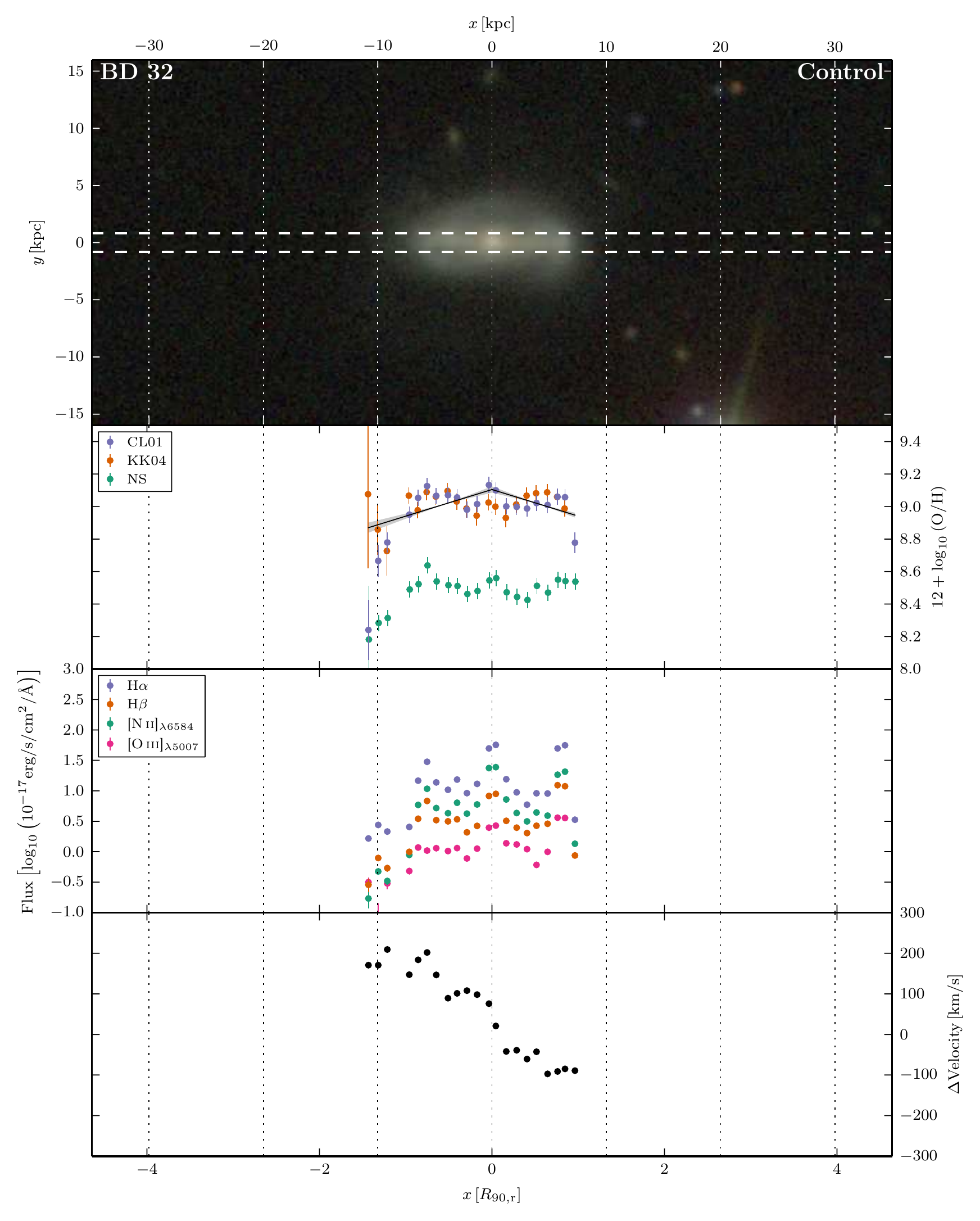}
\caption{Atlas of data for BD 32. See text for details.}
\label{fig:atlas_32}
\end{figure*}

\begin{figure*}
\includegraphics[width=\linewidth]{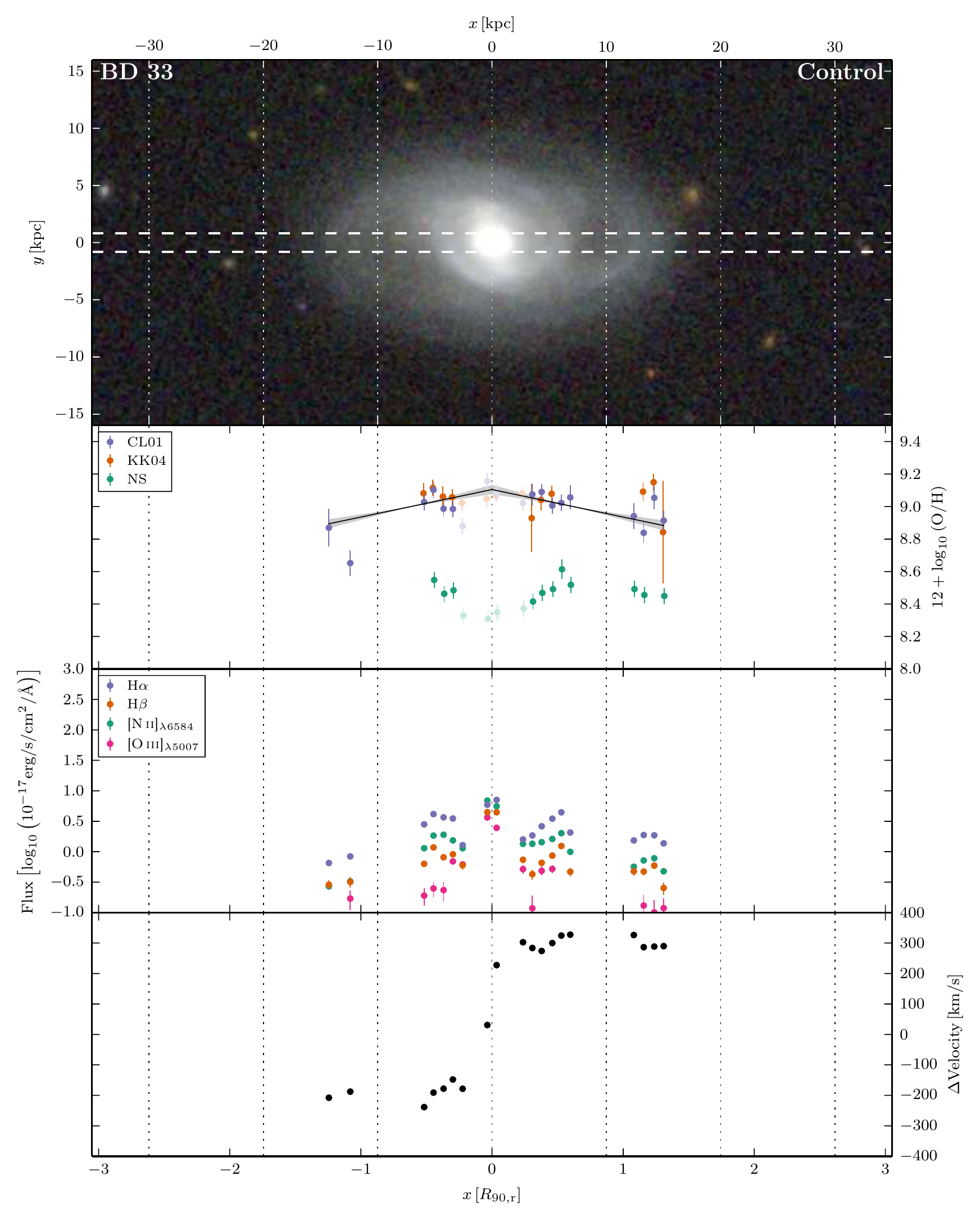}
\caption{Atlas of data for BD 33. See text for details.}
\label{fig:atlas_33}
\end{figure*}

\begin{figure*}
\includegraphics[width=\linewidth]{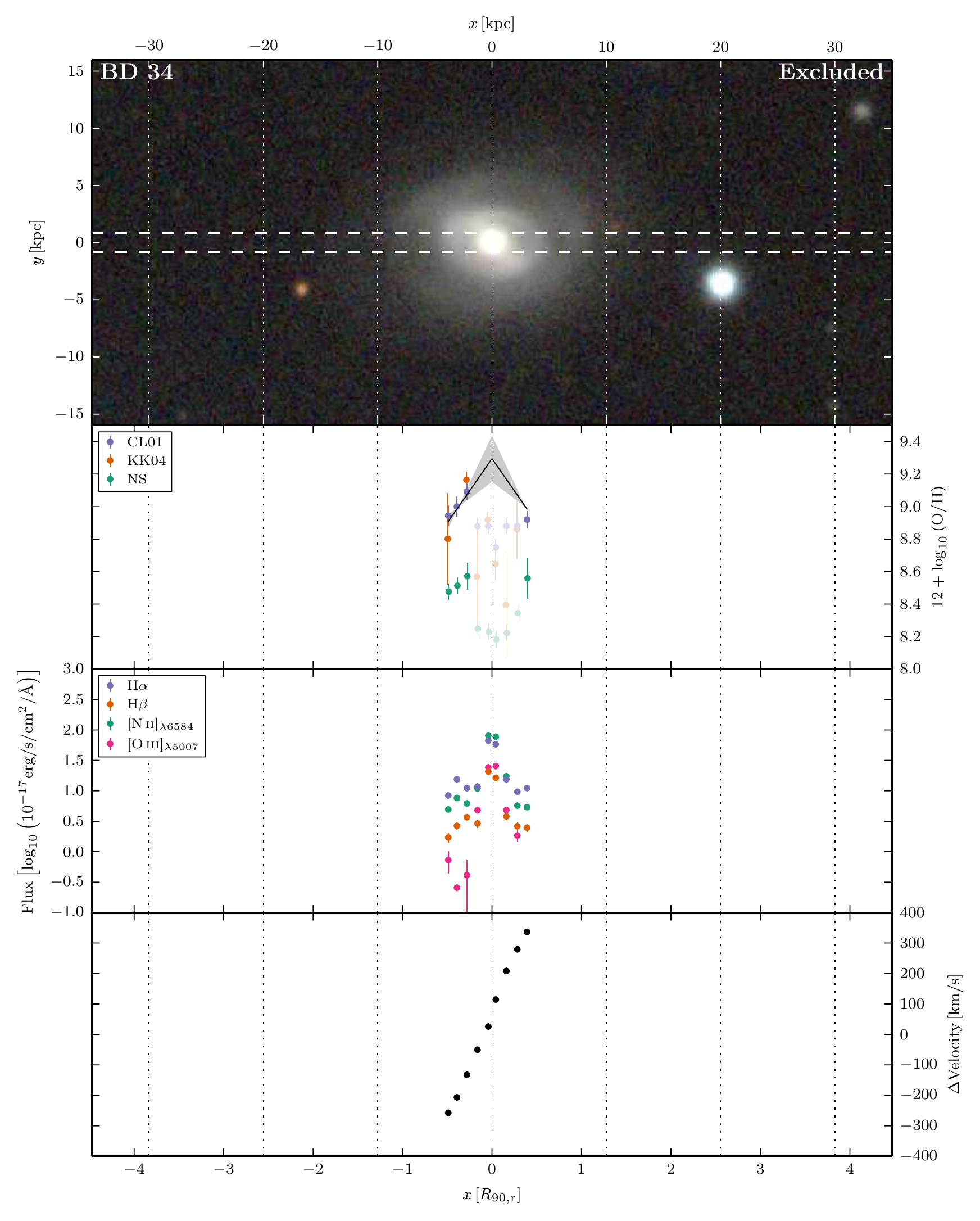}
\caption{Atlas of data for BD 34. See text for details.}
\label{fig:atlas_34}
\end{figure*}

\begin{figure*}
\includegraphics[width=\linewidth]{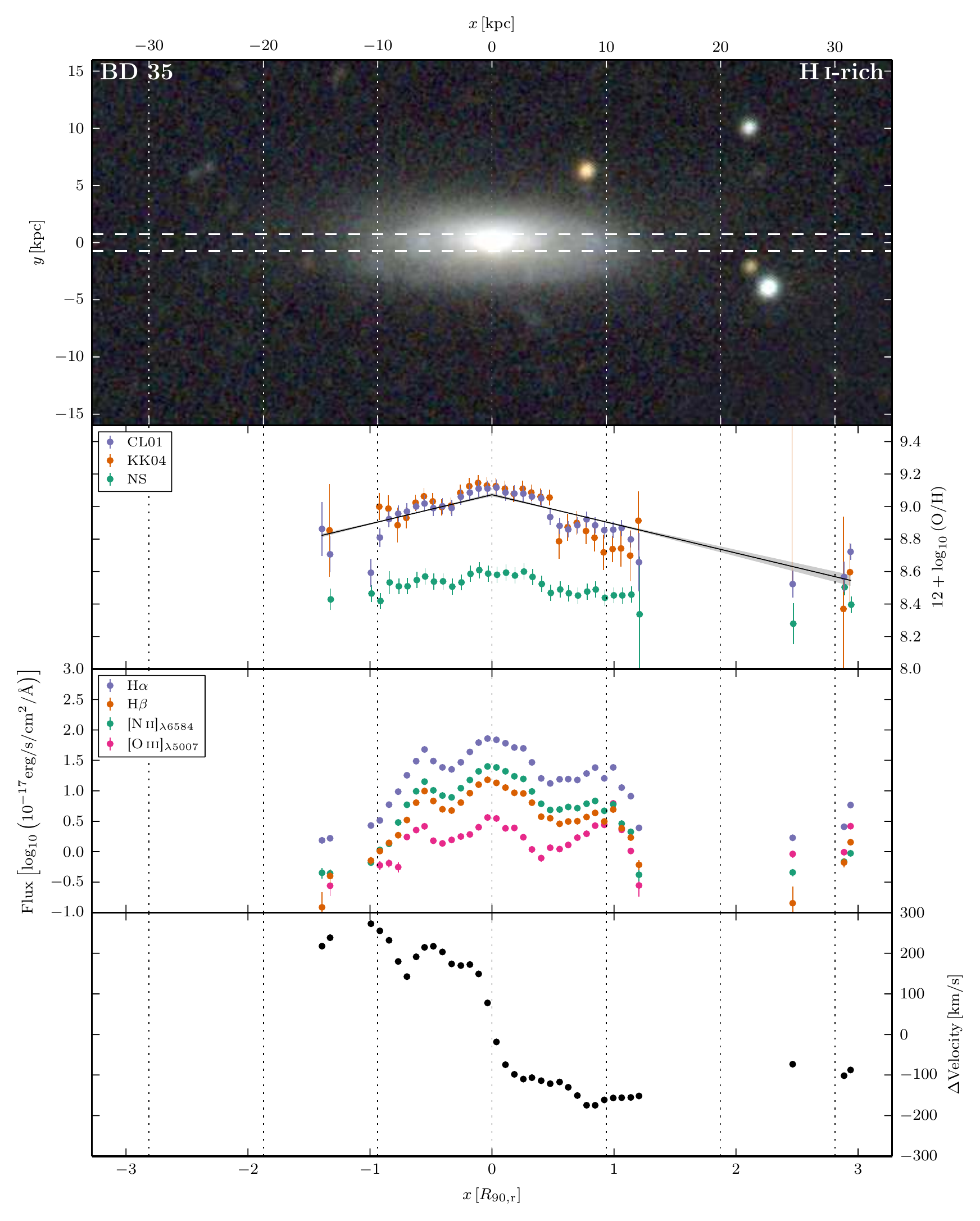}
\caption{Atlas of data for BD 35. See text for details.}
\label{fig:atlas_35}
\end{figure*}

\begin{figure*}
\includegraphics[width=\linewidth]{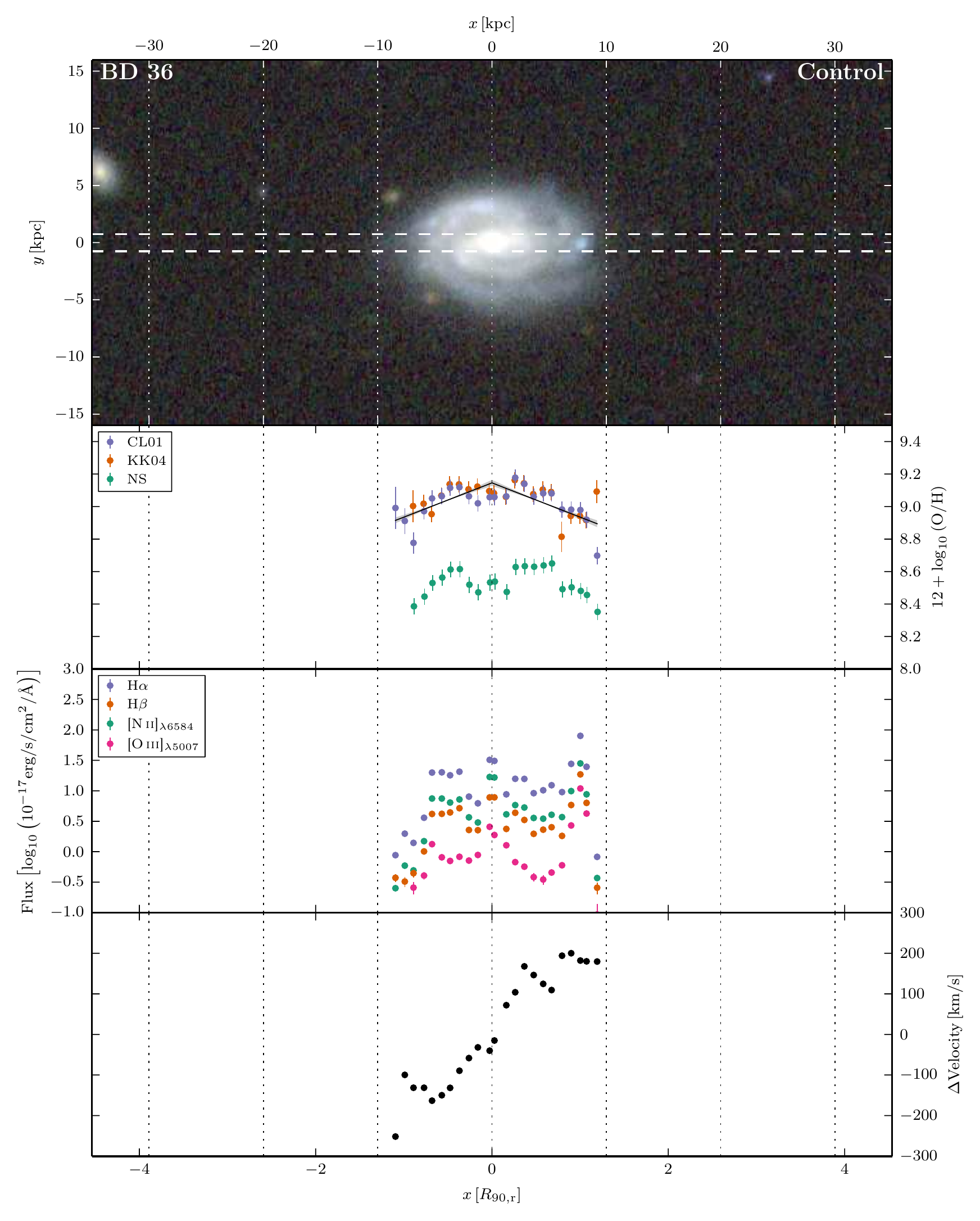}
\caption{Atlas of data for BD 36. See text for details.}
\label{fig:atlas_36}
\end{figure*}

\begin{figure*}
\includegraphics[width=\linewidth]{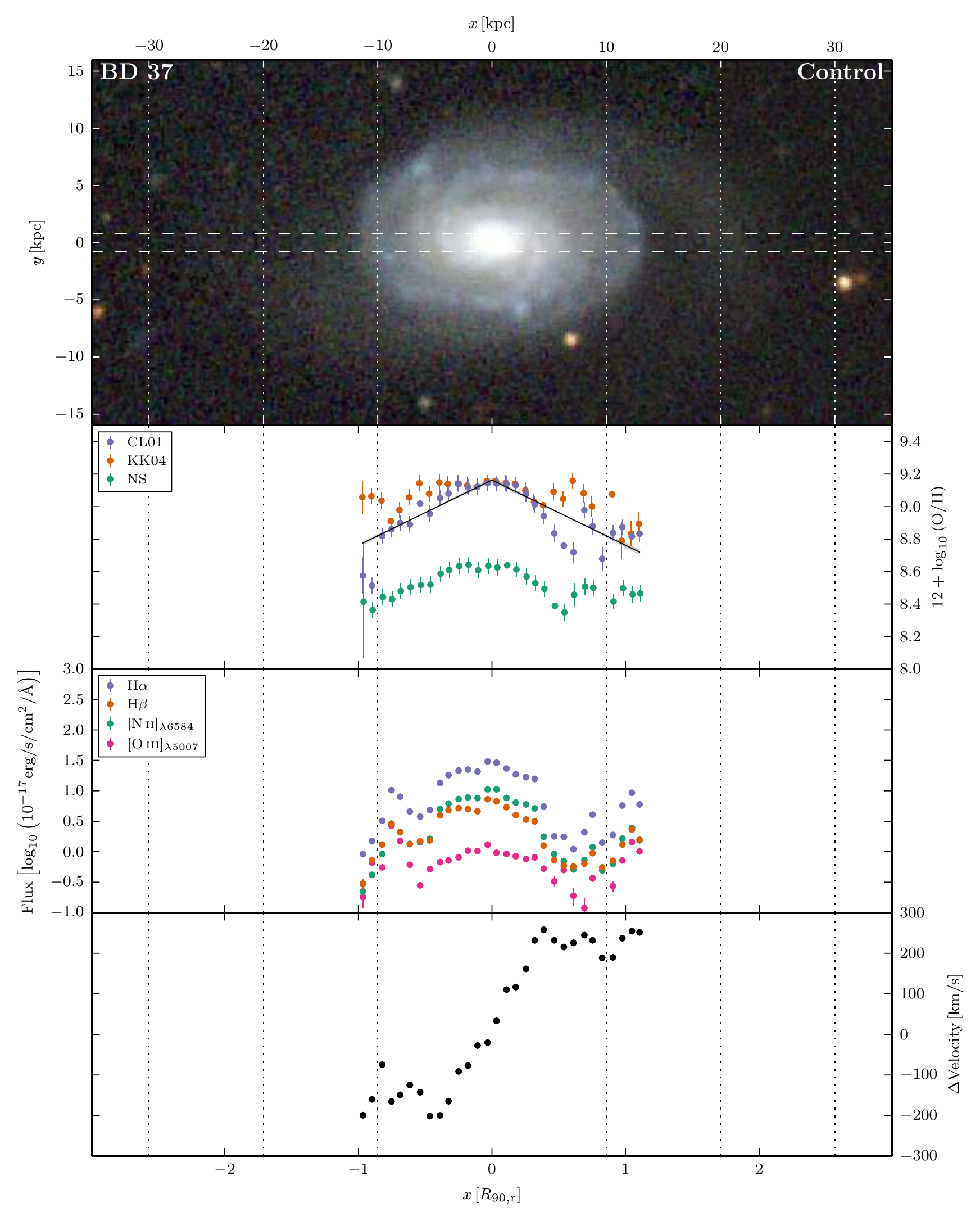}
\caption{Atlas of data for BD 37. See text for details.}
\label{fig:atlas_37}
\end{figure*}

\begin{figure*}
\includegraphics[width=\linewidth]{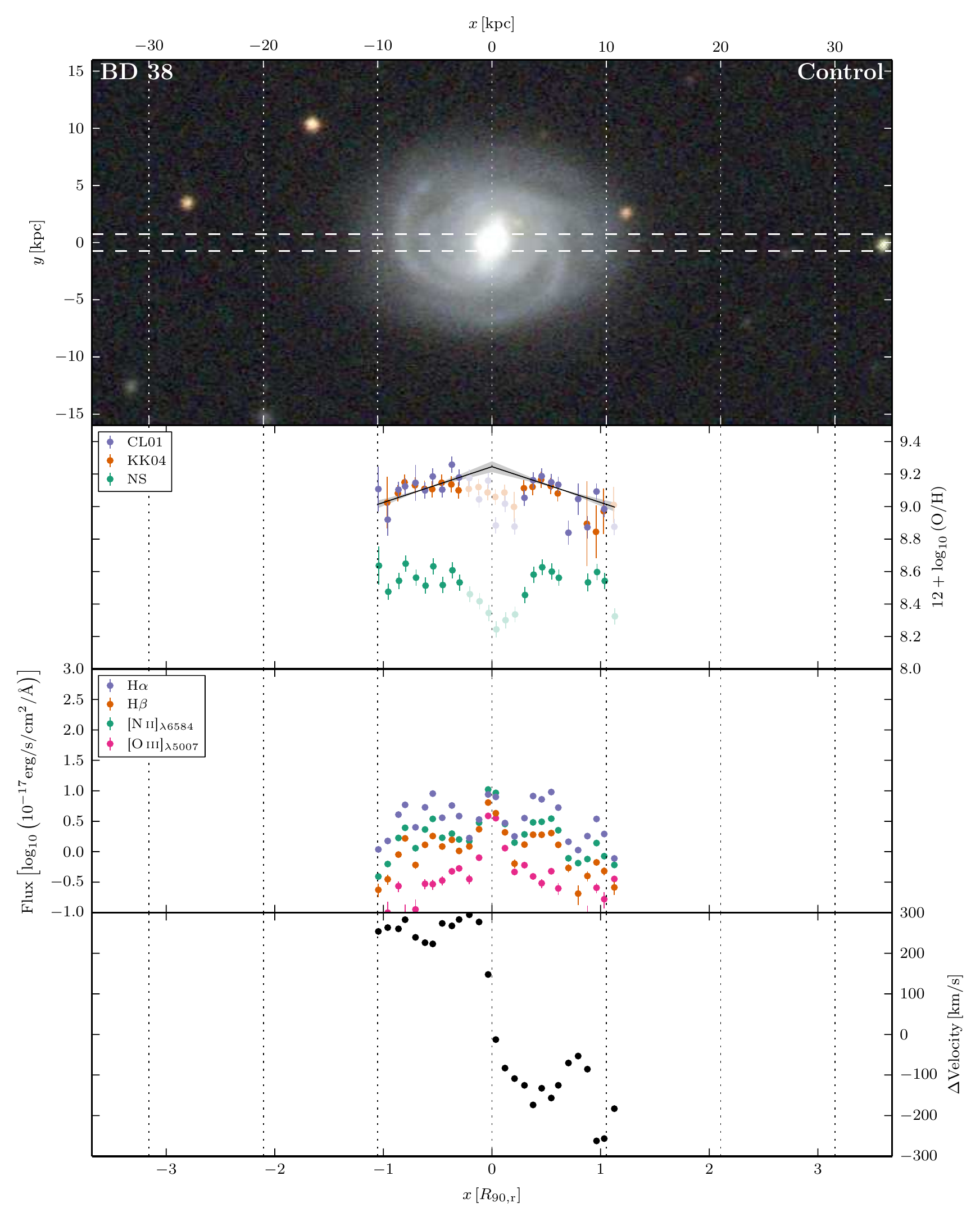}
\caption{Atlas of data for BD 38. See text for details.}
\label{fig:atlas_38}
\end{figure*}

\begin{figure*}
\includegraphics[width=\linewidth]{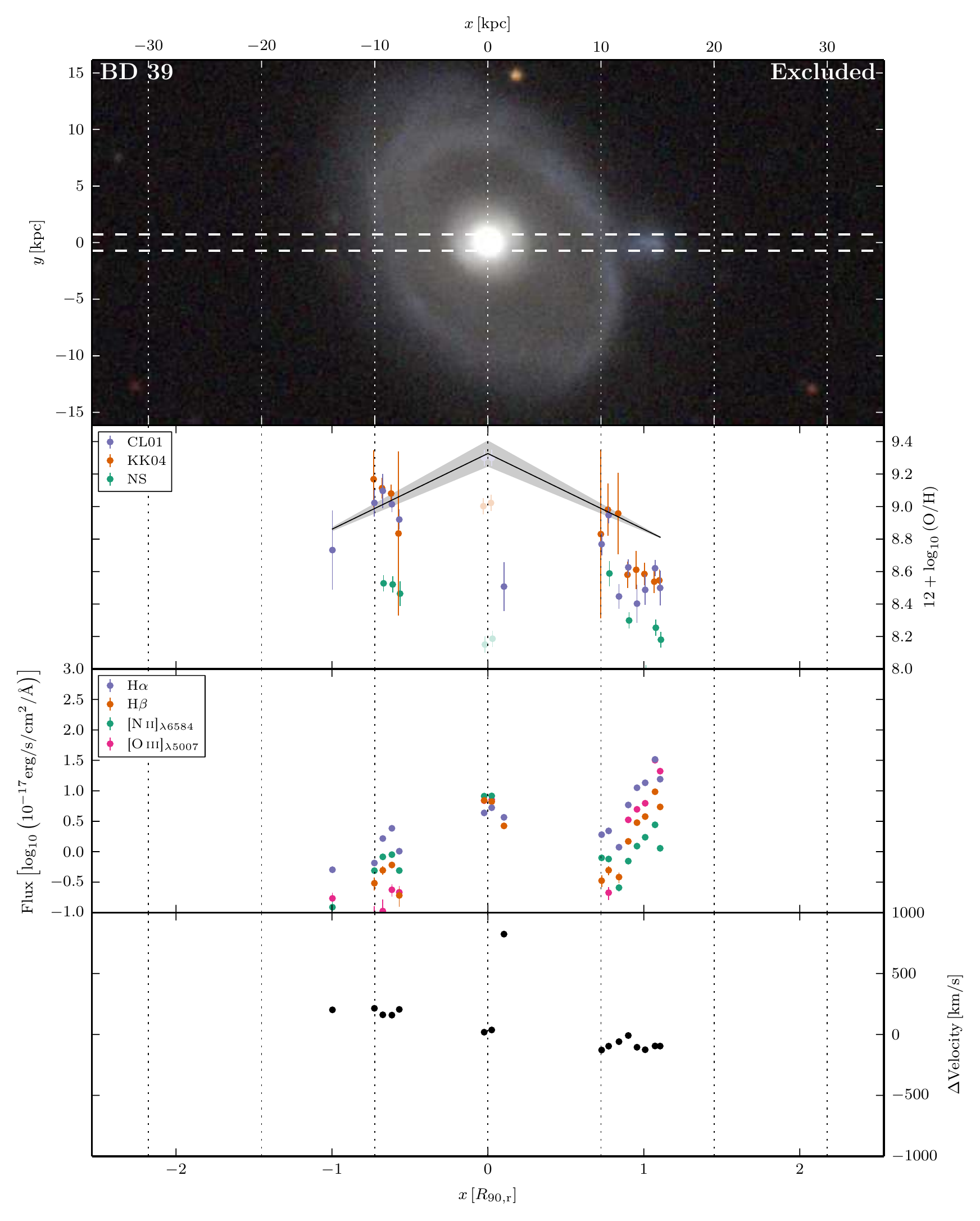}
\caption{Atlas of data for BD 39. See text for details.}
\label{fig:atlas_39}
\end{figure*}

\begin{figure*}
\includegraphics[width=\linewidth]{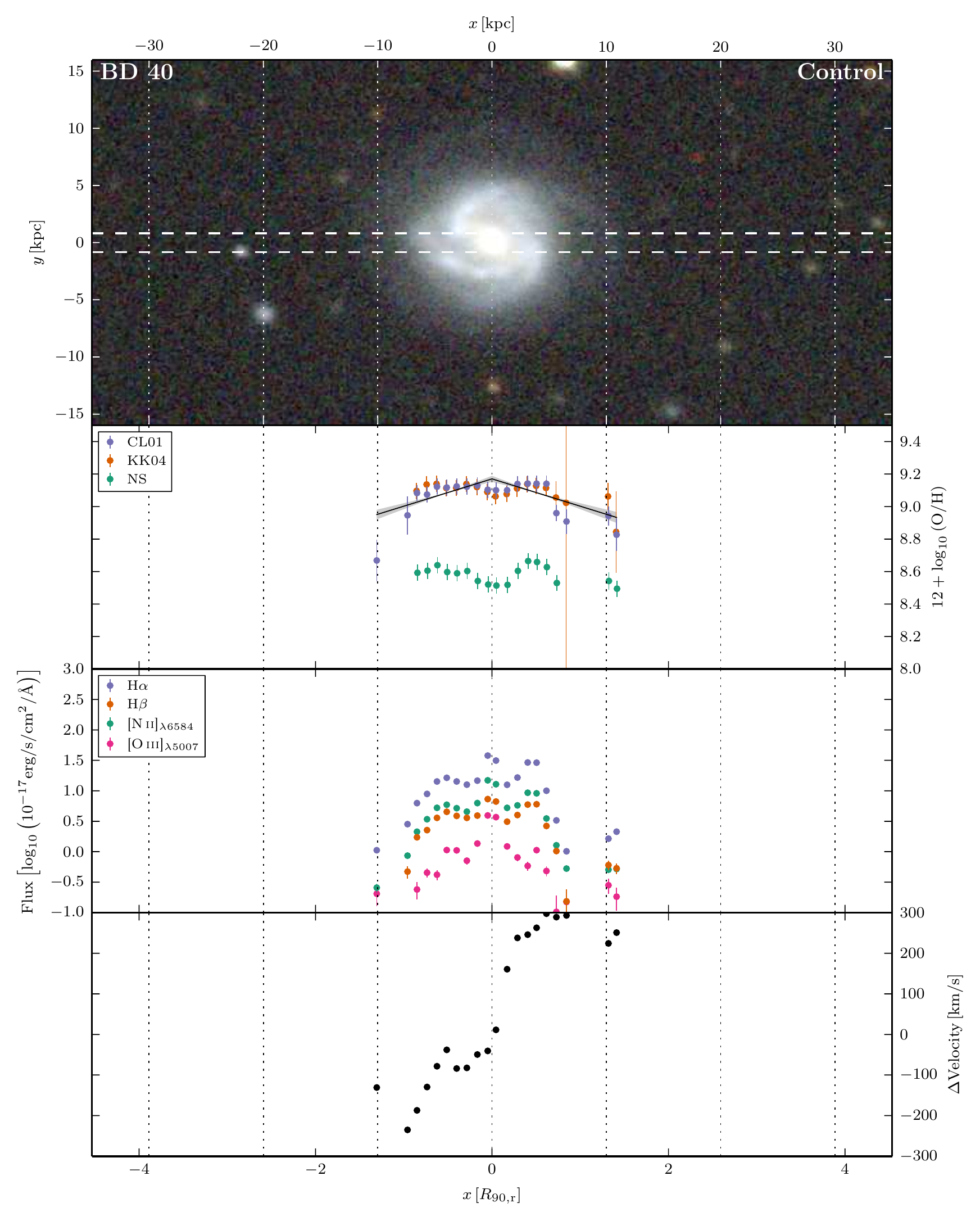}
\caption{Atlas of data for BD 40. See text for details.}
\label{fig:atlas_40}
\end{figure*}

\begin{figure*}
\includegraphics[width=\linewidth]{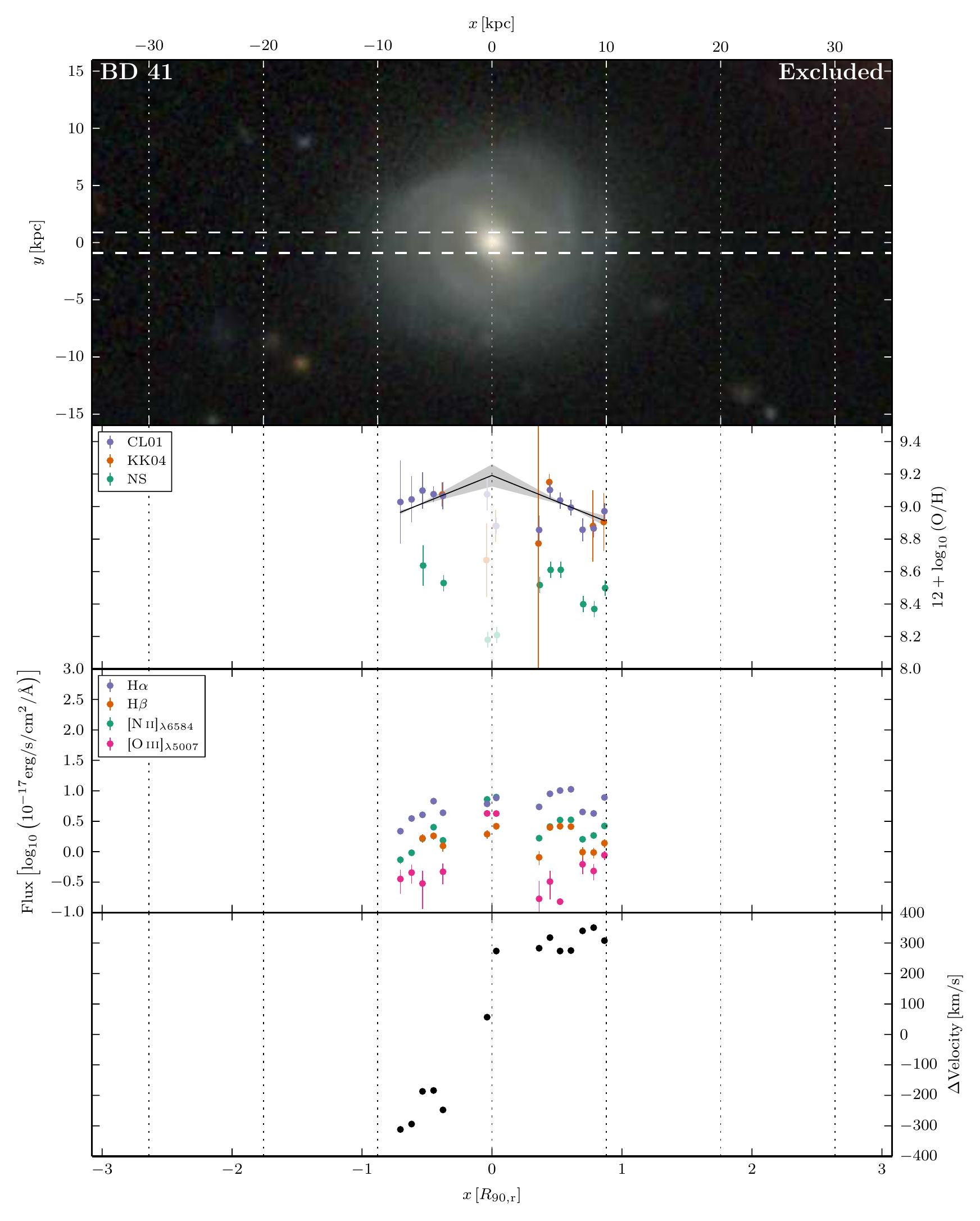}
\caption{Atlas of data for BD 41. See text for details.}
\label{fig:atlas_41}
\end{figure*}

\begin{figure*}
\includegraphics[width=\linewidth]{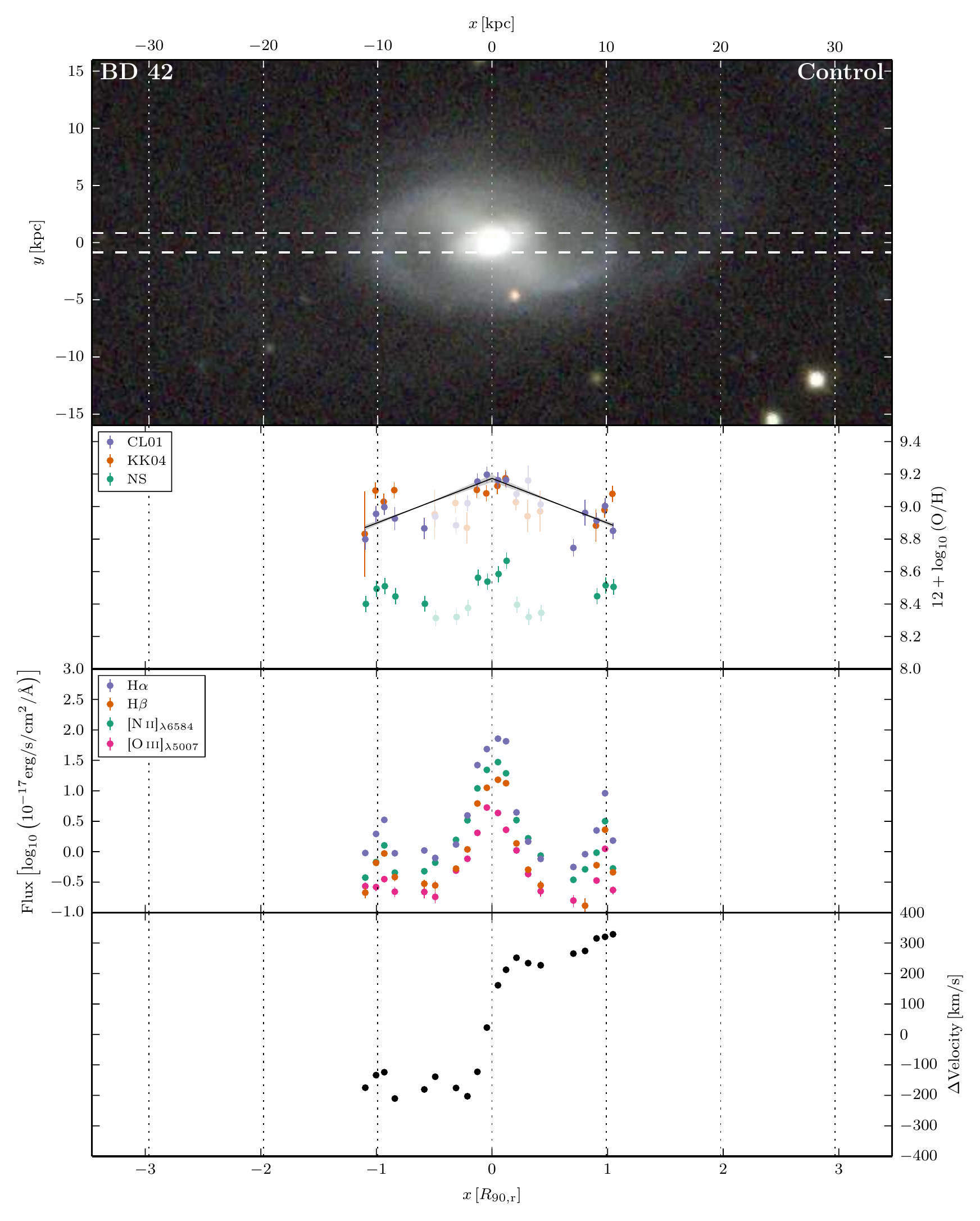}
\caption{Atlas of data for BD 42. See text for details.}
\label{fig:atlas_42}
\end{figure*}

\begin{figure*}
\includegraphics[width=\linewidth]{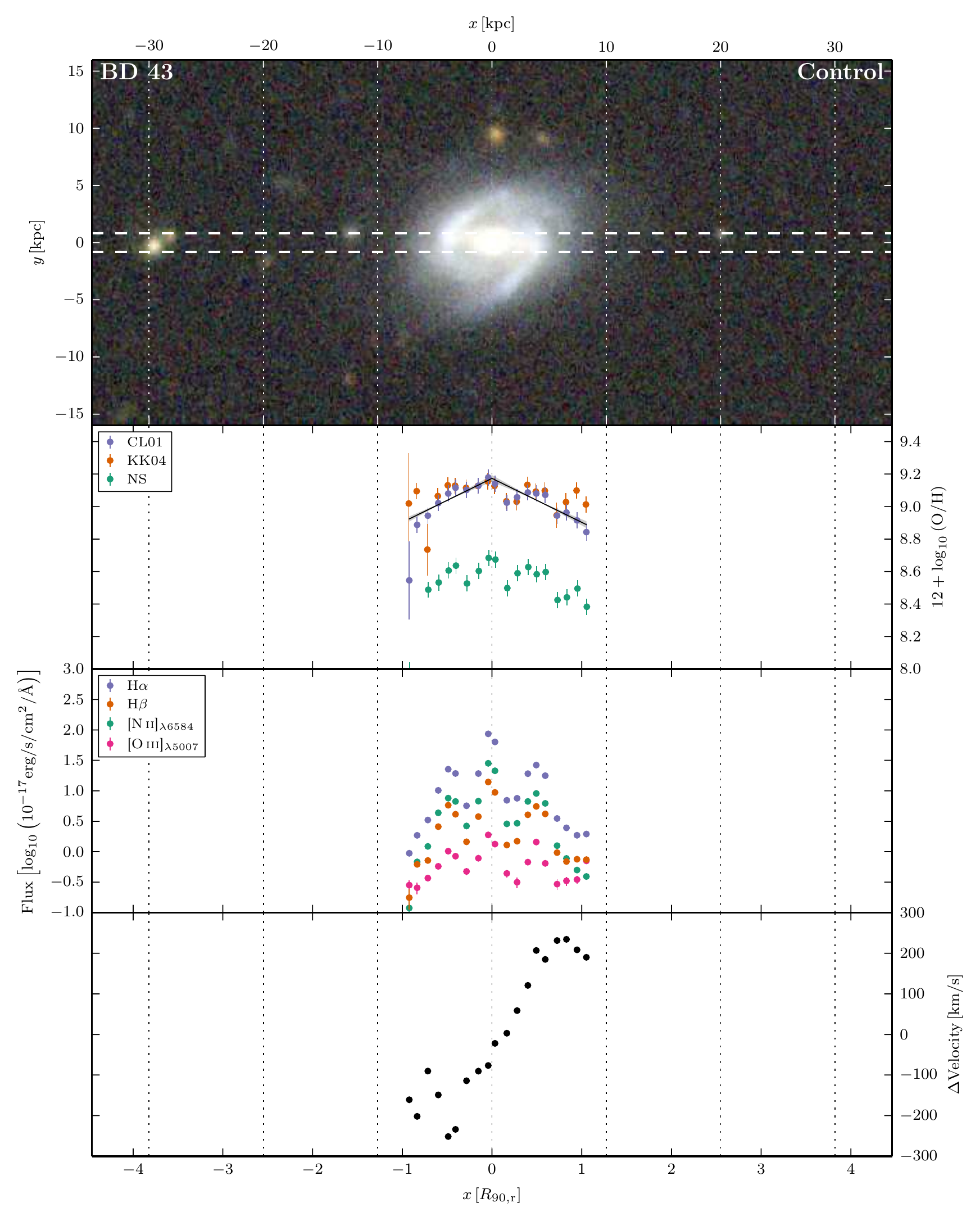}
\caption{Atlas of data for BD 43. See text for details.}
\label{fig:atlas_43}
\end{figure*}

\begin{figure*}
\includegraphics[width=\linewidth]{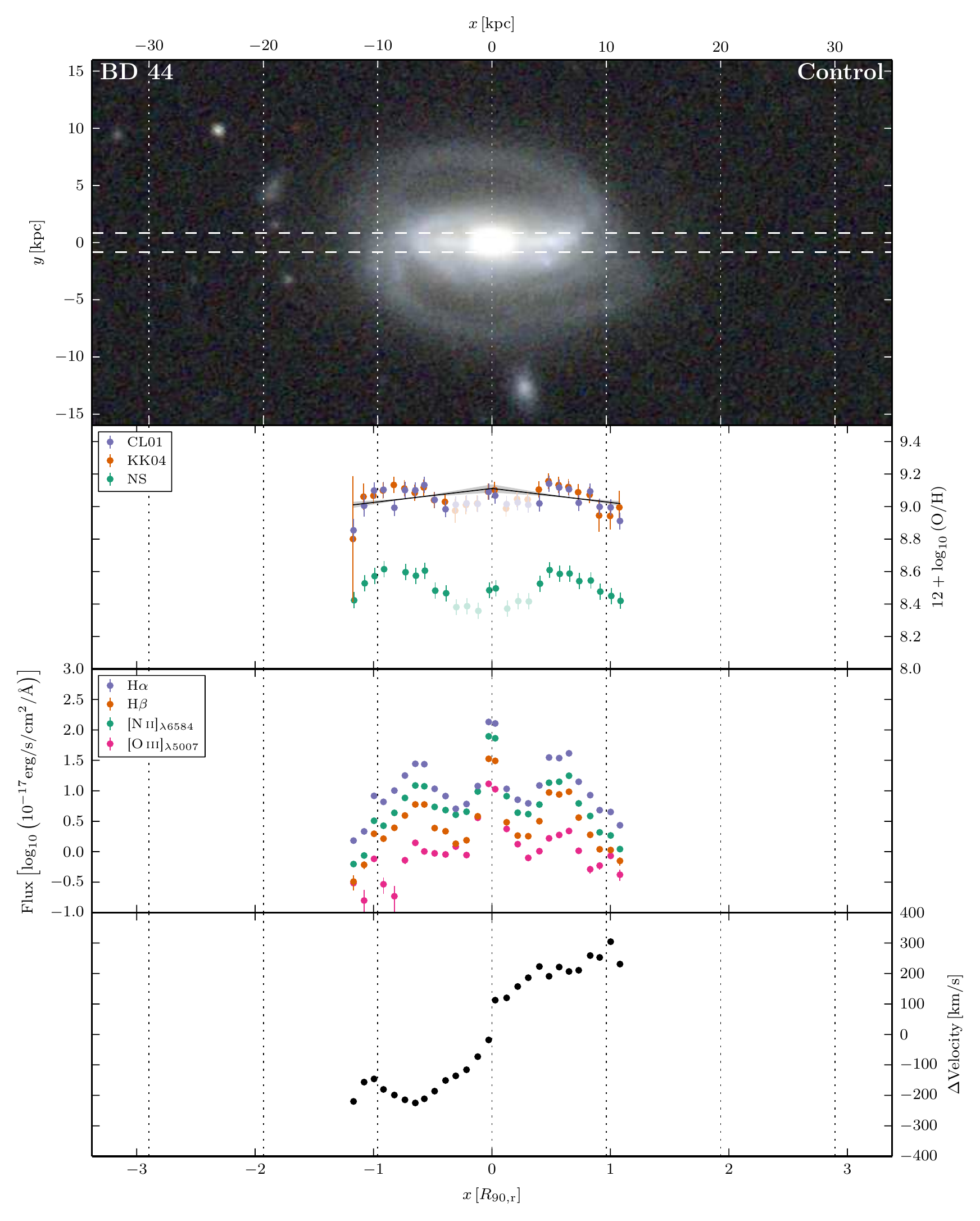}
\caption{Atlas of data for BD 44. See text for details.}
\label{fig:atlas_44}
\end{figure*}

\begin{figure*}
\includegraphics[width=\linewidth]{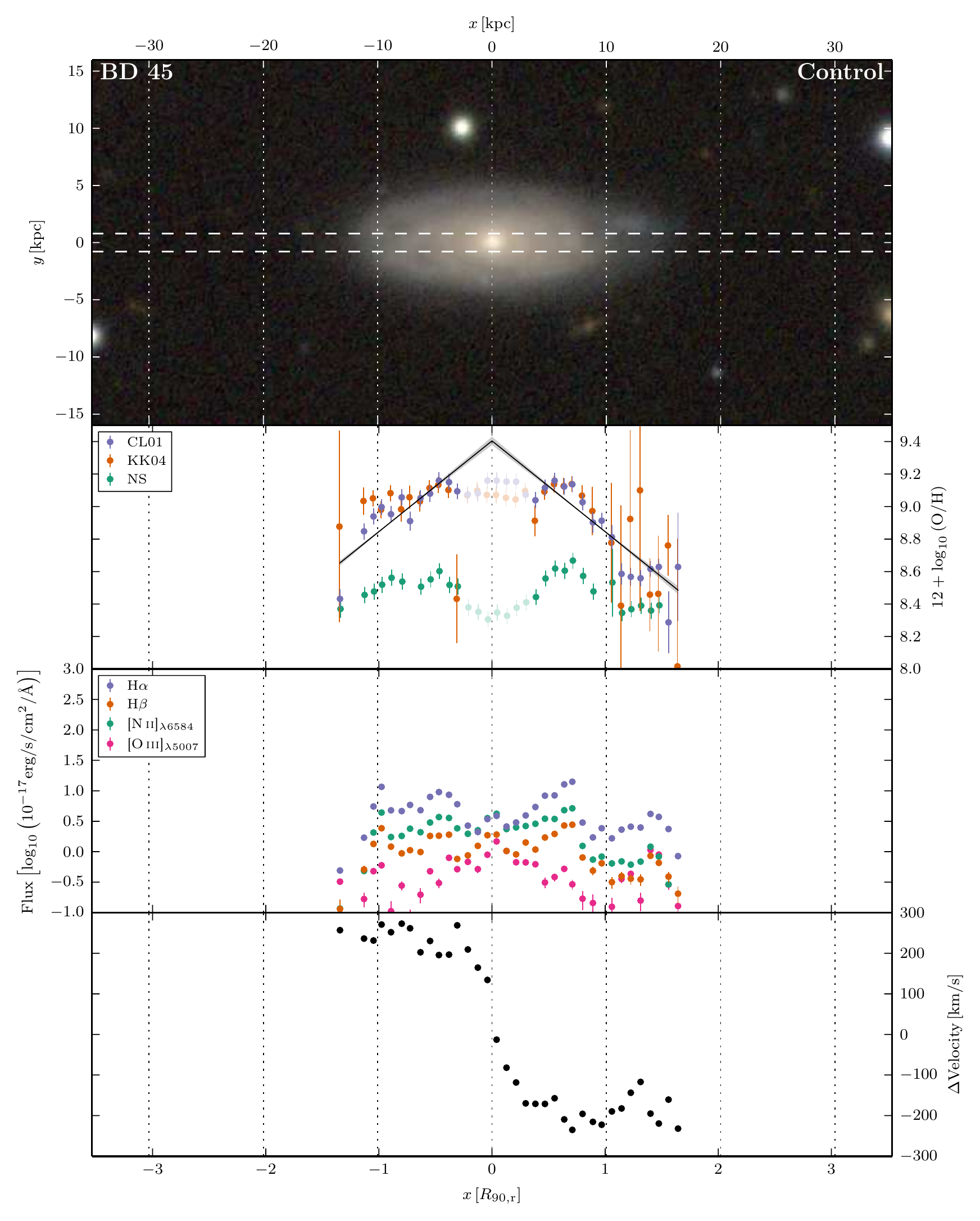}
\caption{Atlas of data for BD 45. See text for details.}
\label{fig:atlas_45}
\end{figure*}

\begin{figure*}
\includegraphics[width=\linewidth]{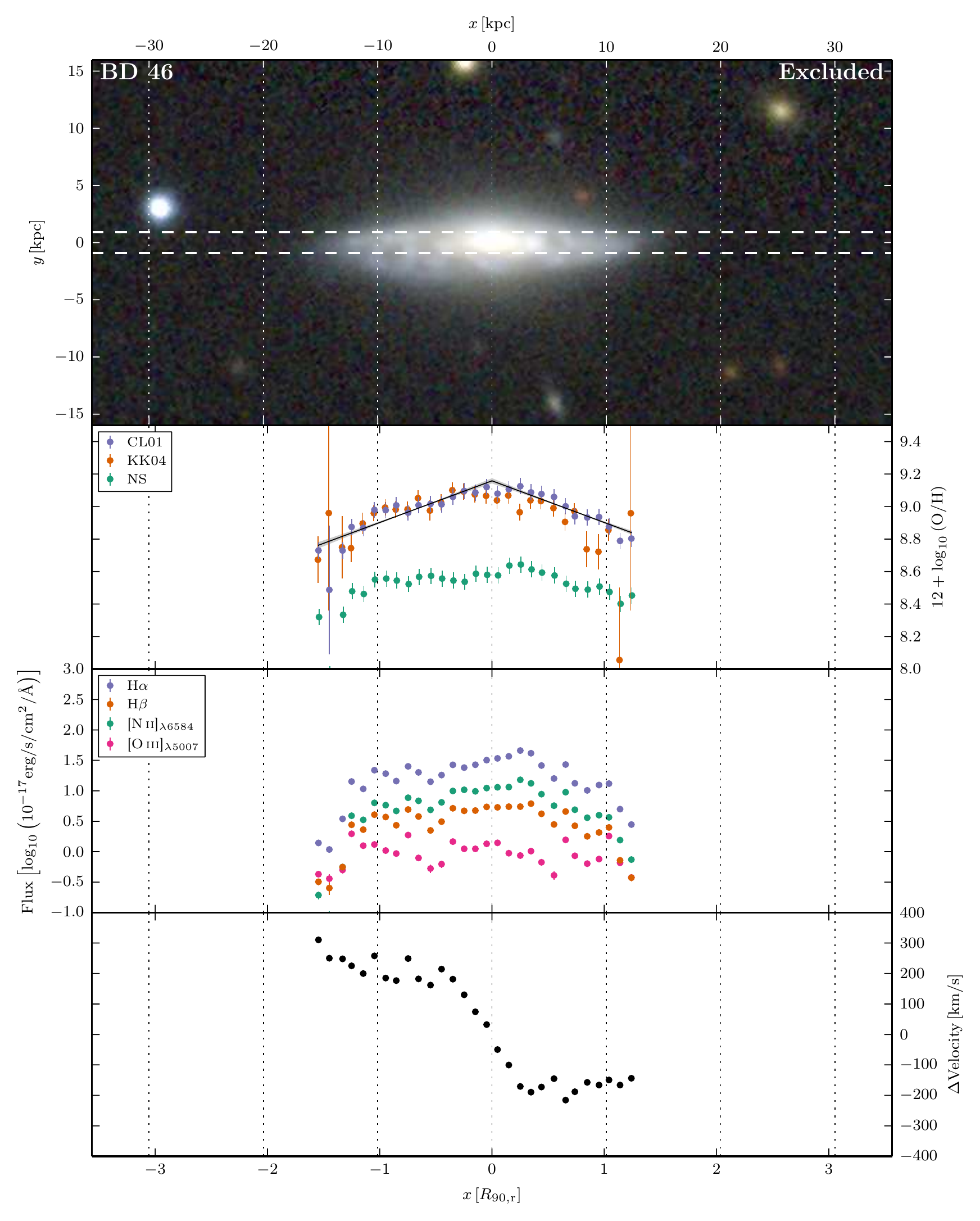}
\caption{Atlas of data for BD 46. See text for details.}
\label{fig:atlas_46}
\end{figure*}

\begin{figure*}
\includegraphics[width=\linewidth]{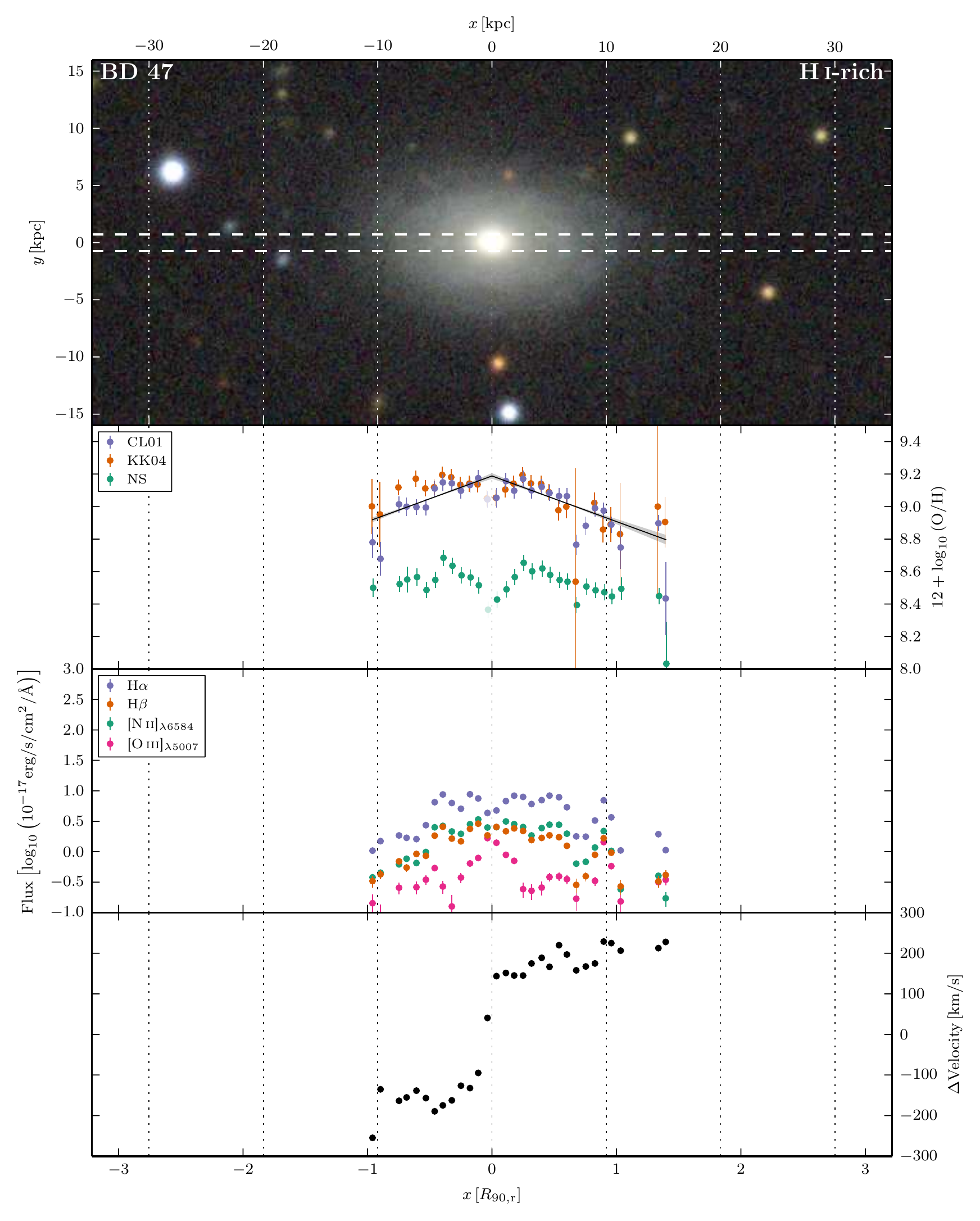}
\caption{Atlas of data for BD 47. See text for details.}
\label{fig:atlas_47}
\end{figure*}

\begin{figure*}
\includegraphics[width=\linewidth]{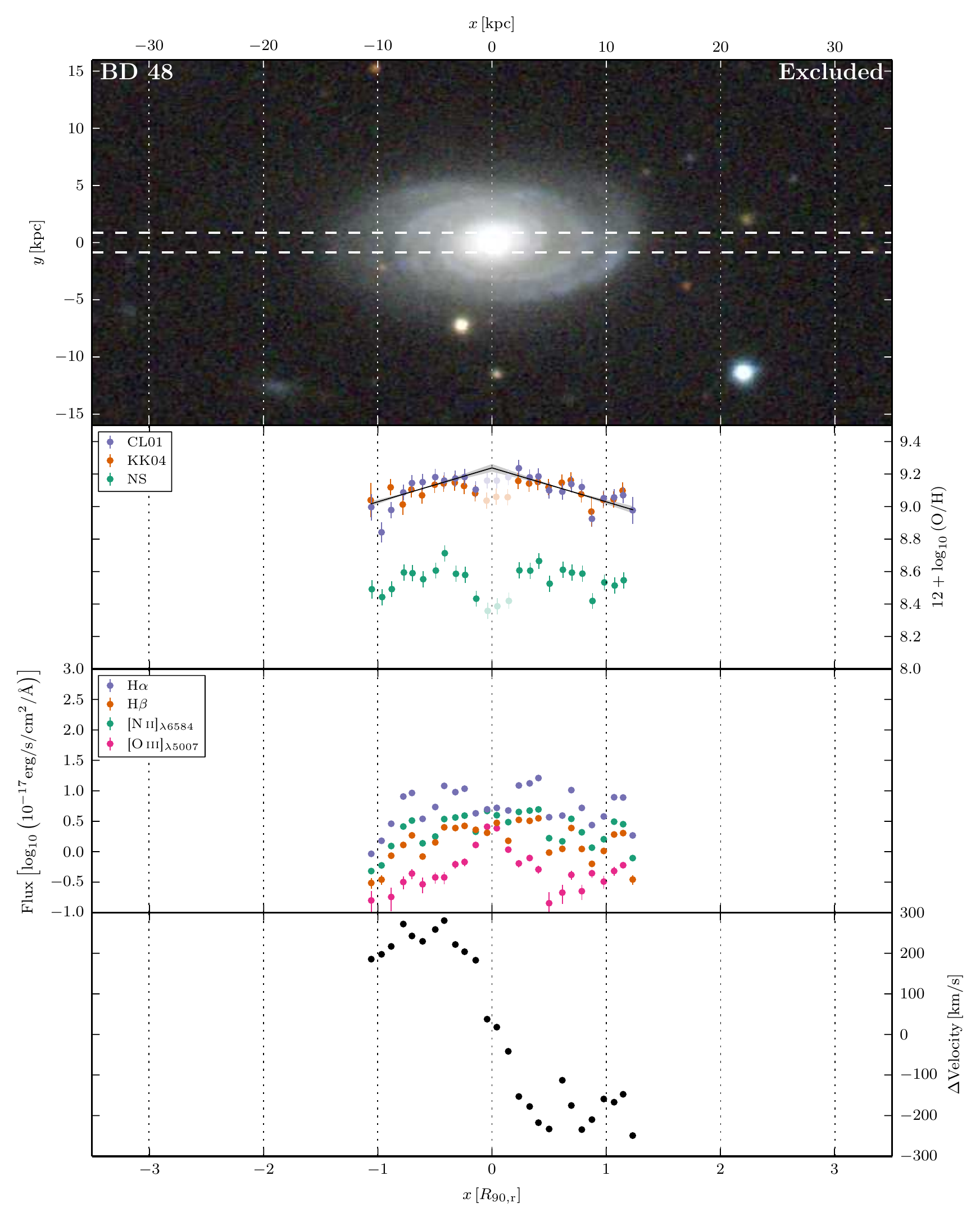}
\caption{Atlas of data for BD 48. See text for details.}
\label{fig:atlas_48}
\end{figure*}

\begin{figure*}
\includegraphics[width=\linewidth]{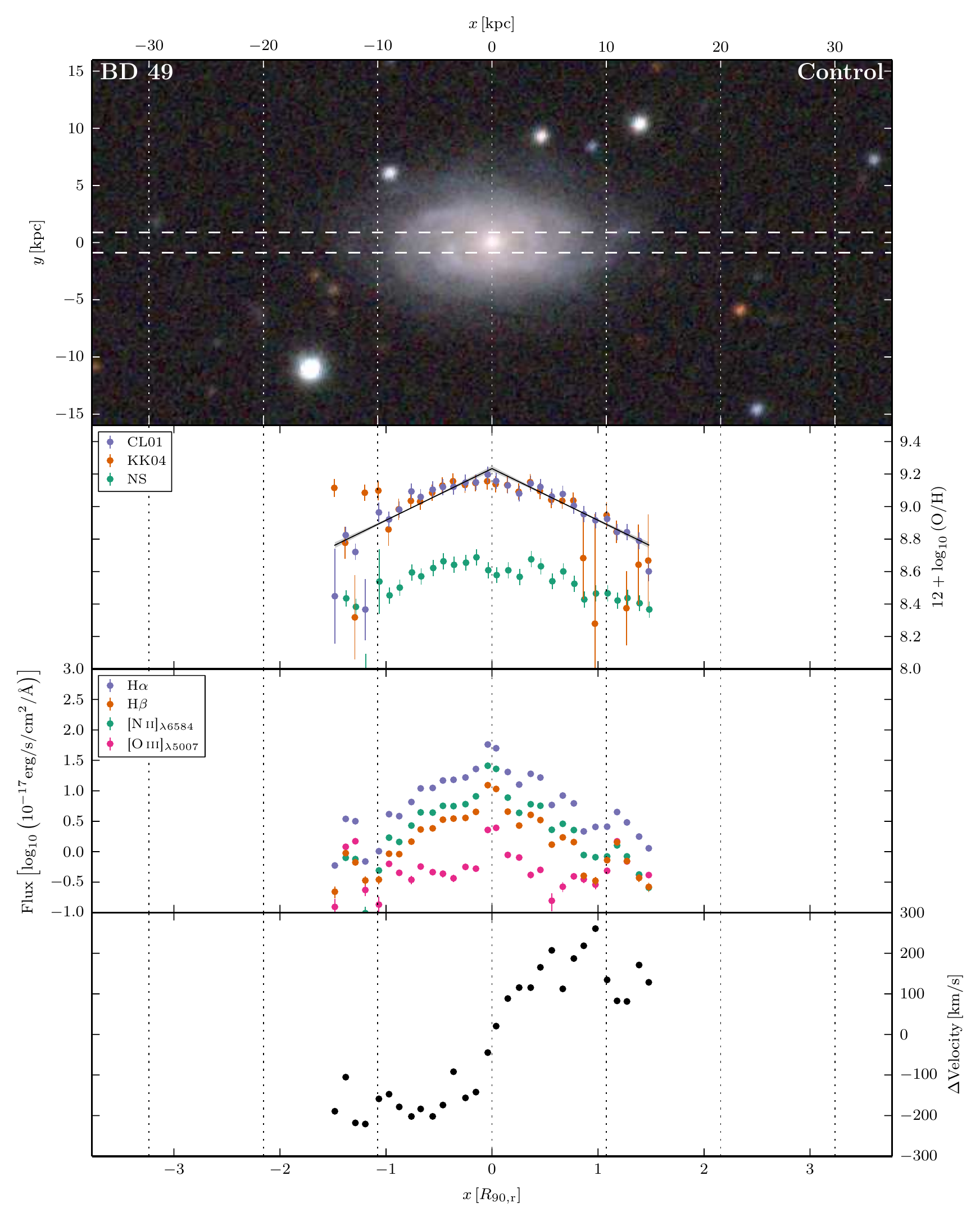}
\caption{Atlas of data for BD 49. See text for details.}
\label{fig:atlas_49}
\end{figure*}

\begin{figure*}
\includegraphics[width=\linewidth]{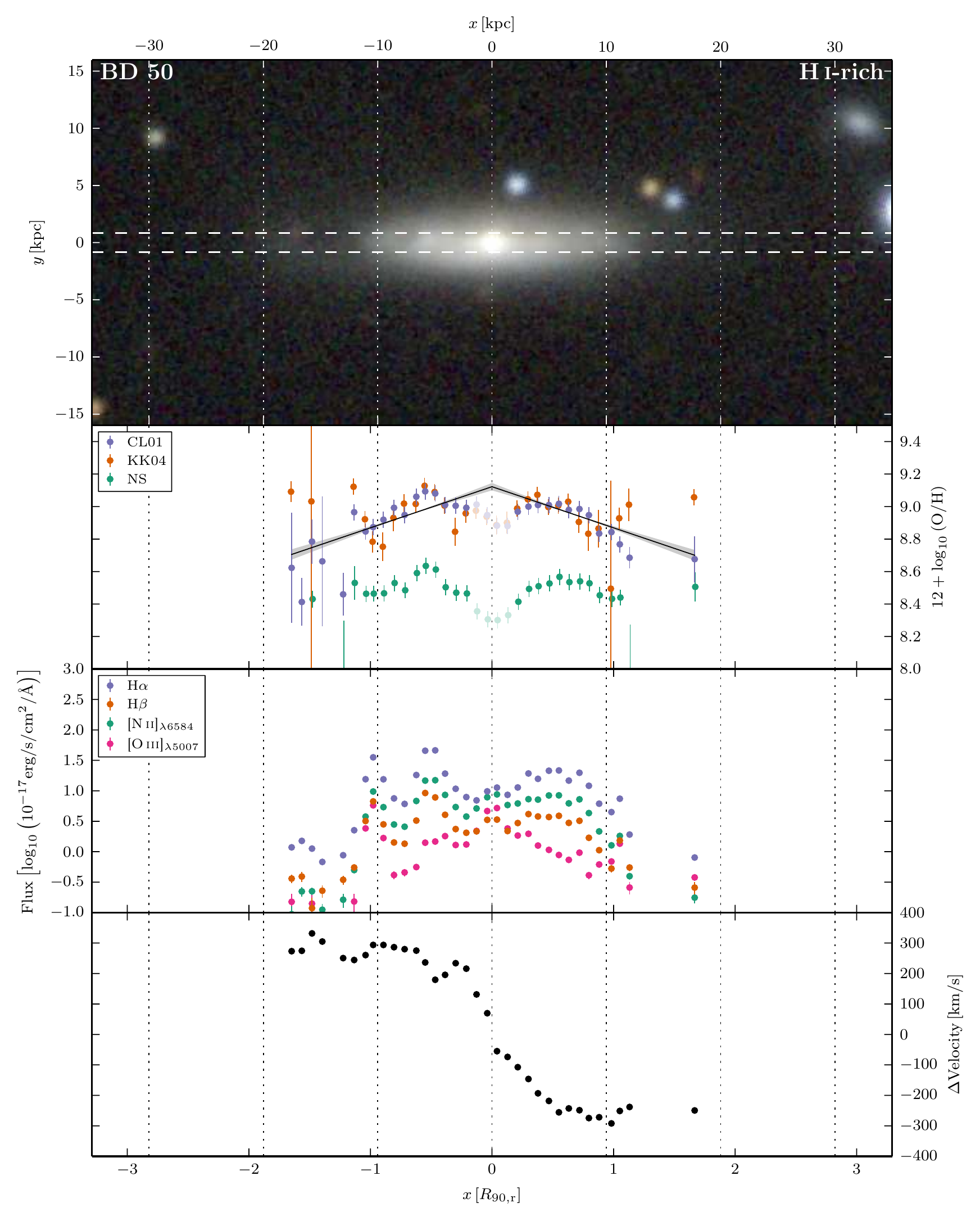}
\caption{Atlas of data for BD 50. See text for details.}
\label{fig:atlas_50}
\end{figure*}

In the online edition of this paper we provide an atlas of the Bluedisk galaxies and observed data in Figs.~\ref{fig:atlas_1}--\ref{fig:atlas_50}.
Each contains four panels, the $x$-axes of which are aligned.

In the first panel shows an SDSS \textit{gri} composite image of the galaxy. Two horizontal dashed lines indicate orientation of the 3\,arcsec spectroscopic slit.
The second panel shows the metallicity profiles as shown in Figs.~\ref{fig:profiles_blue}-\ref{fig:profiles_excl}.
In the third panel we indicate the flux of four emission lines (\Halpha, \Hbeta, \ion{N}{ii} and \ion{O}{iii}).
In the forth panel we plot the inferred velocity shifts of the emission lines from the red-arm of the spectrograph. These velocities have been corrected for the effect of inclination.

\begin{table*}
\begin{minipage}{126mm}
\caption{Positions and orientations of the spectroscopic slit observations: Galaxy identification number, Common name, Right ascension, Declination, and Position angle of slit. Position angle is defined with ($\textrm{North}=0\degr$, $\textrm{East}=90\degr$)}

\begin{tabular}{llllr}
\hline
ID & Name & RA & Dec. & \multicolumn{1}{l}{PA} \\
 & & (degree) & (degree) & \multicolumn{1}{l}{(degree)} \\
\hline
BD 1 & UGC 4283 & 123.591750 & +39.251361 & 133 \\
BD 2 & UGC 4429 & 127.194792 & +40.665889 & 32 \\
BD 3 & MCG+07-18-029 & 129.277167 & +41.456333 & 3 \\
BD 4 & IC 2387 & 129.641667 & +30.798694 & 20 \\
BD 5 & NVSS J084916+360710 & 132.318292 & +36.119806 & 70 \\
BD 6 & NGC 2668 & 132.344042 & +36.710333 & 160 \\
BD 7 & UGC 4615 & 132.356167 & +41.771250 & 15 \\
BD 8 & UGC 4798 & 137.177583 & +44.810667 & 140 \\
BD 9 & NVSS J091458+512138 & 138.743000 & +51.361056 & 18 \\
BD 10 & NGC 2895 & 143.104375 & +57.482889 & 40 \\
BD 11 & NGC 3135 & 152.726583 & +45.950361 & 84 \\
BD 12 & UGC 5534 & 154.042708 & +58.427000 & 98 \\
BD 13 & UGC 6183 & 166.995917 & +35.463278 & 5 \\
BD 14 & UGC 6755 & 176.739750 & +50.702139 & 66 \\
BD 15 & NGC 3897 & 177.247750 & +35.016056 & 135 \\
BD 16 & NVSS J125203+514050 & 193.014792 & +51.680056 & 103 \\
BD 17 & UGC 8205 & 196.806625 & +58.135028 & 107 \\
BD 18 & UGC 8338 & 199.015042 & +35.043528 & 82 \\
BD 19 & NGC 5497 & 212.631833 & +38.893556 & 49 \\
BD 20 & UGC 9429 & 219.499750 & +40.106194 & 112 \\
BD 21 & 2MASX J16073420+3629026 & 241.892583 & +36.484028 & 109 \\
BD 22 & UGC 10523 & 250.793500 & +42.192778 & 173 \\
BD 23 & UGC 10553 & 251.811625 & +40.245083 & 145 \\
BD 24 & MCG+10-24-123 & 259.156042 & +58.411889 & 169 \\
BD 25 & MCG+10-25-046 & 262.156333 & +57.145056 & 92 \\
BD 26 & 2MASX J07274518+4210499 & 111.938042 & +42.180722 & 50 \\
BD 27 & 2MASX J08024061+3431171 & 120.669375 & +34.521444 & 165 \\
BD 28 & MCG+09-14-017 & 123.309125 & +52.458722 & 150 \\
BD 29 & UGC 4427 & 127.312167 & +55.523000 & 180 \\
BD 30 & UGC 4863 & 138.603167 & +40.777917 & 169 \\
BD 31 & MCG+08-17-066 & 139.191875 & +45.813735 & 49 \\
BD 32 & 2MASXI J0918351+321611 & 139.646250 & +32.270000 & 57 \\
BD 33 & UGC 5016 & 141.539292 & +49.310194 & 32 \\
BD 34 & NGC 3013 & 147.539000 & +33.569333 & 70 \\
BD 35 & UGC 5346 & 149.420208 & +45.258667 & 29 \\
BD 36 & 2MASX J09574902+5149162 & 149.454542 & +51.821194 & 107 \\
BD 37 & NGC 3164 & 153.797625 & +56.672083 & 3 \\
BD 38 & 2MASX J10154226+5540030 & 153.926083 & +55.667500 & 26 \\
BD 39 & UGC 5936 & 162.530375 & +36.341833 & 134 \\
BD 40 & MCG+06-25-025 & 168.563542 & +34.154389 & 18 \\
BD 41 & MCG+08-24-070 & 197.879167 & +46.341778 & 68 \\
BD 42 & MCG+08-24-089 & 198.236250 & +47.456667 & 34 \\
BD 43 & MCG+07-28-032 & 203.374583 & +40.529667 & 55 \\
BD 44 & 2MASX J13410027+4225525 & 205.251042 & +42.431417 & 133 \\
BD 45 & IC 1074 & 222.988833 & +51.264889 & 115 \\
BD 46 & 2MFGC 12932 & 241.528958 & +35.981444 & 144 \\
BD 47 & UGC 10312 & 244.382542 & +31.194472 & 112 \\
BD 48 & NGC 6145 & 246.259833 & +40.946639 & 4 \\
BD 49 & 2MASX J17143602+3044011 & 258.650208 & +30.733528 & 160 \\
BD 50 & UGC 10863 & 261.557250 & +62.149472 & 102 \\
\hline
\end{tabular}

\label{tab:slit_descr}

\end{minipage}
\end{table*}

\clearpage
\clearpage

\clearpage

\clearpage

\clearpage

\clearpage

\clearpage

\clearpage

\clearpage

\clearpage

\clearpage

\clearpage

\clearpage

\clearpage

\clearpage

\clearpage

\clearpage

\clearpage

\clearpage

\clearpage

\clearpage

\clearpage

\clearpage

\clearpage

\clearpage

\clearpage

\clearpage

\clearpage

\clearpage

\clearpage

\clearpage

\clearpage

\clearpage

\clearpage

\clearpage

\clearpage

\clearpage

\clearpage

\clearpage

\clearpage

\clearpage

\clearpage

\clearpage

\clearpage

\clearpage

\clearpage

\clearpage

\clearpage

\clearpage

\clearpage

\clearpage

\label{lastpage}

\end{document}